\documentclass[10pt,journal,compsoc]{./IEEEtran}
\ifCLASSOPTIONcompsoc
  \usepackage[nocompress]{cite}
\else
  \usepackage{cite}
\fi

\usepackage{amsfonts}
\usepackage{amsopn}
\usepackage{amssymb}

\usepackage{amsmath}
\usepackage{mathtools}
\usepackage{amsthm}
\usepackage{algpseudocode}

\usepackage{rotating}

\DeclareMathOperator{\adj}{adj}
\DeclareMathOperator{\inc}{inc}
\DeclareMathOperator{\low}{low}
\DeclareMathOperator{\high}{high}

\DeclareMathOperator*{\argmin}{arg\,min}

\newcommand{\true}{\mathrm{true}}
\newcommand{\false}{\mathrm{false}}
\newcommand{\MyIf}{\textbf{if }}
\newcommand{\MyThen}{\textbf{ then}}
\newcommand{\MyAnd}{\textbf{and }}
\newcommand{\MyOr}{\textbf{or }}
\newcommand{\MyBreak}{\textbf{break}}

\theoremstyle{plain}
\newtheorem{algorithm}{Algorithm}

\usepackage{multicol}

\usepackage{makecell}

\usepackage{tikz}
\usetikzlibrary{positioning,matrix,arrows}
\newcommand{\light}[1]{\textcolor{gray}{#1}}
\usepackage{graphicx}
\ifCLASSINFOpdf
  \DeclareGraphicsExtensions{.eps,.pdf,.jpeg,.png}
\else
  \DeclareGraphicsExtensions{.eps}
\fi

\usepackage{url}

\begin{document}
%
\title{Contiguous Graph Partitioning For Optimal Total Or Bottleneck Communication}
%
%
%
%
 
\author{Peter~Ahrens\IEEEcompsocitemizethanks{\IEEEcompsocthanksitem Computer
Science and Artificial Intelligence Laboratory, Massachusetts Institute of
Technology, Cambridge, MA (pahrens@mit.edu). This work was supported by a
Department of Energy Computational Science Graduate Fellowship,
DE-FG02-97ER25308. }}%

%
%

\markboth{
}%
{Ahrens: Contiguous Graph Partitioning For Optimal Total Or Bottleneck Communication}
%



\IEEEtitleabstractindextext{%
\begin{abstract}

Graph partitioning schedules parallel calculations like sparse matrix-vector multiply (SpMV). We consider contiguous partitions, where the $m$ rows (or columns) of a sparse matrix with $N$ nonzeros are split into $K$ parts without reordering. We propose the first near-linear time algorithms for several graph partitioning problems in the contiguous regime.

Traditional objectives such as the simple edge cut, hyperedge cut, or hypergraph connectivity minimize the total cost of all parts under a balance constraint. Our total partitioners use $O(Km + N)$ space. They run in $O((Km\log(m) + N)\log(N))$ time, a significant improvement over prior $O(K(m^2 + N))$ time algorithms due to Kernighan and Grandjean et.  al.

Bottleneck partitioning minimizes the maximum cost of any part. We propose a new bottleneck cost which reflects the sum of communication and computation on each part. Our bottleneck partitioners use linear space.  The exact algorithm runs in linear time when $K^2$ is $O(N^C)$ for $C < 1$. Our $(1 + \epsilon)$-approximate algorithm runs in linear time when $K\log(c_{\high}/(c_{\low}\epsilon))$ is $O(N^C)$ for $C < 1$, where $c_{\high}$ and $c_{\low}$ are upper and lower bounds on the optimal cost. We also propose a simpler $(1 + \epsilon)$-approximate algorithm which runs in a factor of $\log(c_{\high}/(c_{\low}\epsilon))$ from linear time.


We empirically demonstrate that our algorithms efficiently produce high-quality contiguous partitions on a test suite of 42\unskip~test matrices. When $K=8$, our hypergraph connectivity partitioner achieved a speedup of $53\times$\unskip~(mean $15.1\times$\unskip) over prior algorithms. The mean runtime of our bottleneck partitioner was $5.15$\unskip~SpMVs.
\end{abstract}

\begin{IEEEkeywords}
Contiguous, Partitioning, Load Balancing, Communication-Avoiding, Chains-On-Chains, Dominance Counting, Least Weight Subsequence, Quadrangle Inequality, Convex, Concave, Monotonic
\end{IEEEkeywords}}

\maketitle

\IEEEdisplaynontitleabstractindextext

%
\IEEEpeerreviewmaketitle

\IEEEraisesectionheading{\section{Introduction}\label{sec:introduction}}

%
%
%
%
\IEEEPARstart{S}{parse} matrix multiplication is often the most expensive
subroutine in scientific computing applications. When multiplying a dense vector
or matrix, we refer to this kernel as \textbf{SpMV} or \textbf{SpMM}
respectively. Many applications, such as iterative solvers, multiply by the same
sparse matrix repeatedly. Parallelization can increase efficiency, and datasets
can be large enough that distributed memory approaches are necessary.  A common
parallelization strategy is to partition the rows (or similarly, the columns) of
the sparse matrix and corresponding elements of the dense vector(s) into
disjoint parts, each assigned to a separate processor. While there are a myriad
of methods for partitioning the rows of sparse matrices, we focus on the case
where the practitioner does not wish to change the ordering of the rows and the
parts are therefore contiguous.

There are several reasons to prefer contiguous partitioning. The row ordering
may already be carefully optimized for numerical considerations (such as
fill-in), the natural row ordering may already be amenable to partitioning, or
reordering may simply be too costly to implement on the target architecture.
Furthermore, several noncontiguous partitioners (spectral methods or other
heuristics \cite{chan_optimality_1997, aydin_distributed_2019}) work by
producing a one-dimensional embedding of the rows, then using a contiguous
partitioner to subsequently ``round'' the ordering into a partition.  Similarly,
geometric partitioners partition points by splitting a multidimensional
embedding, such as the multidimensional spectral embedding or simply the natural
locations of the corresponding mesh cells. The popular multi-jagged geometric
approach recursively splits one spatial dimension at a time using a
load-balancing contiguous partitioner
\cite{acer_sphynx_2021,deveci_multi-jagged_2016}.  As a final example, acyclic partitioning
is used to schedule execution of directed acyclic computation graphs.  A
prominent approach to acyclic partitioning applies contiguous partitioning to a
topological ordering of the graph that respects data dependence relationships
\cite{herrmann_acyclic_2017,herrmann_multilevel_2019}.

\begin{figure}
    \input{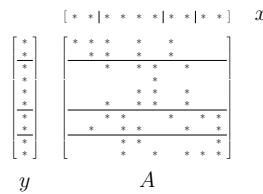}\unskip%
    \begin{minipage}{0.5\linewidth}
    \caption{Our running example matrix, together with an example symmetric
    partition of $x$ and $y$. Nonzeros are denoted with
    $*$.}\label{fig:partition}
    \end{minipage}
\end{figure}

Since noncontiguous partitioning to minimize communication costs is NP-Hard
\cite{karypis_fast_1998, catalyurek_hypergraph-partitioning-based_1999}, one can
view contiguous partitioning as a compromise, where the user is asked to use
domain knowledge or heuristics to produce a good ordering, but the partitioning
can be performed efficiently and optimally.  Our algorithms demonstrate that
contiguous partitioning can be efficient in theory and practice, and support
highly expressive cost models.

We consider two main partitioning objectives, \textbf{total} partitioning and
\textbf{bottleneck} partitioning.  Traditional partitioning objectives minimize
the total communication under the constraint that the work or storage should be
approximately balanced.  However, in many applications of SpMV or SpMM, such as
iterative solvers, all processors wait for the last processor to finish.
Therefore, it may be more accurate to minimize the runtime of the
longest-running (bottleneck) processor. Here, we model  runtime as the combined cost of
communication and computation on a part. This allows us to shift work burdens from
extroverted processors which communicate frequently to introverted processors
which communicate rarely, more efficiently utilizing resources 
\cite{hendrickson_load_2000}.

\subsection{Contributions}

We propose efficient contiguous partitioners for total and bottleneck
objectives. Our algorithms run in linear or near-linear time, representing
asymptotic improvements over the state of the art. We evaluate our algorithms
experimentally\footnote{\url{github.com/peterahrens/ChainPartitioners.jl/tree/2007.16192v4}}
on a suite of \unskip~test matrices.  We show
our algorithms to be efficient in practice. Our partitions are often
$3\times$ the quality of computation-balanced partitions or evenly sized partitions on matrices in
their natural order, spectral order, or Cuthill-McKee order
\cite{cuthill_reducing_1969, pothen_partitioning_1990}.  Our partitioners are
competitive with existing graph partitioning approaches when we account for the
runtime of the partitioners themselves.

We address symmetric and nonsymmetric partitions. When the partition is
nonsymmetric, we address the cases where either the rows or column partition is
considered fixed.  Without loss of generality, we consider partitions of the
rows of an $m \times n$ matrix with $N$ nonzeros to run on $K$ processors.  We
normalize our runtimes to the time it takes to perform a sparse matrix-vector
multiply (SpMV) on the same matrix.

\subsubsection{Contiguous Total Partitioners}
The total contiguous partitioning problem includes the traditional simple edge
cut, hyperedge cut, and hypergraph connectivity objectives. We propose an exact
algorithm for these costs which runs in time \[
    O(Km\log(m)\log(N) + n + N\log(N)),
\] and uses $O(Km + n + N)$ space. This represents a significant improvement
(quadratic to near-linear) over the previous $O(K(m^2 + n + N))$ time algorithms
due to Grandjean et. al. and Kernighan
\cite{grandjean_partitioning_2014, kernighan_optimal_1971}.
When $K=8$ and the work was $10\%$ balanced, our total partitioner for
hypergraph connectivity runs in an average of
$667$\unskip~SpMVs, a
mean speedup of
$15.1\times$\unskip~
over the existing algorithm.

\subsubsection{Contiguous Bottleneck Partitioners}
The bottleneck contiguous partitioning problem includes the traditional
``Chains-On-Chains'' computational load balancing objective. We propose new
cost functions which add hypergraph communication terms to the computational
load. We propose exact and approximate algorithms for our new costs. Our algorithms
use linear space.  Our exact algorithm is parameterized by a constant $H$ and
runs in time \[
    O\left(m + n + H N + K^2\log(m)^2H^2 N^{1/H}\right).
\] This can be made strictly linear when when $K^2$ is $O(N^C)$ for some constant $C <
1$.  Our $(1 + \epsilon)$-approximate algorithm is parameterized by a constant
$H$ and runs in time \[
    O\left(H N + K\log(N)\log\left(\frac{c_{\high}}{c_{\low}\epsilon}\right)H^2 N^{1/H}\right),
\] where $c_{\high}$ and $c_{\low}$ are upper and lower bounds on the optimal
cost. This can be made strictly linear when $K\log(c_{\high}/(c_{\low}\epsilon))$ is
$O(N^C)$ for some constant $C < 1$. Our experiments set $H$ to $3$.  We also
propose a simpler $(1 + \epsilon)$-approximate algorithm that runs in time \[
    O(\log(c_{\high}/(c_{\low}\epsilon)) (m + n + N)),
\] which is only near linear but is faster in practice. If the cost functions
are monotonic increasing and subadditive, we can derive bounds so that
$\log(c_{\high}/(c_{\low}\epsilon))$ is $O(\log(K/\epsilon))$. 
When $K = 8$ and $\epsilon = 10\%$, our exact, approximate, and lazy bottleneck
hypergraph partitioners run in an average of
$18$\unskip,
$17.7$\unskip,
and
$5.15$\unskip~SpMVs
over our symmetric test matrices, making our algorithms practical even in
situations where the matrix is not heavily reused. 

\subsubsection{Core Algorithmic Advances}

Our algorithms perform structured queries of the cost of various potential parts
to determine a good partition.  Our results are enabled by three main
algorithmic advances.

\begin{itemize}
    \item We decompose our novel and traditional partitioning objectives into
    partwise terms, and taxonomize the resulting partwise costs based on the key
    properties of monotonicity, sub- or super-additivity, and convexity or
    concavity. We then reduce the task of computing our costs to two dimensional
    dominance counting (sparse prefix sum) queries.  We generalize an offline
    two-dimensional dominance counting datastructure due to Chazelle to trade
    construction time for query time, allowing our bottleneck partitioners to
    run in linear time \cite{chazelle_functional_1988}.  Our dominance counting
    algorithm may also be of interest for two-dimensional rectilinear load
    balancing problems with linear cost functions \cite{nicol_rectilinear_1994,
    saule_load-balancing_2012, yasar_symmetric_2020}.

    \item Our total partitioner reduces the problem to a sequence of constrained
    convex or concave least-weight-subsequence (LWS) subproblems. We generalize
    Eppstein's constrained convex LWS algorithm from part size constraints to
    arbitrary monotonic constraints \cite{eppstein_sequence_1990}. The
    constrained LWS algorithm (which runs in time $O(m\log(m)\log(N) + n +
    N\log(N))$) may be of independent interest for the purposes of partitioning
    into an arbitrary number of row blocks to minimize total cache misses
    \cite{akbudak_exploiting_2017, abubaker_spatiotemporal_2019}.

    
    \item To minimize the bottleneck cost, we adapt state of the art
    chains-on-chains partitioning algorithms (originally due to Iqbal et. al.
    and Nicol et. al.  \cite{iqbal_approximate_1991, nicol_rectilinear_1994} and
    further improved by Pinar et. al. \cite{pinar_fast_2004}) to support
    arbitrary monotonic increasing and decreasing cost functions. We simplify
    the approximate algorithm by relaxing a binary search to a linear search
    and using an online dominance counter.
\end{itemize}



\section{Background}
\label{sec:background}

We index starting from 1. Consider the $m \times n$ \textbf{matrix} $A$. We
refer to the entry in the $i^{th}$ row and $j^{th}$ column of $A$ as
$a_{ij}$. A matrix is called \textbf{sparse} if most of its entries are zero.
It is more efficient to store sparse matrices in \textbf{compressed} formats
that only store the nonzeros. Let $N$ be the number of nonzeros in $A$.
Without loss of generality, as transposition and format conversion usually
run in linear time, we consider the \textbf{Compressed Sparse Row (CSR)}
format. CSR stores a sorted vector of nonzero column coordinates in each row,
and a corresponding vector of values. This is accomplished with
three arrays $pos$, $idx$, and $val$ of length $m + 1$, $N$, and $N$
respectively. Locations $pos_j$ to $pos_{j + 1} - 1$ of $idx$ hold the sorted
column indices $i$ such that $a_{ij} \neq 0$, and the corresponding entries
of $val$ hold the nonzero values.

A \textbf{$K$-partition} $\Pi$ of the rows of $A$, assigns each row $i$ to a
single part $\pi_k$. Any partition $\Pi$ on $n$ elements must satisfy
coverage and disjointness constraints.
\begin{equation}\label{eq:partition}
    \begin{gathered}
        \bigcup_{k} \pi_k = \{1, ..., n\},\quad \quad 
        \forall k \neq k', \pi_{k} \cap \pi_{k'} = \emptyset.\\
    \end{gathered}
\end{equation}

A partition is \textbf{contiguous} when adjacent elements are assigned to the
same part. Formally, $\Pi$ is contiguous when,
\begin{equation}\label{eq:contiguous}
    \forall k < k', \forall (i, i') \in \pi_k \times \pi_k', i < i'.
\end{equation}
We can describe a contiguous partition $\Pi$ using its \textbf{split points}
$S$, where $i \in \pi_k$ for $S_k \leq i < S_{k + 1}$. Thus, $\pi_k =
S_k{:}(S_{k + 1} - 1)$.

We consider \textbf{graphs} and \textbf{hypergraphs} $G = (V, E)$ on $m$
vertices and $n$ edges. \textbf{Edges} can be thought of as connecting sets
of \textbf{vertices}. We say that a vertex $i$ is \textbf{incident} to edge
$j$ if $i \in e_j$. Note that $i \in e_j$ if and only if $j \in v_i$. Graphs
can be thought of as a specialization of hypergraphs where each edge is
incident to exactly two vertices. The \textbf{degree} of a vertex is the
number of incident edges, and the \textbf{degree} of an edge is the number of
incident vertices.

We distinguish between symmetric and nonsymmetric partitioning regimes. A
\textbf{symmetric partitioning} regime assumes $A$ to be square and that the
input and output vectors will use the same partition $\Pi$. Note that we can
use symmetric partitioning on square, asymmetric matrices. The
\textbf{nonsymmetric partitioning} regime makes no assumption on the size of
$A$, and allows the input vector (columns) to be partitioned according to
some partition $\Phi$ which may differ from the output vector (row) partition
$\Pi$. Since our load balancing algorithms are designed to optimize only one
partition at a time, we alternate between optimizing $\Pi$ or $\Phi$,
considering the other partition to be fixed. When $\Phi$ is considered fixed
and our goal is only to find $\Pi$, we refer to this more restrictive problem
as \textbf{primary alternate partitioning}. When $\Pi$ is considered fixed
and our goal is only to find $\Phi$, we refer to this problem as
\textbf{secondary alternate partitioning}. Alternating partitioning has been
examined as a subroutine in heuristic solutions to nonsymmetric partitioning
regimes, where the heuristic alternates between improving the row partition
and the column partition, iteratively converging to a local optimum
\cite{kolda_partitioning_1998, hendrickson_partitioning_2000}. Similar
alternating approaches have been used for the related two-dimensional
rectilinear partitioning regimes \cite{nicol_rectilinear_1994,
yasar_symmetric_2020}.

\section{Problem Statement}

We seek a contiguous partition $\Pi$ minimizing either
\begin{equation}\label{eq:objective}
    \sum\limits_k f_k(\pi) \text{ or } \max\limits_k f_k(\pi),
\end{equation}
where $f_k$ is a cost function mapping parts (sets of vertices) to the
extended reals (the real numbers and $\pm\infty$). We refer to these problems
as \textbf{total} or \textbf{bottleneck} partitioning, respectively. The
number of parts, $K$, may be fixed or allowed to vary. Because we consider
only contiguous parts, we may use $f_k(i, j)$ as a shorthand for $f_k(i:j -
1)$.

The cost function, $f$, might enjoy certain properties which allow us to make
algorithmic optimizations. We say that $f$ is \textbf{monotonic increasing}
if for any $i' \leq i \leq j \leq j'$,
\begin{equation}\label{eq:monotonic}
    f(i,j) \leq f(i', j'),
\end{equation}
and $f$ is \textbf{monotonic decreasing} when the inequality is reversed.
We say that $f$ is \textbf{superadditive}
if for any $i \leq l \leq j$,
\begin{equation}\label{eq:subadditivity}
    f(i, l) + f(l, j) \leq f(i, j),
\end{equation}
and $f$ is \textbf{subadditive} when the inequality is reversed.
Finally, we say that $f$ is \textbf{concave} when $f$ satisfies the
quadrangle inequality for any $i \leq i' \leq j \leq j'$,
\begin{equation}\label{eq:concavity}
    f(i, j) + f(i', j') \leq f(i, j') + f(i', j),
\end{equation}
and $f$ is \textbf{convex} when the inequality is reversed. Concavity and
convexity imply superadditivity and subadditivity, respectively. When the
inequality in \eqref{eq:subadditivity} becomes an equality, we say $f$ is
\textbf{additive}. Additive functions are subadditive, superadditive, convex,
and concave.

Note that negating an increasing, subadditive, or concave cost function will
produce a decreasing, superadditive, or convex cost function, respectively.
All of our properties are preserved under addition and scaling by positive
constants. Subadditivity is preserved under $\max$, superadditivity is
preserved under $\min$. Monotonicity is preserved under both $\max$ and
$\min$, but neither concavity nor convexity are preserved under either.

Partitioners often use constraints to reflect storage limits or to balance
work. We can use thresholds in our cost functions to reflect such
constraints. If our weight limit is $w_{\max}$, then we define
\[
    \tau_{w_{\max}}(w) = \begin{cases}0\text{ if }w \leq w_{\max}\\1\text{ otherwise}\end{cases}
\]

If $w$ is increasing, then $\tau$ is increasing, superadditive, and concave.
If $w$ is decreasing, then $\tau$ is decreasing, subadditive, and concave.
A simple way to add a weight limit on $w$ to a cost function $f$ would be to
form $f + \tau_{w_{\max}}(w) \cdot \infty$ (we assume that $0$ times $\infty$
is $0$). We can also use thresholds to capture discontinuous phenomena like cache
effects. For example, one might use thresholds to switch to more expensive
cost models when the input or output vectors no longer fit in cache.

\subsection{Traditional Partitioning Objectives}

Traditional graph and hypergraph partitioning problems map rows of $A$ to
vertices of a graph or hypergraph, and use edges to represent communication
costs. The parts are weighted to represent storage or computation costs, and
the objective is to minimize the total communication between a fixed number
of parts $K$ under an $\epsilon$-approximate weight balance constraint like
\begin{equation}\label{eq:balance}
        \forall k, \sum\limits_{i \in \pi_k} |v_i| < \frac{1 + \epsilon}{K}\sum\limits_{i}|v_i|,
\end{equation}
which we might represent as $\tau_{(1 + \epsilon) \cdot N/K}(\sum_{i \in \pi_k}|v_i|)$.

Graph models for symmetric partitioning of symmetric matrices typically use
the \textbf{adjacency representation} $\adj(A)$ of a sparse matrix $A$. If $G
= \adj(A)$, an edge exists between vertices $i$ and $i'$ if and only if
$a_{ii'} \neq 0$. Thus, cut edges (edges whose vertices lie in different
parts) require communication of their corresponding columns. However, this
model overcounts communication costs in the event that multiple cut edges
correspond to the same column, since each column only represents one entry of
$y$ which needs to be sent. Reductions to bipartite graphs are used to extend
this model to the possibly rectangular, nonsymmetric case
\cite{hendrickson_partitioning_2000}. Graph partitioning under the ``edge cut''
seeks to optimize
\begin{equation} \label{eq:edgecut}
    \argmin_\Pi |\{E \cap (\pi_{k} \times \pi_{k'}) : k \neq k'\}|,
\end{equation}
under balance constraints \eqref{eq:balance}, where $G = (V, E) = \adj(A)$.
We can rearrange the edge cut \eqref{eq:edgecut} objective in the form of
Problem \eqref{eq:objective} by subtracting the constant $|E|$,
\[
|\{E \cap (\pi_{k} \times \pi_{k'}) : k \neq k'\}| - |E| = \sum\nolimits_{k} \sum\nolimits_{i \in \pi_k} -|v_i\cap \pi_{k}|,
\]
where $\sum\nolimits_{i \in \pi_k} -|v_i\cap \pi_{k}|$ is monotonic decreasing and convex. 

Inaccuracies in the graph model led to the development of the hypergraph
model of communication. Here we use the \textbf{incidence representation} of
a hypergraph, $\inc(A)$. If $G = \inc(A)$, edges correspond to columns in the
matrix, vertices correspond to rows, and we connect the edge $e_j$ to the
vertex $v_i$ when $a_{ij} \neq 0$. Thus, if there is some edge $e_j$ which is
not cut in a row partition $\Pi$, all incident vertices to $e_j$ must belong
to the same part $\pi_k$, and we can avoid communicating the input $j$ by
assigning it to processor $k$ in our column partition $\Phi$. In this way, we
can construct a secondary column partition $\Phi$ such that the number of cut
edges in a row partition $\Pi$ corresponds exactly to the number of entries
of $y$ that must be communicated, and the number of times an edge is cut is
one more than the number of processors which need to receive that entry of
$y$, since one of these processors has that entry of $y$ stored locally. By
filling in the diagonal of the matrix, this correspondence still holds when
the partition is symmetric
\cite{catalyurek_hypergraph-partitioning-based_1999}. To formalize these cost
functions on a partition $\Pi$, we define $\lambda_j(A,\Pi)$ as the set of
row parts which contain nonzeros in the $j^{th}$ column. Tersely, $\lambda_j
= \{k : i \in \pi_k, a_{ij} \neq 0\}$. Hypergraph partitioning with the
``hyperedge cut'' metric seeks to optimize
\begin{equation}\label{eq:hyperedgecut}
    \argmin_\Pi |\{j : |\lambda_j| > 1\}|,
\end{equation}
and with the ``($\lambda$ - 1) cut'' metric seeks to optimize
\begin{equation}\label{eq:lambdaminusonecut}
    \argmin_\Pi \sum\limits_{e_j \in E}|\lambda_j| - 1,
\end{equation}
under balance constraints \eqref{eq:balance}, where $G = (V, E) = \inc(A)$.
Rearranging again by subtracting $|E|$, we find that minimizing the
``hyperedge cut'' \eqref{eq:hyperedgecut} is equivalent to a partwise
objective in the form of Problem \eqref{eq:objective},
\[
|\{e_j \in E : |\lambda_j| > 1\}| - |E| = \sum\limits_k -|\{e_j \subseteq \pi_k\}|,
\]
because edges entirely contained within a part are unique to that part.
The cost $-|\{e_j \subseteq \pi_k\}|$ is monotonic decreasing and convex.
If we instead add $|E|$ to the ``($\lambda$ - 1) cut''
\eqref{eq:lambdaminusonecut}, we obtain
\[
|E| + \sum\nolimits_{e_j \in E}|\lambda_j| - 1 = \sum\nolimits_{e_j \in E}|\lambda_j| = \sum\nolimits_{k}\left|\bigcup\nolimits_{i \in \pi_k} v_i\right|,
\]
where the last equality comes from counting incidences over parts rather than over edges.
The cost $|\bigcup_{i \in \pi_k} v_i|$ is monotonic increasing and convex.

If we block the rows of a matrix on a serial processor for caching purposes,
$|\bigcup_{i \in \pi_k} v_i|$ also corresponds to the
number of cache misses on $x$ when the row block's portion of $y$ fits in
cache. While hypergraph partitioning is usually applied to distributed
settings with a fixed number of processors $K$ and a balance constraint on
nonzeros, this setting would allow variable $K$ and limit the number of rows
in each part (since each row corresponds to a cached entry of $y$.)
\cite{akbudak_exploiting_2017, abubaker_spatiotemporal_2019}

The hypergraph model better captures communication in our kernel,
but heuristics for noncontiguous hypergraph partitioning problems can be
expensive. For example, state-of-the-art multilevel hypergraph partitioners
recursively merge pairs of similar vertices at each level. Finding similar
pairs usually takes quadratic time
\cite{catalyurek_hypergraph-partitioning-based_1999,devine_parallel_2006}.

Both the graph and hypergraph formulations minimize total communication
subject to a work or storage balance constraint, but it has been observed
that the runtime depends more on the processor with the most work and
communication, rather than the sum of all communication
\cite{hendrickson_partitioning_2000, bader_parallel_2013}. Several approaches
seek to use a two-phase approach to nonsymmetric partitioning where the
matrix is first partitioned to minimize total communication volume, then the
partition is refined to balance the communication volume (and other metrics)
across processors \cite{pinar_partitioning_2001, ucar_encapsulating_2004,
akbudak_partitioning_2018}. Other approaches modify traditional hypergraph
partitioning techniques to incorporate communication balance (and other
metrics) in a single phase of partitioning \cite{deveci_hypergraph_2015,
acer_improving_2016, karsavuran_reduce_2020}. Given a row partition,
Bisseling et. al. consider column partitioning to balance
communication (the secondary alternate partitioning regime)
\cite{bisseling_communication_2005}.

While noncontiguous primary and secondary graph and hypergraph partitioning
are usually NP-Hard \cite{garey_simplified_1976, lengauer_combinatorial_1990,
bisseling_communication_2005}, the contiguous case is much more forgiving.
Kernighan proposed a dynamic programming algorithm which solves the
contiguous graph partitioning problem to optimality in quadratic time, and
this result was extended to hypergraph partitioning by Grandjean and U\c{c}ar
\cite{kernighan_optimal_1971, grandjean_optimal_2012}.

Simplifications to the cost function lead to faster algorithms. If we ignore
communication and minimize only the maximum amount of work in a noncontiguous
partition (where the cost of a part is modeled as the sum of some per-row
cost model), our partitioning problem becomes equivalent to bin-packing,
which is approximable using straightforward heuristics that can be made to
run in log-linear time \cite{dosa_tight_2007}. The ``Chains-On-Chains''
partitioning problem also optimizes a linear model of work, but further
constrains the partition to be contiguous. Chains-On-Chains has a rich
history of study, we refer to \cite{pinar_fast_2004} for a summary. These
problems are often described as ``load balancing'' rather than
``partitioning.'' In Chains-On-Chains partitioning, the work of a part is
typically modeled as directly proportional to the number of nonzeros in that
part. Formally, Chains-On-Chains seeks the best contiguous partition $\Pi$ under
\begin{equation}\label{eq:chainsonchains}
    \argmin_\Pi \max\limits_{k} \sum\limits_{i \in \pi_k} |v_i|
\end{equation}
where $G = (V, E) = \adj(A)$. This objective is already in the form of
Problem \ref{eq:objective}, and the cost model is monotonic increasing and
additive.

This cost model is easily computable, and algorithms for Chains-On-Chains
partitioning can run in sublinear time. Nicol observed that the work terms
for the rows in Chains-On-Chains partitioners can each be augmented by a
constant to reflect the cost of vector operations in iterative linear solvers
\cite{nicol_improved_1991}. Local refinements to contiguous partitions have
been proposed to take communication factors into account, but our work
proposes the first globally optimal linear-time contiguous partitioner with a
communication-aware cost model \cite{ziantz_run-time_1994,
aydin_distributed_2019}.

\subsection{Novel Objectives}\label{sec:novel}

As we will soon see, the techniques used to solve Chains-On-Chains problems can
be applied to any monotonic bottleneck objectives in contiguous settings.
\cite{pinar_fast_2004, ashraf_iqbal_efficient_1995, alpert_splitting_1997}.  In
fact, since the runtime of parallel programs depends on the longest-running
processor, bottleneck partitioning may be more accurate than total partitioning
in parallel settings.  Therefore, we propose several monotonic cost models for
use in bottleneck partitioning.

Most applications of SpMV or SpMM don't have a synchronization point between
communication and computation, so we model the runtime of each processor as
the sum of work and communication. 
The inner loop of the conjugate gradient method, for example, has two
synchronization points where all processors wait for the results of a
global reduction (a dot product) \cite[Chapter 6]{saad_iterative_2003}. This
separates the inner loop into two phases; the dominant phase contains
our SpMV or SpMM. The processors first send and receive the required elements of
their local portions of the input vector $x$, then multiply by $A$ to produce
$y$. This phase also contains some elementwise vector operations. We model the
per-row computation cost (due to dot products, vector scaling, and vector
addition) with the scalar $c_{\textbf{row}}$, and the per-entry computation
costs (due to matrix multiplication) with the scalar $c_{\textbf{entry}}$. Thus,
the computational work on a processor can be described as
\begin{equation}\label{eq:workcost}
         c_{\textbf{row}}|\pi_k| + c_{\textbf{entry}}\sum\nolimits_{i \in \pi_k}|v_i|
\end{equation}
which is monotonically increasing in $\pi_k$.

In distributed memory settings, both the sending and the receiving node must
participate in transmission of the message. We assume that the runtime of a
part is proportional to the sum of local work and the number of received
entries. This differs from the model of communication used by Bisseling Et.
Al., where communication is modeled as proportional to the maximum number of
sent entries or the maximum number of received entries, whichever is larger
\cite{bisseling_communication_2005}. While ignoring sent entries may seem to
represent a loss in accuracy, there are several reasons to prefer such a
model. Processors have multiple threads which can perform local work or
process sends or receives independently. If we assume that the network is not
congested (that sending processors can handle requests when receiving
processors make them), then the critical path for a single processor to
finish its work consists only of receiving the necessary input entries and
computing its portion of the matrix product. We model the cost of receiving
an entry with the scalar $c_{\textbf{message}}$.



\subsubsection{Nonsymmetric Bottleneck Cost Modeling}

Since it admits a more accurate cost model, we consider the nonsymmetric
partitioning regime first. This regime considers only one of the row (output)
space partition $\Pi$ or the column (input) space partition $\Phi$,
considering the other to be fixed.

We model our matrix as an incidence hypergraph. The nonlocal entries
of the input vector which processor $k$ must receive are the edges $j$
incident to vertices $i \in \pi_k$ such that $j \not\in \phi_k$. We can
express this tersely as $(\bigcup_{i \in \pi_k} v_i) \setminus \phi_k$. Thus,
our cost model $f_k(\pi_k, \phi_k)$ for the alternating regime is
\begin{equation}\label{eq:nonsymmetriccost}
         c_{\textbf{row}}|\pi_k| + c_{\textbf{entry}}\sum\nolimits_{i \in \pi_k}|v_i| + c_{\textbf{message}}\left|\bigcup\nolimits_{i \in \pi_k} v_i \setminus \phi_k\right|,
\end{equation}
which is monotonically increasing in $\pi_k$ and decreasing in $\phi_k$.
Intuitively, adding rows to the processor increases work and communication,
while adding columns decreases communication.

While we require the opposite partition to be fixed, we will only require the
partition we are currently constructing to be contiguous. Requiring both
$\Pi$ and $\Phi$ to be contiguous is sometimes but not always desirable;
such a constraint would limit us to matrices whose nonzeros are clustered
near the diagonal. Allowing arbitrary fixed partitions gives us the
flexibility to use other approaches for the secondary alternate partitioning
problem. For example, one might assign each column to an arbitrary incident part
to optimize total communication volume as suggested by \c{C}atalyurek
\cite{catalyurek_hypergraph-partitioning-based_1999}. We also consider the
similar greedy strategy to assign each column to a currently most expensive incident
part, attempting to reduce the cost of the most expensive part.

Of course, since our alternating partitioning regime assumes we have a fixed
$\Pi$ or $\Phi$, we need an initial partition. We propose starting by
constructing $\Pi$ because this partition involves more expensive tradeoffs
between work and communication. Since we would have no $\Phi$ to start with,
we assume no locality, upper-bounding the cost of communication and removing
$\Phi$ from our cost model. In the hypergraph model, processor $k$ receives
at most $|\bigcup_{i \in \pi_{k}} v_i|$ entries of the input vector. Thus,
our cost model would be
\begin{equation}\label{eq:nonsymmetricinitialcost}
    f(\pi_k) = c_{\textbf{row}}|\pi_k| + c_{\textbf{entry}}\sum\limits_{i \in \pi_k}|v_i| + c_{\textbf{message}}\left|\bigcup\limits_{i \in \pi_k} v_i\right|.
\end{equation}
which is monotonically increasing in $\pi_k$.

Note that any column partition $\Phi$ will achieve or improve on the modeled
cost \eqref{eq:nonsymmetricinitialcost}.

\subsubsection{Symmetric Bottleneck Cost Modeling}

The symmetric case asks us to produce a single partition $\Pi$ which will be
used to partition both the row and column space simultaneously.
We do not need to alternate between rows and columns; by adjusting scalar
coefficients, we can achieve an approximation of the accuracy of the
nonsymmetric model by optimizing a single contiguous partition under
objective \eqref{eq:objective}. Replacing $\Phi$ with $\Pi$
in cost \eqref{eq:nonsymmetriccost} yields,

\begin{equation}\label{eq:symmetriccost}
    f(\pi_k) = c_{\textbf{row}}|\pi_k| + c_{\textbf{entry}}\sum\limits_{i \in \pi_k}|v_i| + c_{\textbf{message}}\left|\left(\bigcup\limits_{i \in \pi_k} v_i\right) \setminus \pi_k\right|
\end{equation}

Unfortunately, the factor $|(\bigcup_{i \in \pi_k} v_i) \setminus \pi_k|$ is
not monotonic. However, notice that $|(\bigcup_{i \in \pi_k} v_i) \setminus
\pi_k| + |\pi_k| = |(\bigcup_{i \in \pi_k} v_i) \cup \pi_k|$ is monotonic.
Assume that each row of the matrix has at least $w_{\min}$ nonzeros (in
linear solvers, $w_{\min}$ should be at least two, or less occupied rows
could be trivially computed from other rows.) We rewrite cost
\eqref{eq:symmetriccost} in the following form,
\begin{multline}\label{eq:monotonicsymmetriccost}
    f(\pi_k) = (c_{\textbf{row}} + w_{\min} \cdot c_{\textbf{entry}} - c_{\textbf{message}})|\pi_k| + \\
    c_{\textbf{entry}}\sum\limits_{i \in \pi_k}(|v_i| - w_{\min}) + c_{\textbf{message}}\left|\left(\bigcup\limits_{i \in \pi_k} v_i\right) \cup \pi_k\right|.
\end{multline}
This function is monotonic when the coefficients on all
terms are positive. We therefore require
\begin{equation}\label{eq:monotoniccoefficients}
    c_{\textbf{row}} + w_{\min} c_{\textbf{entry}} \geq c_{\textbf{message}}
\end{equation}

Informally, constraint \eqref{eq:monotoniccoefficients} asks that the rows
and dot products hold ``enough'' local work to rival communication costs.
These conditions roughly correspond to situations where it is cheaper to
communicate an entry of $y = A \cdot x$ than it is to compute it if the
relevant entries of $x$ were stored locally. These constraints are most
suitable to matrices with heavy rows, because increasing the number of
nonzeros in a row increases the amount of local work and the communication
footprint, but not the cost to communicate the single entry of output
corresponding to that row.

Depending on the sparsity of our matrix, we may approximate the modeled cost
of the matrix by assuming $w_{\min}$ to be larger than it really is. If there
are at most $m'$ ``underfull'' rows with less than $w_{\min}$ nonzeros, then
cost \eqref{eq:symmetriccost} will be additively inaccurate by at most
$m'\cdot w_{\min} c_{\textbf{entry}}$.


\section{Computing Costs}\label{sec:footprint}


Our contiguous partitioners will probe the cost of many potential parts, and
rely on datastructures that efficiently answer such queries. Our goal is to
create datastructures to compute costs \eqref{eq:edgecut},
\eqref{eq:hyperedgecut}, \eqref{eq:lambdaminusonecut},
\eqref{eq:chainsonchains}, \eqref{eq:nonsymmetriccost},
\eqref{eq:nonsymmetricinitialcost}, and \eqref{eq:monotonicsymmetriccost}. 

In \eqref{eq:nonsymmetriccost}, we compute $|\bigcup_{i \in \pi_k} v_i
\setminus \phi_{k}|$ as $|\bigcup\nolimits_{i \in \pi_k} v_i| -
|\bigcup\nolimits_{i \in \pi_k} v_i \cap \phi_k|$. In
\eqref{eq:monotonicsymmetriccost}, since $|\bigcup\nolimits_{i \in \pi_k} v_i
\cup \pi_k|$ is just $|\bigcup\nolimits_{i \in \pi_k} v_i|$ with the diagonal
of the matrix filled in, we only consider the latter term. One can fill
the diagonal explicitly, or compute the result lazily to avoid a copy of the
matrix.

Therefore, apart from constants, we need only compute the unique terms
$|\pi_k|$, $\sum\nolimits_{i \in \pi_k} |v_i|$, $\sum_{i \in \pi_k} |v_i \cap
\pi_{k}|$, $|\{e_j \subseteq \pi_k\}|$, $|\bigcup\nolimits_{i \in \pi_k}
v_i|$, and $|\bigcup\nolimits_{i \in \pi_k} v_i \cap \phi_k|$.

Our algorithms only make contiguous queries of the form $\pi_k = i:(i'-1)$ or
$\phi_k = j:(j'-1)$, although the fixed partition $\Phi$ or $\Pi$ may not be
contiguous, respectively.

Our first term, $|\pi_k| = i' - i$, is easy to compute in constant time.
Similarly, since the $pos$ vector in CSR format is a prefix sum (cumulative
sum) of the number of nonzeros in each row $|v_i|$, we can compute $\sum_{i
\in \pi_k} |v_i| = pos_{i'} - pos_{i}$ in constant time. If our matrix is
not stored in CSR format, we can usually construct $pos$ in linear time.

The only term with an explicit dependence on $\phi_k$ is
$|\bigcup_{i \in \pi_k} v_i \cap \phi_k|$. When $\Pi$ is
fixed, we can construct sorted list representations of each set $\bigcup_{i
\in \pi_k} v_i$ in linear time and space (using, for example, a histogram
sort and deleting adjacent duplicates). This allows us to evaluate
$|\bigcup_{i \in \pi_k} v_i \cap \phi_k|$ in $O(\log(m))$ time by searching
our list for the boundaries of the contiguous region of elements which are
also in $\phi_k$.

What remains to be shown is how to compute the remaining terms for various
contiguous $\pi_k$ when $\Phi$ is held constant.

\subsection{Reduction to Dominance Counting}

In this section, we reduce computing our remaining terms to two-dimensional
dominance counting, a classic computational geometry problem. Consider points
of the form $(i, i')$ in an integer grid $\mathbb{N}^2$. We say that a point
$(i_1, i'_1)$ \textbf{dominates} a point $(i_2, i'_2)$ if $i_1 \geq i_2$ and
$i'_1 \geq i'_2$. The \textbf{dominance counting} problem in two dimensions
asks for a data structure to count the number of points dominated by each
query point. Note that by negating either the first or second coordinate, we
can reverse the first or second inequality to ask for containment
constraints.

We start with the term $\sum_{i \in \pi_k} |v_i \cap \pi_{k}|$, which counts
the number of nonzero entries $A_{j, j'}$ where $j, j' \in \pi_k$. When $j <
j'$ we represent it with the point $(-j, j')$, otherwise we represent it with
the point $(-j', j)$. Since $\pi_k = i:(i' - 1)$, the number of points
dominated by $(-i, i' - 1)$ is the number of nonzeros where both coordinates
are contained in $\pi_k$.

We handle the term $|\{e_j \subseteq \pi_k\}|$ similarly. In linear time, we
compute $(i_j, i'_j)$, the smallest and largest elements of each $e_j$. This
corresponds to the position of the first and last nonzero in each column of
$A$. Since $\pi_k = i:(i' - 1)$, the number of points $(-i_j, i'_j)$ dominated by
$(-i, i' - 1)$ is the number of columns $e_j$ whose nonzeros are completely contained
by $\pi_k$, or $|\{e_j \subseteq \pi_k\}|$.

\begin{figure}
\[
    \begin{array}{cc}
    \vcenter{\hbox{\hspace{-0.6em}\begin{tikzpicture}
\matrix (A) [matrix of math nodes,%
    nodes={inner sep=0.162em, outer sep=0},%
    inner sep = 0em,%
    row sep = {0.162em},%
    column sep = {0.27em},%
    left delimiter  = {[},%
    right delimiter = {]}] at (0,0)
{%
\light{*} & \light{*} & \light{*} &   & \light{*} &   & \light{*} &   &   &  \\\\
  & \light{*} & \light{*} &   & \light{*} &   & \light{*} &   &   &  \\\hline\\
  &   & \boldsymbol{*} &   & \boldsymbol{*} & \boldsymbol{*} &   & \boldsymbol{*} &   &  \\\\
  &   &   &   &   & \boldsymbol{*} &   &   &   &  \\\\
  &   &   &   & \boldsymbol{*} & \boldsymbol{*} &   & \boldsymbol{*} &   &  \\\\
  &   & \boldsymbol{*} &   & \boldsymbol{*} & \boldsymbol{*} &   & \boldsymbol{*} &   &  \\\hline\\
  &   & \light{*} & \light{*} &   &   & \light{*} &   & \light{*} & \light{*}\\\\
  & \light{*} &   & \light{*} & \light{*} &   &   & \light{*} &   & \light{*}\\\hline\\
  &   &   & \light{*} & \light{*} &   &   &   &   & \light{*}\\\\
  &   &   & \light{*} &   & \light{*} &   & \light{*} & \light{*} & \light{*}\\\\
};
\draw[line width = 0.27pt, gray]  (A-1-2) -- (A-3-2);
\draw[line width = 0.27pt, gray]  (A-3-2) -- (A-15-2);
\draw[line width = 0.27pt, gray]  (A-1-3) -- (A-3-3);
\draw[line width = 0.27pt, gray]  (A-3-3) -- (A-5-3);
\draw[line width = 0.54pt, black] (A-5-3) -- (A-11-3);
\draw[line width = 0.27pt, gray]  (A-11-3) -- (A-13-3);
\draw[line width = 0.27pt, gray]  (A-13-4) -- (A-15-4);
\draw[line width = 0.27pt, gray]  (A-15-4) -- (A-17-4);
\draw[line width = 0.27pt, gray]  (A-17-4) -- (A-19-4);
\draw[line width = 0.27pt, gray]  (A-1-5) -- (A-3-5);
\draw[line width = 0.27pt, gray]  (A-3-5) -- (A-5-5);
\draw[line width = 0.54pt, black] (A-5-5) -- (A-9-5);
\draw[line width = 0.54pt, black] (A-9-5) -- (A-11-5);
\draw[line width = 0.27pt, gray]  (A-11-5) -- (A-15-5);
\draw[line width = 0.27pt, gray]  (A-15-5) -- (A-17-5);
\draw[line width = 0.54pt, black] (A-5-6) -- (A-7-6);
\draw[line width = 0.54pt, black] (A-7-6) -- (A-9-6);
\draw[line width = 0.54pt, black] (A-9-6) -- (A-11-6);
\draw[line width = 0.27pt, gray]  (A-11-6) -- (A-19-6);
\draw[line width = 0.27pt, gray]  (A-1-7) -- (A-3-7);
\draw[line width = 0.27pt, gray]  (A-3-7) -- (A-13-7);
\draw[line width = 0.54pt, black] (A-5-8) -- (A-9-8);
\draw[line width = 0.54pt, black] (A-9-8) -- (A-11-8);
\draw[line width = 0.27pt, gray]  (A-11-8) -- (A-15-8);
\draw[line width = 0.27pt, gray]  (A-15-8) -- (A-19-8);
\draw[line width = 0.27pt, gray]  (A-13-9) -- (A-19-9);
\draw[line width = 0.27pt, gray]  (A-13-10) -- (A-15-10);
\draw[line width = 0.27pt, gray]  (A-15-10) -- (A-17-10);
\draw[line width = 0.27pt, gray]  (A-17-10) -- (A-19-10);
\end{tikzpicture}\hspace{-0.6em}}} & \vcenter{\hbox{\begin{tikzpicture}
\draw[thick, gray, ->] (0, 0) -- (11.440000000000001em, 0) node[midway, below, black] {};
\draw[thick, gray, ->] (0, 0) -- (0, 11.440000000000001em) node[midway, left, black] {};
\draw[step=1.04em,gray,very thin] (0, 0) grid (10.4em, 10.4em);
\draw[dashed, black, ->] (6.24em, 7.28em) -- (-1.04em, 7.28em);
\draw[dashed, black, ->] (6.24em, 7.28em) -- (6.24em, -1.04em);
\filldraw [gray, inner sep = 2.08pt] (10.4em, 2.08em) circle (1.56pt) node[below left, black] {$\light{1}$};
\filldraw [black, inner sep = 2.08pt] (6.24em, 7.28em) circle (1.56pt) node[below left, black] {$\mathbf{1}$};
\filldraw [gray, inner sep = 2.08pt] (3.12em, 8.32em) circle (1.56pt) node[below left, black] {$\light{2}$};
\filldraw [black, inner sep = 2.08pt] (6.24em, 5.2em) circle (1.56pt) node[below left, black] {$\mathbf{3}$};
\filldraw [gray, inner sep = 2.08pt] (9.36em, 2.08em) circle (1.56pt) node[below left, black] {$\light{3}$};
\filldraw [gray, inner sep = 2.08pt] (8.32em, 4.16em) circle (1.56pt) node[below left, black] {$\light{2}$};
\filldraw [gray, inner sep = 2.08pt] (8.32em, 8.32em) circle (1.56pt) node[below left, black] {$\light{1}$};
\filldraw [black, inner sep = 2.08pt] (5.2em, 6.24em) circle (1.56pt) node[below left, black] {$\mathbf{1}$};
\filldraw [gray, inner sep = 2.08pt] (10.4em, 4.16em) circle (1.56pt) node[below left, black] {$\light{1}$};
\filldraw [gray, inner sep = 2.08pt] (8.32em, 3.12em) circle (1.56pt) node[below left, black] {$\light{2}$};
\filldraw [gray, inner sep = 2.08pt] (10.4em, 3.12em) circle (1.56pt) node[below left, black] {$\light{1}$};
\filldraw [gray, inner sep = 2.08pt] (10.4em, 1.04em) circle (1.56pt) node[below left, black] {$\light{2}$};
\filldraw [gray, inner sep = 2.08pt] (7.28em, 4.16em) circle (1.56pt) node[below left, black] {$\light{1}$};
\filldraw [gray, inner sep = 2.08pt] (2.08em, 9.36em) circle (1.56pt) node[below left, black] {$\light{4}$};
\filldraw [gray, inner sep = 2.08pt] (7.28em, 8.32em) circle (1.56pt) node[below left, black] {$\light{1}$};
\filldraw [black, inner sep = 2.08pt] (5.2em, 7.28em) circle (1.56pt) node[below left, black] {$\mathbf{2}$};
\filldraw [black, inner sep = 2.08pt] (4.16em, 7.28em) circle (1.56pt) node[below left, black] {$\mathbf{1}$};
\end{tikzpicture}}}\\
    A & \text{The Links}
    \end{array}
\]
\caption{Links of our example matrix $A$ are illustrated as line segments
connecting elements of $A$ on left, and as points (with labeled
multiplicities) on right. Links residing entirely
within part $2$ are shown in bold. Part 2 contains two links starting at $i =
3$ and terminating at $i = 5$, and three links starting at $i = 5$ and
terminating at $i=6$. In total, part 2 contains $1 + 2 + 1 + 1 + 3 = 8$
links, which is equal to the number of points dominated by our dotted region
representing the partition split points.} \label{fig:links}
\end{figure}
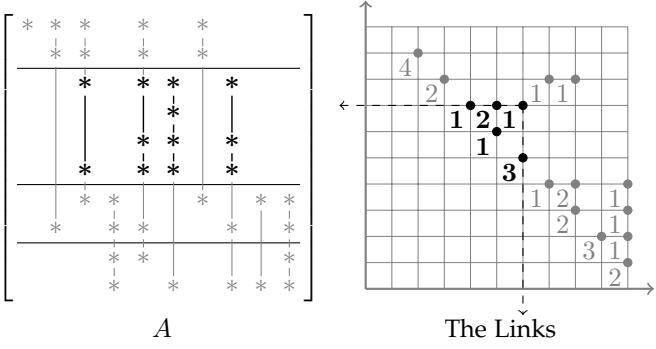

The final two terms, $|\bigcup_{i \in \pi_k} v_i|$ and $|\bigcup_{i \in
\pi_k} v_i \cap \phi_{k}|$, present more of a challenge. These quantities
concern the size of the set of distinct nonzero column locations in some row
part. We know that $\sum_{i \in \pi_k} |v_i|$ is an easy upper bound on
$|\bigcup_{i \in \pi_k} v_i|$, but it overcounts columns for each row $v_i$
they are incident to. If we were somehow able to count only the first
appearance of a nonzero column in our part, we could compute $|\bigcup_{i \in
\pi_k} v_i|$. We refer to the pair of a nonzero entry and the next
(redundant) nonzero entry in the column as a \textbf{link}. If the
$l^{th}$ nonzero occurs at row $i_l$ in some column and the closest following
nonzero occurs at $i'_l$ in that column, we call this a $(i_l, i_l')$ link, and
represent it with the point $(-i_l, i_l')$. Thus,
\begin{equation}
    \left|\bigcup\nolimits_{i \in \pi_k} v_i\right| = \sum\nolimits_{i \in \pi_k} \left|v_i\right| - |\{l : (i_l, i'_l) \in \pi_k\}|.
\end{equation}
We have already shown how to compute the first term from the $pos$ array. The
second term is the number of points dominated by $(-i, i'-1)$. Figure
\ref{fig:links} illustrates this relationship. Our reduction is almost
equivalent to the reduction from multicolored one-dimensional dominance
counting to two-dimensional standard dominance counting described by Gupta
et. al., but our reduction requires only one dominance query, while Gupta's
requires two \cite{gupta_further_1995}. We have re-used the values in the
$pos$ array of the link matrix in CSR format to avoid the second dominance
query (dominance queries can be expensive in practice).

Finally, to compute $|\bigcup\nolimits_{i \in \pi_k} v_i \cap \phi_k|$ when
$\Phi$ is fixed, we can logically split $A$ into $K$ separate matrices where
$A^{(k)}$ is $A$ where all columns other than $\phi_k$ are zeroed out. Note
that the total number of nonzeros is still $N$. Computing
$|\bigcup\nolimits_{i \in \pi_k} v_i|$ on $A^{(k)}$ gives us
$|\bigcup\nolimits_{i \in \pi_k} v_i \cap \phi_k|$.

To avoid $K$ separate dominance counting problems for each column part
$\phi_k$, we concatenate the $K$ separate problems into one problem and then
transform our problem into \textbf{rank space} \cite{gabow_scaling_1984,
chazelle_functional_1988}. Recall that our points $(i_1, j_1), ..., (i_N,
j_N)$ are given to us in $i$-major, $j$-minor order. We transform to rank
space by resorting the points into a $j$-major, $i$-minor order
$(i_{\sigma(1)}, j_{\sigma(1)}), ..., (i_{\sigma(N)}, j_{\sigma(N)})$. Our
transformation maps a point $(i_q, j_q)$ to its position pair $(q,
\sigma(q))$. Notably, $i_{q} < i_{q'}$ if and only if $q < q'$ and $j_{q} <
j_{q'}$ if and only if $\sigma(q) < \sigma(q')$, so our dominance counts are
preserved under our transformation. If we store our two orderings, we can
use binary search to find the transformed point at query time.

We can sort the links first by their column part, then into $i$-major
and $j$-major orders in linear time with, for example, two histogram sorts.
By storing the boundaries of where each part's subproblem begins and ends,
we can then resolve dominance queries within the region corresponding to the
target part.

\subsection{Online Dominance Counting}\label{sec:online_dominance}

First, we consider a simple direct method for incremental dominance counting,
adapted from \cite{ahrens_optimal_2020}, the most directly applicable method
of its kind \cite{grandjean_optimal_2012, ziantz_run-time_1994,
aydin_distributed_2019, alpert_multiway_1995}. The datastructure requires $n
+ 1$ words, can be constructed in $O(n)$ time, and given we have just
answered a query for $\pi_k = i:(i' - 1)$, we can answer a query for $i:i'$
in $O(|v_{i'}|)$ time, and a query for $(i - 1):(i' - 1)$ or $i':i'$ in
constant time.

Our datastructure starts at $0:0$ with a counter $c$ of dominated points and
a vector $\Delta$ initialized to $0$. To increment $i'$, we iterate over each
point $(i_l, i')$ with right coordinate $i'$, incrementing $\Delta_{i_l}$ by
one each time, and incrementing $c$ when $i_l \leq i'$. To decrement $i$, we
simply add $\Delta_{i - 1}$ to $c$. To set $i$ to $i'$, we need only set $c$
to $\Delta_{i'}$.

When links are needed, we can use a vector $h$ of size $n$ to record the last
time we have seen a particular nonzero column as $i'$ advances from the top
of the matrix to the bottom. The next time we see a nonzero in the same
column, we can report the corresponding link.

When we compute $|\bigcup\nolimits_{i \in \pi_k} v_i \cap \phi_k|$ for various
$\pi_k$, we maintain $K$ separate counters for the range $i':i'$ for each
column part, so that we may increment $k$ in constant time.

The online approach is efficient in settings when we wish to compute the
cost of parts for many starting points corresponding to a fixed end point, or
for many end points corresponding to a fixed start point. It is also a simple
way to compute the cost of a single partition once.

\subsection{Offline Dominance Counting}

Offline dominance counting is more appropriate when our partitioner makes
fewer queries of more arbitrary intervals. The two dimensional problem has been
the subject of intensive theoretical study \cite{chazelle_functional_1988,
jaja_space-efficient_2004, chan_adaptive_2016}. However, because of the focus
on only query time and storage, little attention has been given to
construction time, which is always superlinear. Therefore, we modify
Chazelle's algorithm to allow us to trade construction time for query time
\cite{chazelle_functional_1988}.

Dominance counting (and the related semigroup range sum problem
\cite{chazelle_lower_1990, alstrup_new_2000, jaja_space-efficient_2004}) are
roughly equivalent to the problem of computing prefix sums on sparse matrices
and tensors in the database community. These data structures are called
``Summed Area Tables'' or ``Data Cubes,'' and they value dynamic update
support and low or constant query time at the expense of storage and
construction time. We refer curious readers to \cite{shekelyan_sparse_2019}
for an overview of existing approaches, with the caution that most of these
works reference the $o(m^2)$ size of a naive dense representation of the
summed area table when they use words like ``sublinear'' and ``superlinear.''

Chazelle's dominance counting algorithm uses linear space in the number of
points to be stored, requiring log-linear construction time and logarithmic
query time. We adapt this structure to a small integer grid, allowing us to
trade off construction time and query time. Whereas Chazelle's algorithm can
be seen as searching through the wake of a merge sort, our algorithm can be
seen as searching through the wake of a radix sort. Our algorithm can also be
thought of as a decorated transposition of a CSR matrix. Because the
algorithm is so detail-oriented, we give a high-level description, but leave
the specifics to the pseudocode presented in Algorithms \ref{alg:construct}
and \ref{alg:query}.

\begin{algorithm}
    Construct the dominance counting data structure over $N$ points $(i_1,
    j_1), ..., (i_N, j_N)$, ordered on $i$ initially. We assume we are given
    the $j$ coordinates in a vector $idx$. We require that $2^{H b} > n$.
    \label{alg:construct}
    \begin{algorithmic}
        \Function{ConstructDominance}{$N$, $idx$}
            \State $qos \gets \text{zeroed vector of length $n + 2$}$
            \State $qos_1 \gets 1$
            \State $qos_{n + 2} \gets N + 1$
            \State $tmp \gets \text{uninitialized vector of length $2^b + 1$}$
            \State $cnt \gets \text{zeroed 3-tensor of size $2^b + 1 \times \lfloor N/2^{b'} \rfloor + 1 \times H$}$
            \State $byt \gets \text{uninitialized vector of length $N$}$
            \For{$h \gets H, H - 1, ..., 1$}
                \For{$J \gets 1, 1 + 2^{h b}, ..., n + 1$}
                    \State Fill $tmp$ with zeros.
                    \For{$q \gets qos_J, qos_J + 1, ..., qos_{J + 2^{h b}}$}
                        \State $d \gets key_h(idx_q)$
                        \State $tmp_{d + 1} \gets tmp_{d + 1} + 1$
                    \EndFor
                    \State $tmp_1 \gets qos_J$
                    \For{$d \gets 1, 2, ..., 2^b$}
                        \State $tmp_{d + 1} \gets tmp_{d} + tmp_{d + 1}$
                    \EndFor
                    \For{$q \gets qos_J, qos_J + 1, ..., qos_{J + 2^{h b}}$}
                        \State $d \gets key_h(idx_q)$
                        \State $q' \gets tmp_d$
                        \State $byt_{q'} \gets 2^{hb}\lfloor idx_{q'}/2^{hb} \rfloor + idx_q \bmod 2^{hb}$
                        \State $tmp_d \gets q' + 1$
                    \EndFor
                    \For{$d \gets 1, 2, ..., 2^b$}
                        \State $qos_{J + d 2^{(h - 1)b}} \gets tmp_d$
                    \EndFor
                \EndFor
                \State Fill $tmp$ with zeros.
                \For{$Q \gets 1, 1 + 2^{b'}, ..., N$}
                    \For{$q \gets Q, Q + 1, ..., Q + 2^{b'}$}
                        \State $d \gets key_h(idx_q)$
                        \State $tmp_d \gets tmp_d + 1$
                    \EndFor
                    \For{$d \gets 1, 2, ..., 2^b$}
                        \State $cnt_{(d + 1) Q h} \gets tmp_d + cnt_{d Q h}$
                    \EndFor
                \EndFor
                \State $(idx, byt) \gets (byt, idx)$
            \EndFor
            \State $byt \gets idx$
            \State \Return $(qos, byt, cnt)$
        \EndFunction
    \end{algorithmic}
\end{algorithm}

\begin{algorithm}
    Query the dominance counting data structure for the number of points
    dominated by $(i, j)$.
    \label{alg:query}
    \begin{algorithmic}
        \Function{QueryDominance}{$i$, $j$}
            \State $\Delta q \gets pos_{i + 1} - 1$
            \State $c \gets 0$
            \For{$h \gets H, H - 1, ..., 1$}
                \State $j' \gets 2^{hb}\lfloor j/2^{hb} \rfloor$
                \State $q_1 \gets qos_{j'} - 1$
                \State $q_2 \gets q_1 + \Delta q$
                \State $d \gets key_{h}(j)$
                \State $Q_1 \gets \lfloor q_1/2^{b'} \rfloor + 1$
                \State $Q_2 \gets \lfloor q_2/2^{b'} \rfloor + 1$
                \State $c \gets cnt_{d Q_2 h} - cnt_{d Q_1 h}$
                \State $\Delta q \gets (cnt_{(d + 1)Q_2 h} - cnt_{d Q_2 h})$
                \State $\Delta q \gets \Delta q - (cnt_{(d + 1)Q_1 h} - cnt_{d Q_1 h})$
                \For{$q \gets 2^{b'}(Q_1 - 1) + 1, 2^{b'}(Q_1 - 1) + 2, ..., q_1$}
                    \State $d' \gets key_h(byt_q)$
                    \If{$d' < d$}
                        \State $c \gets c - 1$
                    \ElsIf{$d' = d$}
                        \State $\Delta q \gets \Delta q - 1$
                    \EndIf
                \EndFor
                \For{$q \gets 2^{b'}(Q_2 - 1) + 1, 2^{b'}(Q_2 - 1) + 2, ..., q_2$}
                    \State $d' \gets key_h(byt_q)$
                    \If{$d' < d$}
                        \State $c \gets c + 1$
                    \ElsIf{$d' = d$}
                        \State $\Delta q \gets \Delta q + 1$
                    \EndIf
                \EndFor
            \EndFor
            \State \Return $c$
        \EndFunction
    \end{algorithmic}
\end{algorithm}

In this section, assume we have been given $N$ points $(i_1, j_1), ..., (i_N,
j_N)$ in the range $[1, ..., m] \times [1, ..., n]$. Since these
points come from CSR matrices or our link construction, we assume that our
points are initially sorted on their $i$ coordinates ($i_q \leq i_{q + 1}$)
and we have access to an array $pos$ to describe where the points
corresponding to each value of $i$ starts. If this is not the case, either
the matrix or the following algorithm can be transposed, or the points can be
sorted with a histogram sort in $O(m + N)$ time.

The construction phase of our algorithm successively sorts the points on
their $j$ coordinates in rounds, starting at the most significant digit and
moving to the least. We refer to the ordering at round $h$ as $\sigma_h$. We
use $H$ rounds, each with $b$ bit digits, where $b$ is the smallest integer
such that $2^{H\cdot b} \geq m$. Let $key_h(j)$ refer to the $h^{th}$ most significant
group of $b$ bits (the $h^{th}$ digit). Formally,
\begin{equation*}
    key_h(j) = \lfloor j / 2^{(h - 1)b}\rfloor \bmod 2^b
\end{equation*}

At each round $h$, our points will be sorted by the top $h\cdot b$ bits of
their $j$ coordinates using a histogram sort in each bucket formed by the
previous round. We use an array $qos$ (similar to $pos$) to store the
starting position of each bucket in the current ordering of points. Formally,
$qos_j$ will record the starting position for points $(i_q, j_q)$ where $j_q
\geq j$. Note that $qos$ is of size $n + 2$ instead of $n + 1$, as one might
expect, to allow $0$ as a possible value for $j$ during query time.

Although we can interpret the algorithm as resorting the points several
times, each construction phase only needs access to its corresponding bit
range of $j$ coordinates (the keys) in the current ordering. The query phase
needs access to the ordering of keys before executing each phase. Thus, the
algorithm iteratively constructs a vector $byt$, where the $h^{th}$ group of
$b$ bits in $byt$ corresponds to the $h^{th}$ group of $b$ bits in current
ordering of $j$ coordinates ($key_h(byt_q) = key_h(j_{\sigma_h(q)})$). As the
construction algorithm proceeds, we can use the lower bits of $byt$ to store
the remaining $j$ coordinate bits to be sorted.

Each phase of our algorithm needs to sort $\lceil n/2^{h b} \rceil$
buckets. Our histogram sort uses a scratch array of size $2^b$ to
sort a bucket of $N'$ points in $O(N' + 2^b)$ time. Thus, we can sort the buckets
of level $h$ in $O(2^b\lceil n/2^{h b} \rceil + N)$ time, and bucket sorting
takes $O(n + H N)$ time in total over all levels.

A query requests the number of points in our data structure dominated by $(i,
j)$. In the initial ordering, $i_q < i$ is equivalent to $q <
pos_i$. Thus, the dominating points reside within the first $pos_i - 1$
positions of the initial ordering. Our algorithm starts by counting the
number of points such that $key_1(j_q) < key_1(j)$ and $q < pos_i$. All
remaining dominating points satisfy $key_1(j_q) = key_1(j)$, so let $q'$ be
the number of points $key_1(j_q) = key_1(j)$ and $q < pos_i$. After our first
sorting round, the set of points in the initial ordering where 
$key_1(i_q) = key_1(i)$ would have been stored contiguously, and therefore the
first $q'$ of them satisfy $i_{\sigma_h(q)} < i$. We can then apply our procedure
recursively within this bucket to count the number of dominating points.

We have left out an important aspect of our algorithm. Our query procedure
needs to count the number of dominating points that satisfy
$key_h(j_{\sigma_h(q)}) < h$ within ranges of $q$ that agree on the top $h\cdot b$
bits of each $j$. While $qos$ stores the requisite ranges of $q$, we still need
to count the points. In $O(N + 2^b)$ time, for a particular value of $h$, we
can walk $byt$ from left to right, using a scratch vector of size $2^b$ to
count the number of points we see with each value of $key_h(byt_q)$. If
we cache a prefix sum of our scratch vector once every $2^{b'}$ points (the prefix
sum takes $O(2^b)$ time), we can use the cache to jump-start the
counting process at query time. During a query, after checking our cached
count in constant time, we only need to count a maximum of $2^{b'}$ points at each level to
obtain the correct count. Our cache is a 3-tensor $cnt$, where
$cnt_{h q d}$ stores the number of points $q'$ such that $q' < cq$ and
$key_h(j_{\sigma_h(q')}) < d$. If we cache every $2^{b'}$ points, computing
$cnt$ takes $O(2^b\lceil N/2^{b'} \rceil) = O(N2^{b-b'})$ time per phase. The
$pos$ vector uses $m + 1$ words, the $qos$ vector uses $n + 2$ words, the
$byt$ vector uses $N$ words, and the $cnt$ tensor uses at most $H N 2^{b -
b'}$ words. The runtime and storage of our offline algorithms are summarized
in Table \ref{tbl:offline_dominance}.

We consider two ways to set $H$, $b$, and $b'$. Chazelle proposed setting $b
= 2$, $H = \lceil\log_2(n)\rceil$, and $b' = \lceil\log_2(H)\rceil$. When $b
= 2$, each key is one bit, and Chazelle suggested we ``transpose'' the $byt$
array by storing a bit-packed vector for each level. Because the size of a
word bounds the size of the input, we can count the bits at each level of the
query step with a constant number of bit-counting instructions, saving a
factor of $\log(n)$ at query time. While storage would be linear and query
time would be logarithmic, constructing a dominance counter with these
settings would require $\log_2(n)$ passes. In rank space, this would require
an onerous $20$ passes for just $1,048,576$ nonzeros.

We also consider setting $H$, the height of the tree and the number of
passes, to a small constant like $2$ or $3$, while keeping storage costs
linear, since storage is often a critical resource in scientific computing.
For correctness, we minimize $b$ subject to $2^{H b} \geq n$. We minimize
$b'$ subject to $2^{b'} \geq H 2^b$ to ensure that the footprint of our
dominance counter is at most four times the size of $A$. These settings do
not permit the bit-counting optimization at query time because the digits
become larger than a single bit.

When our points come from a rank space transformation, then $m$ and $n$ become
$N$. Note that transforming to rank space simplifies our algorithm because $pos$
and $qos$ become the identity and we no longer need to store them. However, we
do need to store our orderings of $i$ and $j$ values, so the storage requirement
is the same and $m = n = N$. Although we need to perform binary search for
rank-space queries, the $O(\log(N))$ runtime is dominated by the query time in
both parameterizations that we consider.

\begin{table}
\caption{Runtime of offline dominance counting parameterizations for points
in an $m \times n$ grid. For simplicity, we assume that $N \geq m$ and $N
\geq n$. When we apply a link-based reduction or the matrix is symmetric, $n
= m$. When points are transformed to rank space, $m = n = N$.}
\label{tbl:offline_dominance}
\setlength{\tabcolsep}{2pt}
\begin{tabular}{cccc}
    Settings & Generic & \makecell{$b = 2$ \\ $H = \lceil\log_2(n)\rceil$ \\ $b' = \lceil\log_2(H)\rceil$} & \makecell{$H$ constant \\$b = \lceil \log_2(n)/H\rceil$ \\ $b' = b + \lceil \log_2(H) \rceil$}\\\\
    Construct & $O(n + H{\cdot}N(1 + 2^{b-b'}))$ & $O(N\log_2(n))$ & $O(n + H{\cdot}N)$\\
    Query & $O(H 2^{b'})$ & $O(\log_2(n))$ & $O(H^2n^{1/H})$ \\
    Storage & $m + n + N + H{\cdot}N 2^{b - b'}$ & $m + n + N$ & $m + n + N$ \\
\end{tabular}
\end{table}

\section{Bottleneck Partitioners}\label{sec:partitionmonotonic}

Pinar et. al. present a multitude of algorithms for optimizing linear cost
functions \cite{pinar_fast_2004}. We examine both an approximate and an
optimal algorithm, and point out that with fairly minor modifications they
can be modified to optimize arbitrary monotonic increasing or decreasing cost
functions, given an oracle to compute the cost of a part. We chose the
approximate ``$\epsilon$-BISECT+'' algorithm (originally due to Iqbal et. al.
\cite{iqbal_approximate_1991}) and the exact ``NICOL+'' algorithm (originally
due to Nicol et. al. \cite{nicol_rectilinear_1994}) because they are easy to
understand and enjoy strong guarantees, but use dynamic split point bounds
and other optimizations based on problem structure, resulting in
empirically reduced calls to the cost function.

Since the approximate algorithm introduces many key ideas which are expanded
upon in the exact algorithm, we start with our adaptation of the
``$\epsilon$-BISECT+'' partitioner, which produces a $K$-partition within
$\epsilon$ of the optimal cost when it lies within the given bounds.

If our cost is monotonic increasing, the optimal $K$ partition of a
contiguous subset of rows cannot cost more than the optimal $K$ partition of
the whole matrix, since we could simply truncate a partition of the entire
set of rows to the subset in question and achieve the same or lesser cost.
Therefore, if we know that there exists a $K$-partition of cost $c$, and we
can set the endpoint of the first part to the largest $i'$ such that $f_1(1,
i') \leq c$, there must exist $K-1$-partition of the remaining columns that
starts at $i'$ and costs at most $c$. This observation implies a procedure
that determines whether a partition of cost $c$ is feasible by attempting to
construct the partition, maximizing split points at each part in turn. This
procedure is called a \textproc{Probe}. Clearly, using linear search for
split points, \textproc{Probe} only uses $O(m)$ evaluations of the cost
function. If \textproc{Probe} uses binary search for split points, it only
needs $K\log_2(m)$ evaluations of the cost function. We can repeatedly call
\textproc{Probe} to search the space of possible costs, stopping when our
lower bound on the cost is within $\epsilon$ of the upper bound. Note that if
we had the optimal value of a $K$-partition, a single call to \textproc{Probe}
can recover the split points. While Pinar et. al. use this fact to simplify
their algorithms and return only the optimal partition value, our cost
function is more expensive to evaluate than theirs, so our algorithms have
been modified to compute the split points themselves without increasing the
number of evaluations of the cost function \cite{pinar_fast_2004}.

The intuition in the monotonic decreasing case is the opposite of the
monotonic increasing case. Instead of searching for the last split point less
than a candidate cost, we search for the first. Instead of looking for a
candidate partition which can reach the last row without exceeding the
candidate cost, we look for a candidate partition which does not need to
exceed the last row to achieve the candidate cost. The majority of changes
reside in the small details, which we leave to the pseudocode.

Our adapted bisection algorithm is detailed in Algorithm
\ref{alg:bisectpartition}. Algorithm \ref{alg:bisectpartition} differs from
the algorithm presented by Pinar et. al. in that it allows for decreasing
functions, is stated in terms of possibly different $f$ for each part, does
not assume $f_k(i, i)$ to be zero, allows for an early exit to the probe
function, returns the partition itself instead of the best cost (this avoids
extra probes needed to construct the partition from the best cost), and
constructs the dynamic split index bounds in the algorithm itself, instead of
using more complicated heuristics (which may not apply to all cost functions)
to initialize the split index bounds. Note that we assume $0 < c_{\low}$ only
for the purposes of providing a relative approximation guarantee.

Algorithm \ref{alg:bisectpartition} considers at most
$\log_2(c_{\high}/(c_{\low}\epsilon))$ candidate partition costs.
If we use a linear search and online (rank space) dominance counters,
\textproc{Probe} runs in linear time and Algorithm \ref{alg:bisectpartition}
runs in time
\begin{equation}\label{eq:linear_bisect_time}
    O\left(\log\left(\frac{c_{\high}}{c_{\low}\epsilon}\right)N\right).
\end{equation}
If we use binary search and offline (rank space) dominance counters
with constant $H$, Algorithm \ref{alg:bisectpartition} runs in time
\begin{equation}\label{eq:binary_bisect_time}
    O\left(H \cdot N + K\log(m)\log\left(\frac{c_{\high}}{c_{\low}\epsilon}\right)H^2 N^{1/H}\right),
\end{equation}
While the binary-searching algorithm has a better asymptotic runtime, the
linear-search algorithm can take advantage of the simpler linear dominance
counters, and may perform better on smaller problems when $\epsilon$ is high
and therefore the number of probes is low.

The key insight made by Nicol et. al. which allows us to improve our
bisection algorithm into an exact algorithm was that there are only $m^2$
possible costs which could be a bottleneck in our partition, corresponding to
$m^2$ possible pairs of split points that might define a part
\cite{nicol_rectilinear_1994}. Thus, Nicol's algorithm searches the split
points instead of searching the costs, and achieves a strongly polynomial
runtime. We will reiterate the main idea of the algorithm, but refer the
reader to \cite{pinar_fast_2004} for more detailed analysis.

Assume that we know the starting split point of processor $k$ to be $i$.
Consider the ending point $i'$ in a partition of minimal cost. If $k$ were a
bottleneck (longest running processor) in such a partition, then $f_k(i, i')$
would be the overall cost of the partition, and we could use this cost to
bound that of all other processors. If $k$ were not a bottleneck, then
$f_k(i, i')$ should be strictly less than the minimum feasible cost of a
partition, and it would be impossible to construct a partition of cost
$f_k(i, i')$. Thus, Nicol's algorithm searches for the first bottleneck
processor, examining each processor in turn. When we find a processor where
the cost $f_k(i, i')$ is feasible, and less than the best feasible cost seen
so far, we record the resulting partition in the event this was the first
bottleneck processor. Then, we set $i'$ so that $f_k(i, i')$ is the greatest
infeasible cost and continue searching, assuming that processor $k$ was not a
bottleneck.

We have made similar modifications in our adaptation of ``NICOL+'' as we did
for our adaptation of ``$\epsilon$-BISECT+.'' Primarily, the algorithm now
handles monotonic decreasing functions. We also phrase our algorithm in terms
of potentially multiple $f$, construct our dynamic split point bounds inside
the algorithm instead of using additional heuristics, make no assumptions on
the value of $f_k(i, i)$, allow for early exits to the probe function, and
return a partition instead of an optimal cost. We also consider bounds on the
cost of a partition to be optional in this algorithm. Our adaptation of
``NICOL+'' (\cite{pinar_fast_2004}) for general monotonic part costs is
presented in Algorithm \ref{alg:nicolpartition}.

Although ``NICOL+'' uses outcomes from previous searches to bound the split
points in future searches, a simple worst-case analysis of the algorithm
shows that the number of calls to the cost function is bounded by
\begin{equation}\label{eq:nicol_quries}
    K^2\log_2(m)^2.
\end{equation}
Using offline (rank space) dominance counters with constant $H$, Algorithm
\ref{alg:bisectpartition} runs in time
\begin{equation}\label{eq:nicol_time_h}
    O\left(H\cdot N + K^2\log(N)^2H^2 N^{1/H}\right).
\end{equation}

We say that $f$ grows polynomially slower than $g$ when $f$ is $O(g^C)$ for
some constant $C < 1$. Thus, the approximate partitioner runs in linear time if
$K\log(c_{\high}/(c_{\low}\epsilon))$ grows polynomially slower than $N^{1 -
1/H}$. If costs are subadditive, then we only need $K\log(K/\epsilon)$, and
therefore $K\log(1/\epsilon)$, to grow polynomially slower than $N^{1 -
1/H}$. The exact partitioner runs in linear time if $K^2$ grows polynomially
slower than $N^{1 - 1/H}$. Our algorithms can run in linear time precisely
when our partitioners use polynomially sublinear queries, since we are able
to offset polynomial query time decreases with logarithmic construction time
increases.

For our practical choice of $H = 3$, $K\log(c_{\high}/(c_{\low}\epsilon))$
needs to grow polynomially slower than $N^{2/3}$ for linear time approximate partitioning
and $K$ needs to grow polynomially slower than $N^{1/3}$ for linear time exact
partitioning. However, both algorithms use dynamic bounds on split indices to
reduce the number of probes, so they are likely to outperform these
worst-case estimates. Furthermore, $K$, the number of processors, is often a
relatively small constant.

\begin{figure*}
\begin{multicols}{2}
\begin{minipage}{\linewidth}

\begin{algorithm}
    Given a monotonic increasing (or decreasing) cost function
    $f_k$ defined on pairs of split points, a starting split point $i$, and a
    maximum cost $c$, find the greatest (least, respectively) $i'$ such
    that $i \leq i'$, $f_k(i, i') \leq c$, and $i'_{\low} \leq i' \leq
    i'_{\high}$. Returns $\max(i, i'_{\low}) - 1$ (returns
    $i'_{\high} + 1$, respectively) if no cost at most $c$ can be found. Changes needed for
    decreasing functions are marked with $\triangleright$.
    \label{alg:search}
    \begin{algorithmic}
        \Function{Search}{$f_k$, $i$, $i'_{\low}$, $i'_{\high}$, $c$}
            \State $i'_{\low} \gets \max(i, i'_{\low})$
            \While{$i'_{\low} \leq i'_{\high}$}
                \State $i' = \lfloor (i'_{\low} + i'_{\high})/2 \rfloor$
                \If{$f_k(i, i') \leq c$}
                    \State $i'_{\low} = i' + 1$ \Comment $i'_{\high} = i' - 1$
                \Else
                    \State $i'_{\high} = i' - 1$ \Comment $i'_{\low} = i' + 1$
                \EndIf
            \EndWhile
            \State \Return $i'_{\high}$ \Comment \Return $i'_{\low}$
        \EndFunction
    \end{algorithmic}
\end{algorithm}
\begin{algorithm}[BISECT Partitioner]
    Given monotonic increasing cost function(s) $f$ defined on pairs of split
    points, find a contiguous $K$-partition $\Pi$ which minimizes
    \[
        c = \max\limits_{k} f_k(s_k, s_{k + 1})
    \]
    to a relative accuracy of $\epsilon$ within the range $0 < c_{\low} \leq
    c \leq c_{\high}$, if such a partition exists. This is an adaptation of
    the ``$\epsilon$-BISECT+'' algorithm by Pinar et. al.
    \cite{pinar_fast_2004}, which was a heuristic improvement on the
    algorithm proposed by Iqbal et. al. \cite{iqbal_approximate_1991}. We
    assume that \textproc{Probe} shares scope with
    \textproc{bisectpartition}. Changes needed for the case where all $f$ are
    monotonic decreasing are marked with $\triangleright$.

    \label{alg:bisectpartition}
    \begin{algorithmic}
        \Function{BisectPartition}{$f$, $m$, $K$, $c_{\low}$, $c_{\high}$, $\epsilon$}
            \State $({s_{\high}}_1, ..., {s_{\high}}_{K + 1}) \gets (1, m + 1, ..., m + 1)$
            \State $({s_{\low}}_1, ..., {s_{\low}}_{K + 1}) \gets (1, ..., 1, m + 1)$
            \State $(s_1, ..., s_{K + 1}) \gets (1, \#, ..., \#, m + 1)$

            \While{$c_{\low}(1 + \epsilon) < c_{\high}$}
                \State $c \gets (c_{\low} + c_{\high})/2$
                \If{$\Call{Probe}{c}$}
                    \State ${c_{\high}} \gets c$
                    \State ${S_{\high}} \gets S$ \Comment ${S_{\low}} \gets S$
                \Else
                    \State ${c_{\low}} \gets c$
                    \State ${S_{\low}} \gets S$ \Comment ${S_{\high}} \gets S$
                \EndIf
            \EndWhile
            \State \Return $S_{\high}$ \Comment \Return $S_{\low}$
        \EndFunction

        \Function{Probe}{$c$}
            \For{$k = 1, 2, ..., K - 1$}
                \State $s_{k + 1} \gets \Call{Search}{f_k, s_{k}, {s_{\low}}_{k + 1}, {s_{\high}}_{k + 1}, c}$
                \If{$s_{k + 1} < s_k$} \Comment \MyIf $s_{k + 1} > m + 1$ \MyThen
                    \State $s_{k + 1}, ..., s_{K} \gets s_k$ \Comment $s_{k + 1}, ..., s_{K} \gets m + 1$
                    \State \Return $\false$
                \EndIf
            \EndFor
            \State \Return $f_K(s_K, s_{K + 1}) \leq c$
        \EndFunction
    \end{algorithmic}
\end{algorithm}
\end{minipage}
\newpage
\begin{minipage}{\linewidth}
\begin{algorithm}[NICOL Partitioner]
    Given monotonic increasing cost function(s) $f$ defined on pairs of split
    points, find a contiguous $K$-partition $\Pi$ which minimizes
    \[
        c = \max\limits_{k} f_k(s_k, s_{k + 1})
    \]
    within the range $c_{\low} \leq c \leq c_{\high}$, if such a partition exists.
    This is an adaptation of the ``NICOL+'' algorithm by Pinar et. al.
    \cite{pinar_fast_2004}, which was a heuristic improvement on the
    algorithm proposed by Nicol et. al. \cite{nicol_rectilinear_1994}.
    We assume that \textproc{ProbeFrom} shares scope with
    \textproc{NicolPartition}. Changes needed for the case where all $f$ are
    monotonic decreasing are marked with~$\triangleright$.
    \label{alg:nicolpartition}
    \begin{algorithmic}
        \Function{NicolPartition}{$f$, $m$, $K$, $c_{\low}$, $c_{\high}$}
            \State $({s_{\high}}_1, ..., {s_{\high}}_{K + 1}) \gets (1, m + 1, ..., m + 1)$
            \State $({s_{\low}}_1, ..., {s_{\low}}_{K + 1}) \gets (1, ..., 1, m + 1)$
            \State $(s_1, ..., s_{K + 1}) \gets (1, \#, ..., \#, m + 1)$

            \For{$k \gets 1, 2, ..., K$}
                \State $i \gets s_k$
                \State $i'_{\high} \gets {s_{\high}}_{k + 1}$
                \State $i'_{\low} \gets \max(s_k, {s_{\low}}_{k + 1})$
                \While{$i'_{\low} \leq i'_{\high}$}
                    \State $i' \gets \lfloor (i'_{\low} + i'_{\high})/2 \rfloor$
                    \State $c \gets f_k(i, i')$
                    \If{$c_{\low} \leq c < c_{\high}$}
                        \State $s_{k + 1} \gets i'$
                        \If{$\Call{ProbeFrom}{c, k}$}
                            \State ${c_{\high}} \gets c$
                            \State ${i'_{\high}} \gets i' - 1$ \Comment ${i'_{\low}} \gets i' + 1$
                            \State ${S_{\high}} \gets S$ \Comment ${S_{\low}} \gets S$
                        \Else
                            \State ${c_{\low}} \gets c$
                            \State ${i'_{\low}} \gets i' + 1$ \Comment ${i'_{\high}} \gets i' - 1$
                            \State ${S_{\low}} \gets S$ \Comment ${S_{\high}} \gets S$
                        \EndIf
                    \ElsIf{$c \geq c_{\high}$}
                        \State $i'_{\high} = i' - 1$ \Comment $i'_{\low} = i' + 1$
                    \Else
                        \State $i'_{\low} = i' + 1$ \Comment $i'_{\high} = i' - 1$
                    \EndIf
                \EndWhile

                \If{$i'_{\high} < s_{k}$} \Comment \MyIf $i'_{\low} > m + 1$ \MyThen
                    \State \MyBreak
                \EndIf

                \State $s_{k + 1} \gets i'_{\high}$ \Comment $s_{k + 1} \gets i'_{\low}$
            \EndFor
            \State \Return $S_{\high}$ \Comment \Return $S_{\low}$
        \EndFunction
        \Function{ProbeFrom}{$c$, $k$}
            \For{$k' = k + 1, k + 2, ..., K - 1$}
                \State $s_{k' + 1} \gets \Call{Search}{f_{k'}, s_{k'}, {s_{\low}}_{k' + 1}, {s_{\high}}_{k' + 1}, c}$
                \If{$s_{k' + 1} < s_k'$} \Comment \MyIf $s_{k' + 1} > m + 1$ \MyThen
                    \State $s_{k' + 1}, ..., s_{K} \gets s_{k'}$ \Comment $s_{k' + 1}, ..., s_{K} \gets m + 1$
                    \State \Return $\false$
                \EndIf
            \EndFor
            \State \Return $f_K(s_K, s_{K + 1}) \leq c$
        \EndFunction
    \end{algorithmic}
\end{algorithm}
\end{minipage}
\end{multicols}
\end{figure*}

\subsection{Bounding the Costs} 

Algorithm \ref{alg:bisectpartition}, our approximate bottleneck partitioner,
requires upper and lower bounds on the cost function, and both Algorithm
\ref{alg:bisectpartition} and \ref{alg:nicolpartition} should perform better
when given better initial bounds.

Given a monotonic increasing cost function $f$ over a set of vertices $V$,
$\max_i f(\{i\})$ and $f(V)$ are naive lower and upper bounds on the
bottleneck cost of a $K$-partition. These become upper and lower bounds when
$f$ is decreasing, respectively.

However, we can use subadditivity in increasing cost functions to increase
the lower bound on the bottleneck cost of a $K$-partition. Using
subadditivity, $\sum_k f(\pi_{k}) \geq f(V)$, and therefore,
\begin{equation}\label{eq:lower_subadditive}
    \max\limits_{k} f(\pi_{k}) \geq \frac{f(V)}{K}.
\end{equation}

Finding good lower bounds on partition costs can be difficult. Costs like
\eqref{eq:nonsymmetricinitialcost} and \eqref{eq:symmetriccost} are
subadditive and use the same cost for each part, so they obey
\eqref{eq:lower_subadditive}. However, other costs such as \eqref{eq:nonsymmetriccost}
use different functions $f_k$ on each part, and therefore we cannot apply
equation \eqref{eq:lower_subadditive}.

We can work around this limitation by lower-bounding all the functions by the
same subadditive one, then applying equation \eqref{eq:lower_subadditive}. For example,
one might lower bound only the work terms in the cost function, since these
are uniform across partitions. As another example, one might assume that all
threshold functions are zero when calculating a lower bound.

Subadditivity does not help us bound monotonic decreasing cost functions. For
these costs, it may make sense to simply use $\max_{k, i} f_k(i, i)$ and
$\max_{k} f_k(1, m)$ as upper and lower bounds on the cost.

\section{Total Partitioners}

Total partitioning under Problem \eqref{eq:objective} is quite similar to the
\textbf{least weight subsequence} problem, which we phrase as evaluating the
dynamic program
\begin{multline}\label{eq:lws}
c_{i'} = \min\limits_{1 \leq i < i'} d_i + f(i:i' - 1)\\
p_{i'} = \argmin\limits_{1 \leq i < i'} d_i + f(i:i' - 1)
\end{multline}
for $1 < i' < n$ when $d_i$ is readily computable from $c_i$.
When there is no constraint on the number of parts, we can minimize the sum
\eqref{eq:objective} by setting $d_i = c_i$, and following the back pointers $p$
to construct $\pi$. If, however, we wish to constrain the number of parts to be
$K$, then we must apply Problem \eqref{eq:lws} $K$ times, producing $K$ separate
cost and pointer vectors $c_k$ and $p_k$, where in the $k^{th}$ application of
\eqref{eq:lws}, $d_i = c_{k - 1, i}$. The naive dynamic programming algorithm
for variable $K$ makes $O(m^2)$ queries to the cost function, and runs in $O(m^2
+ N)$ time using our online dominance counters in rank space (we assume the
inner loop over $i$ is evaluated in reverse). The naive dynamic programming
algorithm for fixed $K$ makes $O(Km^2)$ queries and runs in $O(K(m^2 + N))$
time using our online dominance counters in rank space. If the cost function $f$
is uniform across parts, we can reduce this to $O(Km^2 + N)$ by evaluating
$f$ once for each interval $i:i'$ and using that value to update $K$
simultaneous LWS recurrences. We refer to this as the ``simultaneous'' dynamic
programming approach.

Previous approaches point out that when the weight of each part is monotonic
and constrained, we can use lower and upper bounds on split points in our
dynamic program to obtain heuristic improvements
\cite{grandjean_optimal_2012, ahrens_optimal_2020}. First, when evaluating
the recurrence of \eqref{eq:lws}, if we evaluate $i$ starting at $i' - 1$ and
work backwards towards $1$, the weight of the candidate part will only
increase, and we can stop the recurrence as soon as the weight limit is
exceeded. Note also that if $i'_1 > i'_2$, the first feasible $i_1$ will be
greater than the first feasible $i_2$.

Additionally, when $K$ is fixed, we can bound the range of feasible $i'$ by
considering the least and greatest $i'$ might be given that the previous $k$
parts and following $K - k$ parts are weight constrained. If $s_{\low, k} <
s_k < s_{\high, k}$ are the bounds on the split points for the $k^{th}$
partition, then we can define $s_{\low,k}$ as the least index satisfying
$w(s_{\low,k}:s_{\low, k + 1} - 1) \leq w_{\max}$, and $s_{\high,k}$ as the
greatest index satisfying $w(s_{\high,k + 1}:s_{\high, k} - 1) \leq
w_{\max}$. We can compute these split point bounds in linear time using
online dominance counters and simply walking the matrix once from top to
bottom and once from bottom to top. If the weight is simply a limit
$w_{\max}$ on the number of rows in each part, these heuristic improvements
would bring the running time down to $O(w_{\max} * m + N)$ and
$O(K(w_{\max} * m + N))$, respectively. The ``simultaneous'' version would run in
time $O(K(w_{\max} * m) + N)$.

Unfortunately, these techniques still require roughly quadratic time in
general. We now show how the techniques described in this paper can bring the
runtime down to log-linear time using convexity or concavity in the cost
function.

Efficient algorithms exist to solve \eqref{eq:lws} when $f$ is concave or
convex \cite{galil_speeding_1989, klawe_almost_1990, wilber_concave_1988}.
In this work, we focus on linearithmic solutions because they are simple to
implement and work well in practice. The solutions of Galil and Giancarlo can
solve \eqref{eq:lws} for convex or concave cost functions in $m\log(m)$
steps, using only one query to the cost function in each step. Unfortunately,
these queries are unstructured, so we muse use offline dominance counters.
Using Chazelle's parameterization in rank space where $b = 2$ in Table
\ref{tbl:offline_dominance}, the runtime of the resulting least weight
subsequence algorithm becomes $O(N\log(N) + m\log(m)\log(N))$, and when $K$
is fixed, the runtime becomes $O(N\log(N) + K\cdot m\log(m)\log(N))$.

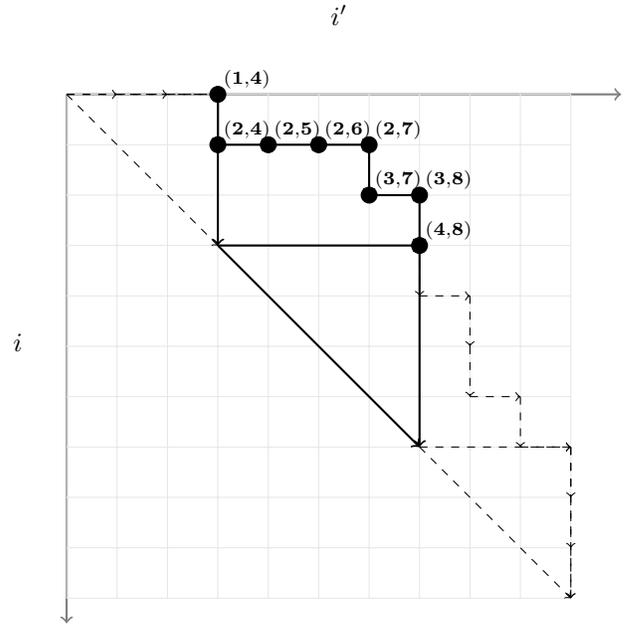
\begin{figure}
\[
    \begin{array}{cc} & i' \\\\ i & \vcenter{\hbox{\begin{tikzpicture}
    \draw[thick, gray, ->] (0, 20em) -- (22em, 20em) node[midway, below, black] {};
    \draw[thick, gray, ->] (0, 20em) -- (0, -1em) node[midway, left, black] {};
    \draw[step=2em,gray!20,very thin] (0, 0) grid (20em, 20em);
    \draw[dashed, black, ->] (0em, 20em) -- (2em, 20em);
    \draw[dashed, black, ->] (2em, 20em) -- (4em, 20em);
    \draw[dashed, black, ->] (4em, 20em) -- (6em, 20em);
    \draw[thick, black, ->] (6em, 20em) -- (6em, 18em);
    \draw[thick, black, ->] (6em, 18em) -- (8em, 18em);
    \draw[thick, black, ->] (8em, 18em) -- (10em, 18em);
    \draw[thick, black, ->] (10em, 18em) -- (12em, 18em);
    \draw[thick, black, ->] (12em, 18em) -- (12em, 16em);
    \draw[thick, black, ->] (12em, 16em) -- (14em, 16em);
    \draw[thick, black, ->] (14em, 16em) -- (14em, 14em);
    \draw[dashed, black, ->] (14em, 14em) -- (14em, 12em);
    \draw[dashed, black, ->] (14em, 12em) -- (16em, 12em);
    \draw[dashed, black, ->] (16em, 12em) -- (16em, 10em);
    \draw[dashed, black, ->] (16em, 10em) -- (16em, 8em);
    \draw[dashed, black, ->] (16em, 8em) -- (18em, 8em);
    \draw[dashed, black, ->] (18em, 8em) -- (18em, 6em);
    \draw[dashed, black, ->] (18em, 6em) -- (20em, 6em);
    \draw[dashed, black, ->] (20em, 6em) -- (20em, 4em);
    \draw[dashed, black, ->] (20em, 4em) -- (20em, 2em);
    \draw[dashed, black, ->] (20em, 2em) -- (20em, 0em);
    "\filldraw [black, inner sep = 2pt] (6em, 20em) circle (3.0pt) node[above right, black, font=\scriptsize] {$\mathbf{(1{,}4)}$};
"
    "\filldraw [black, inner sep = 2pt] (6em, 18em) circle (3.0pt) node[above right, black, font=\scriptsize] {$\mathbf{(2{,}4)}$};
"
    "\filldraw [black, inner sep = 2pt] (8em, 18em) circle (3.0pt) node[above right, black, font=\scriptsize] {$\mathbf{(2{,}5)}$};
"
    "\filldraw [black, inner sep = 2pt] (10em, 18em) circle (3.0pt) node[above right, black, font=\scriptsize] {$\mathbf{(2{,}6)}$};
"
    "\filldraw [black, inner sep = 2pt] (12em, 18em) circle (3.0pt) node[above right, black, font=\scriptsize] {$\mathbf{(2{,}7)}$};
"
    "\filldraw [black, inner sep = 2pt] (12em, 16em) circle (3.0pt) node[above right, black, font=\scriptsize] {$\mathbf{(3{,}7)}$};
"
    "\filldraw [black, inner sep = 2pt] (14em, 16em) circle (3.0pt) node[above right, black, font=\scriptsize] {$\mathbf{(3{,}8)}$};
"
    "\filldraw [black, inner sep = 2pt] (14em, 14em) circle (3.0pt) node[above right, black, font=\scriptsize] {$\mathbf{(4{,}8)}$};
"
    \draw[dashed, black, ->] (0em, 20em) -- (6em, 20em);
    \draw[thick, black, ->] (6em, 20em) -- (6em, 14em);
    \draw[dashed, black, ->] (0em, 20em) -- (6em, 14em);
    \draw[thick, black, ->] (6em, 14em) -- (14em, 14em);
    \draw[thick, black, ->] (14em, 14em) -- (14em, 6em);
    \draw[thick, black, ->] (6em, 14em) -- (14em, 6em);
    \draw[dashed, black, ->] (14em, 6em) -- (20em, 6em);
    \draw[dashed, black, ->] (20em, 6em) -- (20em, 0em);
    \draw[dashed, black, ->] (14em, 6em) -- (20em, 0em);

\end{tikzpicture}}}\end{array}\\
\]
\caption{The feasible regions of setup and cleanup phases under a weight
limit of 12 nonzeros per part. The first (nontrivial) setup phase (irregular
upper region) and cleanup phase (triangular lower region) are displayed in
bold. The pairs $(\sigma_1, \sigma'_1), ... (\sigma_8, \sigma'_8)$ are shown
from the upper left to the lower right of the feasible region of the setup
phase. Compare this figure to the strictly triangular phases of \cite[Figure
2]{eppstein_sequence_1990}\label{fig:feasible}.}
\end{figure}

However, while concave cost functions are still concave after adding monotonic
balance constraints, convex functions are no longer convex after adding such
constraints. Since most of our atoms are convex, we must modify our least weight
subsequence algorithm to handle such cases. We take a similar approach to
Eppstein, who formulated an algorithm for cost functions which are alternately
convex, then concave, then convex, et cetera \cite{eppstein_sequence_1990}.
While the alternations considered by Eppstein were constrained to fixed width
intervals, we consider only the case where we have a convex cost subject to an
upper limit on a monotonic weight function. The width of the feasible region may
therefore be irregular. While we will not consider an arbitrary number of
alternations in the cost, our transition from a convex cost to the concave
constraint is more general than the fixed-width intervals considered by
Eppstein.

Let $p_{\low, i'}$ be the least index such that $w(p_{\low, i'} : i' - 1) \leq
w_{\max}$. Notice that if $i'_1 \leq i'_2$, $p_{\low, i'_1} \leq p_{\low,
i'_2}$. Consider the sequence of at most $m$ split points $t_1, t_2, ...$
such that $t_l = p_{\low, t_{l + 1}}$. We can construct $t$ in reverse
starting at $m$ and following the pointers $p_{\low}$. Our algorithm to
compute \eqref{eq:lws} works in batches corresponding to each range $t_k : t_{k + 1} - 1$.

Given $c_1, ..., c_{t_l}$, consider the problem of computing
\[
c_{i'} = \min\limits_{1 \leq i < i'} d_i + f(i:i' - 1)
\]
for $i' \in \{t_l, ..., t_{l + 1} - 1\}$. We split the feasible range of the
indices into two subproblems, a more complicated setup phase,
\[
c_{\text{setup}, i'} = \min\limits_{p_{\low, i'} \leq i < t_{l}} d_i + f(i:i' - 1),
\]
and a simpler cleanup phase,
\[
c_{i'} = \min(c_{\text{setup}, i'}, \min\limits_{t_{l} \leq i < i'} d_i + f(i:i' - 1)).
\]
The cleanup phase can be computed with any standard algorithm for
\eqref{eq:lws}, but we need to transform the setup phase first before
reducing to \eqref{eq:lws}. If $a, b = t_l, t_{l + 1}$, let $\sigma, \sigma'$
be defined as
\begin{align*}
    \sigma_1 &= p_{\low, a}, & \sigma'_1 &= a, \\
    \sigma_2 &=  p_{\low, a} + 1, & \sigma'_2 &= a, \\
    \vdots &\quad\quad \vdots & \vdots &\quad \vdots \\
    \sigma_{...} &=  p_{\low, a + 1}, & \sigma'_{...} &= a + 1, \\
    \sigma_{...} &=  p_{\low, a + 1} + 1, & \sigma'_{...} &= a + 1, \\
    \vdots &\quad\quad \vdots & \vdots &\quad \vdots \\
    \sigma_{2(b - a)} &=  p_{\low, b - 1}, & \sigma'_{2(b - a)} &= b - 1, \\
\end{align*}
Since $p_{\low, b - 1} < a$, every element of $\sigma$ is less than every
element of $\sigma'$. When $j \leq j'$, the interval $\sigma_j : \sigma'_j$
is feasible. Finally, if $a \leq i' < b$ and $i: i'$ is feasible, there exists
$j < j'$ where $\sigma_j = \sigma_j$. Thus, we can compute the setup phase
with the transformed least weight subsequence problem
\[
c_{\text{transformed}, j'} = \min\limits_{1 \leq j \leq 2(b - a)} d_{\sigma_{i}} + f(\sigma_{i}:\sigma'_{i'} - 1),
\]
and finally setting $c_{\text{setup}, i'} = c_{\text{transformed}, j'}$,
where $j'$ is the greatest index such that $\sigma_j' = i'$. We pick the greatest
such index so that the full feasible range of $i$ is included.

Each least weight subsequence subproblem can be computed with Galil and
Giancarlo's algorithm using $O(l\log(l))$ steps, where $l$ is the length of
the problem. Thus, the setup phase of each segment of length $(b - a)$ can be
computed in $O(2(a - b)\log(2(a - b)))$ steps and the cleanup phase takes
$O((b - a)\log(b - a))$ steps. Since $b - a < m$ and there are at most $m$
segments, the overall algorithm again runs in $O(m\log(m))$ steps. Each step
involves a constant number of calls to the cost function. While linearithmic
LWS algorithms obviate the need for feasible range optimizations within each
LWS subproblem, we can still use the feasible range to restrict the size of each
LWS subproblem in our $K$-step solution to the fixed-$K$ partitioning problem.

\section{Results}

We compare our partitioners on a set of 42
sparse matrices. We include matrices used by Pinar and Aykanat to evaluate
solutions to the Chains-On-Chains problem \cite{pinar_fast_2004} (minimizing the
maximum work without reordering), matrices used by \c{C}atalyurek and Aykanat to
evaluate multilevel hypergraph partitioners (minimizing the sum of
communication) \cite{catalyurek_hypergraph-partitioning-based_1999}, and
matrices used by Grandjean and U\c{c}ar to evaluate contiguous hypergraph
partitioners (minimizing the sum of communication without reordering)
\cite{grandjean_optimal_2012}. Since we intend to partition without reordering,
we also chose some additional matrices of our own to introduce more variety in
the tested sparsity patterns. We exclude ``Qaplib/lp\_nug30'' as KaHyPar could
not partition it in under 24 hours. The test suite is summarized in Table
\ref{tbl:matrices}.

\begin{table}
    \centering
    \begin{multicols}{2}
        \centering
    \caption{Symmetric and asymmetric test matrices used in our test suite. The SI prefixes
    ``$\mu$,'' ``m,'' ``K,'' and ``M'' represent factors of $10^{-6}$,
    $10^{-3}$, $10^3$, and $10^6$, respectively.} \label{tbl:matrices}
        \vfill
        Symmetric Matrices
        
        \setlength{\tabcolsep}{2pt}\begin{tabular}{rrr}\hline
\multicolumn{1}{l}{\textbf{{\scriptsize \makecell[l]{Group /\\ \quad Matrix}}}} & \multicolumn{1}{r}{\textbf{$m \times n$}} & \multicolumn{1}{r}{\textbf{$N$}}\\
\hline
{\tiny \makecell[l]{Boeing/\\ \quad ct20stif}} & $52.3$K$\times$$52.3$K & $2.7$M\\
{\tiny \makecell[l]{Boeing/\\ \quad pwtk}} & $218$K$\times$$218$K & $11.6$M\\
{\tiny \makecell[l]{Chen/\\ \quad pkustk03}} & $63.3$K$\times$$63.3$K & $3.13$M\\
{\tiny \makecell[l]{Chen/\\ \quad pkustk14}} & $152$K$\times$$152$K & $14.8$M\\
{\tiny \makecell[l]{Cunningham/\\ \quad qa8fk}} & $66.1$K$\times$$66.1$K & $1.66$M\\
{\tiny \makecell[l]{Gupta/\\ \quad gupta2}} & $62.1$K$\times$$62.1$K & $4.25$M\\
{\tiny \makecell[l]{HB/\\ \quad bcsstk30}} & $28.9$K$\times$$28.9$K & $2.04$M\\
{\tiny \makecell[l]{HB/\\ \quad bcsstk32}} & $44.6$K$\times$$44.6$K & $2.01$M\\
{\tiny \makecell[l]{HB/\\ \quad dwt\_607}} & $607$$\times$$607$ & $5.13$K\\
{\tiny \makecell[l]{HB/\\ \quad sherman3}} & $5$K$\times$$5$K & $20$K\\
{\tiny \makecell[l]{Hamm/\\ \quad bcircuit}} & $68.9$K$\times$$68.9$K & $376$K\\
{\tiny \makecell[l]{Mulvey/\\ \quad finan512}} & $74.8$K$\times$$74.8$K & $597$K\\
{\tiny \makecell[l]{Nasa/\\ \quad nasasrb}} & $54.9$K$\times$$54.9$K & $2.68$M\\
{\tiny \makecell[l]{PARSEC/\\ \quad H2O}} & $67$K$\times$$67$K & $2.22$M\\
{\tiny \makecell[l]{Rothberg/\\ \quad cfd1}} & $70.7$K$\times$$70.7$K & $1.83$M\\
{\tiny \makecell[l]{Schenk\_IBMNA/\\ \quad c-64}} & $51$K$\times$$51$K & $718$K\\
{\tiny \makecell[l]{Simon/\\ \quad venkat01}} & $62.4$K$\times$$62.4$K & $1.72$M\\
{\tiny \makecell[l]{TKK/\\ \quad smt}} & $25.7$K$\times$$25.7$K & $3.75$M\\
\end{tabular}
 
        \columnbreak
        \vspace*{\fill}

        Asymmetric Matrices
        \setlength{\tabcolsep}{2pt}\begin{tabular}{rrr}\hline
\multicolumn{1}{l}{\textbf{{\scriptsize \makecell[l]{Group /\\ \quad Matrix}}}} & \multicolumn{1}{r}{\textbf{$m \times n$}} & \multicolumn{1}{r}{\textbf{$N$}}\\
\hline
{\tiny \makecell[l]{ATandT/\\ \quad onetone1}} & $36.1$K$\times$$36.1$K & $341$K\\
{\tiny \makecell[l]{ATandT/\\ \quad onetone2}} & $36.1$K$\times$$36.1$K & $228$K\\
{\tiny \makecell[l]{Averous/\\ \quad epb3}} & $84.6$K$\times$$84.6$K & $464$K\\
{\tiny \makecell[l]{Bomhof/\\ \quad circuit\_4}} & $80.2$K$\times$$80.2$K & $308$K\\
{\tiny \makecell[l]{Grund/\\ \quad bayer01}} & $57.7$K$\times$$57.7$K & $278$K\\
{\tiny \makecell[l]{HB/\\ \quad gemat11}} & $4.93$K$\times$$4.93$K & $33.2$K\\
{\tiny \makecell[l]{Hollinger/\\ \quad g7jac180}} & $53.4$K$\times$$53.4$K & $747$K\\
{\tiny \makecell[l]{Hollinger/\\ \quad mark3jac140sc}} & $64.1$K$\times$$64.1$K & $400$K\\
{\tiny \makecell[l]{LPnetlib/\\ \quad lp\_cre\_b}} & $9.65$K$\times$$77.1$K & $261$K\\
{\tiny \makecell[l]{LPnetlib/\\ \quad lp\_cre\_d}} & $8.93$K$\times$$73.9$K & $247$K\\
{\tiny \makecell[l]{LPnetlib/\\ \quad lp\_ken\_11}} & $14.7$K$\times$$21.3$K & $49.1$K\\
{\tiny \makecell[l]{LPnetlib/\\ \quad lp\_ken\_13}} & $28.6$K$\times$$42.7$K & $97.2$K\\
{\tiny \makecell[l]{LPnetlib/\\ \quad lpi\_gosh}} & $3.79$K$\times$$13.5$K & $100$K\\
{\tiny \makecell[l]{Mallya/\\ \quad lhr07}} & $7.34$K$\times$$7.34$K & $157$K\\
{\tiny \makecell[l]{Mallya/\\ \quad lhr14}} & $14.3$K$\times$$14.3$K & $308$K\\
{\tiny \makecell[l]{Mallya/\\ \quad lhr17}} & $17.6$K$\times$$17.6$K & $382$K\\
{\tiny \makecell[l]{Mallya/\\ \quad lhr34}} & $35.2$K$\times$$35.2$K & $764$K\\
{\tiny \makecell[l]{Mallya/\\ \quad lhr71c}} & $70.3$K$\times$$70.3$K & $1.53$M\\
{\tiny \makecell[l]{Meszaros/\\ \quad co9}} & $10.8$K$\times$$22.9$K & $110$K\\
{\tiny \makecell[l]{Meszaros/\\ \quad cq9}} & $9.28$K$\times$$21.5$K & $96.7$K\\
{\tiny \makecell[l]{Meszaros/\\ \quad mod2}} & $34.8$K$\times$$66.4$K & $200$K\\
{\tiny \makecell[l]{Meszaros/\\ \quad nl}} & $7.04$K$\times$$15.3$K & $47$K\\
{\tiny \makecell[l]{Meszaros/\\ \quad world}} & $34.5$K$\times$$67.1$K & $199$K\\
{\tiny \makecell[l]{Shyy/\\ \quad shyy161}} & $76.5$K$\times$$76.5$K & $330$K\\
\end{tabular}

    \end{multicols}
\end{table}

In addition to partitioning natural orderings, we also evaluate our algorithms
on two of the most popular bandwidth-reducing algorithms, the Reverse
Cuthill-McKee (RCM) \cite{cuthill_reducing_1969} and spectral
\cite{pothen_partitioning_1990} orderings, which have a history of application
to partitioning \cite{kolda_partitioning_1998, manguoglu_pspike_2009}. The RCM
ordering uses a breadth-first traversal of the graph, prioritizing lower degree
vertices first. We follow the linear-time implementation described in
\cite{chan_linear_1980}. The spectral reordering orders vertices by their values
in the Fiedler vector, or the second-smallest eigenvector of the Laplacian
matrix. We used the default eigensolver given in the Laplacians.jl library
(\url{github.com/danspielman/Laplacians.jl}), the only eigensolver we tested
that was able to solve all our largest, worst-conditioned problems in a
reasonable time. In order to apply these reorderings to asymmetric matrices, we
instead reorder a bipartite graph where nodes correspond to rows and columns in
our original matrix, which are connected by edges when nonzeros lie at the
intersection of a row and a column, as described by Berry et. al.
\cite{berry_sparse_1996}.

Several results are presented as performance profiles, allowing us to compare
our partitioners on the entire dataset at once \cite{dolan_benchmarking_2002}.
Partitioners are first measured by the relative deviation from the best
performing partitioner for each matrix individually. We then display, for each
partitioner, the fraction of test instances that achieve each target deviation
from the best partitioner.

Our partitioners optimize different cost models, but we use the same
coefficients whenever possible. Since the coefficients should depend on which
linear solver is to be partitioned for which parallel machine, we use powers of
10 which are easy to understand and correspond to the likely orders of magnitude
involved. We assume that the cost of processing one nonzero in an SpMV is
$c_{\textbf{entry}} = 1$. There are overheads to starting the multiplication
loop in each row and several per-row costs incurred by dot products in linear
solvers. We assume that the per-row cost is an order of magnitude larger;
$c_{\textbf{row}} = 10$. The peak MPI bandwidth on Cori, a supercomputer at the
National Energy Research Scientific Computing Center, has been measured at
approximately 8 GB/s \cite{doerfler_evaluating_2018}. Cori uses two
Intel\textregistered Xeon\textregistered Processor E5-2698 v3 CPUs on each node
(one per socket).  The peak computational throughput of one CPU is advertised as
1.1 TB/s (256-bit SIMD lane at 2.3 GHz on each of 16 cores). Since this is
approximately two orders of magnitude faster than the communication bandwidth,
we set $c_{\textbf{message}} = 100$.

The partitioners we tested are described in Table \ref{tbl:partitioners}. Our
asymmetric alternating partitioners partition the primary dimension first,
then partition the secondary dimension. We found that continued alternation
did not significantly improve solution quality, so we do not consider
partitioners which attempt to refine the initial partitioning.

All experiments were run on a single core of an Intel\textregistered
Xeon\textregistered Processor E5-2695 v2 CPU running at $2.4$GHz with $30$MB of
cache and $128$GB of memory. We implemented our partitioners in Julia 1.5.1 and
normalize against the Julia Standard Library SpMV; we do not expect much
variation across memory-bound single-core SpMV implementations. To measure
runtime, we warm up by running the kernel first, then take the minimum of at
most 10,000 executions or 5 seconds worth of sampling, whichever happens first.
Since some of our algorithms use randomization, we measure the average quality
over 100 trials. 


Partitioning represents a tradeoff between partitioning time and partition
quality. Since the runtime of the partitioner must compete with the runtime
of the solver, we normalize the measured serial runtime of our partitioners
against the measured serial runtime of an SpMV on the same matrix. This also
allows us to use aggregate runtime statistics over the whole test set.


\begin{table}
    \caption{Partitioner components, described under their shorthand labels.
    Algorithms proposed in this work are underlined. Steps may be combined, with
    the convention that the first partitioning step is performed on rows, and
    the second on columns. As an example, ``CuthillMcKee, Balance Conn., Balance
    Conn.'' corresponds to reordering the matrix, then optimally partitioning the
    rows under \eqref{eq:nonsymmetricinitialcost}, then the columns under
    \eqref{eq:nonsymmetriccost}. Labels in parenthesis differentiate which
    algorithm was used, when applicable.}\label{tbl:partitioners}
    \paragraph*{Split Equally}
        Assign an equal number of rows and columns to each part. Contiguous.
    \paragraph*{Balance Work}
        Bottleneck partition to balance load \eqref{eq:workcost}. Contiguous.
    \paragraph*{Assign Local}
        Assign each column to the part of a randomly selected incident row. Noncontiguous.
    \paragraph*{Assign Greedy Conn.}
        Working in random order, assign each column to a part to greedily
        balance bottleneck load $+$ communication \eqref{eq:nonsymmetriccost}. Noncontiguous.
    \paragraph*{\underline{Balance Mono. Conn.}}
        Bottleneck partition to balance monotonized symmetric load $+$
        communication \eqref{eq:monotonicsymmetriccost}.  Contiguous.
    \paragraph*{\underline{Balance Conn.}}
        Bottleneck partition to balance load $+$ communication
        \eqref{eq:nonsymmetricinitialcost} or \eqref{eq:nonsymmetriccost}.
        Contiguous.
    \paragraph*{\underline{Minimize Simple Cut}}
        Total partition to minimize simple edge cut \eqref{eq:edgecut}. Contiguous.
    \paragraph*{\underline{Minimize Hyper. Cut}}
        Total partition to minimize hyperedge cut cost \eqref{eq:hyperedgecut}. Contiguous.
    \paragraph*{\underline{Minimize Conn.}}
        Total partition to minimize $\lambda - 1$ cut cost \eqref{eq:lambdaminusonecut}. Contiguous.
    \paragraph*{\underline{Block Equally}}
        Split equally into parts of cardinality $C$. Contiguous.
    \paragraph*{\underline{Block Conn.}}
        Split into parts of cardinality at most $C$ to minimize total cache misses \eqref{eq:nonsymmetriccost}. Contiguous.
    \paragraph*{Metis}
        Use direct, $K$-way Metis with sorted heavy edge matching to optimize
        total edge cut \eqref{eq:edgecut} with a maximum of $10\%$ work
        imbalance under \eqref{eq:workcost} \cite{karypis_fast_1998}.  If matrix is
        asymmetric, use the symmetrized bipartite representation of matrix to
        produce separate row and column partitions. Noncontiguous.
    \paragraph*{KaHyPar}
        Use direct, $K$-way KaHyPar to optimize $\lambda - 1$ cut
        \eqref{eq:lambdaminusonecut} with a maximum of $10\%$ nonzero imbalance
        \cite{gottesburen_advanced_2020}. Noncontiguous.
    \paragraph*{Spectral}
        Spectrally reorder. Noncontiguous.
    \paragraph*{CuthillMcKee}
        Reorder with Cuthill-McKee. Noncontiguous.
    \paragraph*{\underline{(Exact)}}
        Algorithm \ref{alg:nicolpartition} (Nicol's Algorithm) was used to exactly
        bottleneck partition, and an offline dominance counter with $H=3$ was employed to calculate costs.
    \paragraph*{\underline{(Approx)}}
        Algorithm \ref{alg:bisectpartition} (Bisection) was used to $10\%$-approximately
        bottleneck partition, and an offline dominance counter with $H=3$ was employed to calculate costs.
    \paragraph*{\underline{(Lazy)}}
        Algorithm \ref{alg:bisectpartition} (Bisection) was used to
        $10\%$-approximately bottleneck partition, and a fused online dominance
        counter was employed to calculate costs.
    \paragraph*{(Dynamic)}
        Dynamic programming with split point bounds and constraint-based early
        search termination was used to solve least-weight subsequence problems
        for each part separately. An online dominance counter was used to
        calculate costs.
    \paragraph*{(Dynamic')}
        Dynamic programming with split point bounds and constraint-based early
        search termination was used to solve $K$ simultaneous least-weight
        subsequence problems.  An online dominance counter was used to calculate
        costs.
    \paragraph*{\underline{Quadrangle}}
        Our quadrangle-accelerated algorithm with split point bounds was used to
        solve least-weight-subsequence subproblems for each part separately, and
        an offline dominance counter with $H=3$ was employed to calculate costs.
\end{table}

\begin{figure*}
    \centering
    {\large Contiguous Partition Quality Improvements}

    \centering
    Bottleneck Load $+$ Connectivity On Asymmetric Matrices
    \resizebox{\linewidth}{!}{%
    \includegraphics[]{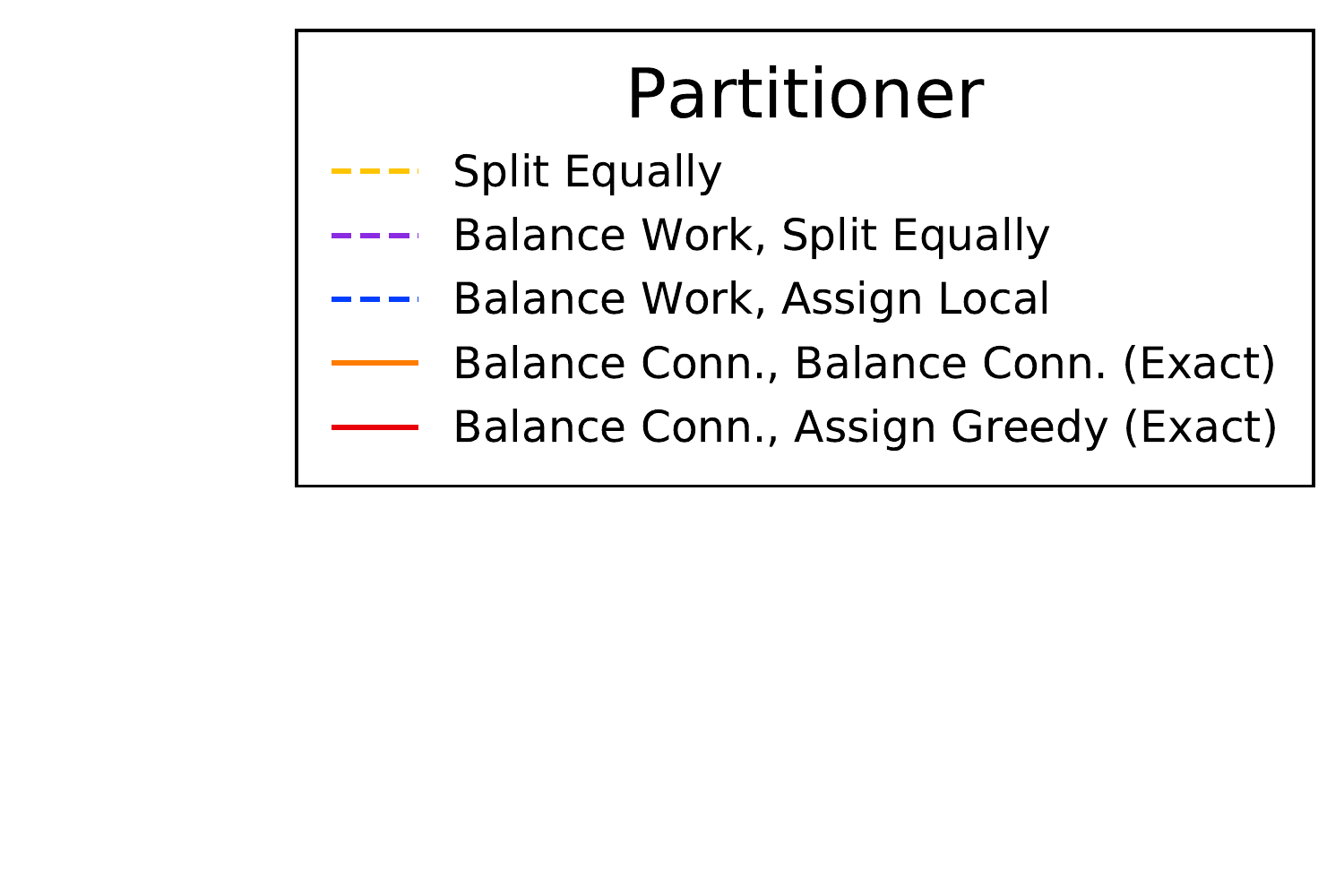}%
    \includegraphics[]{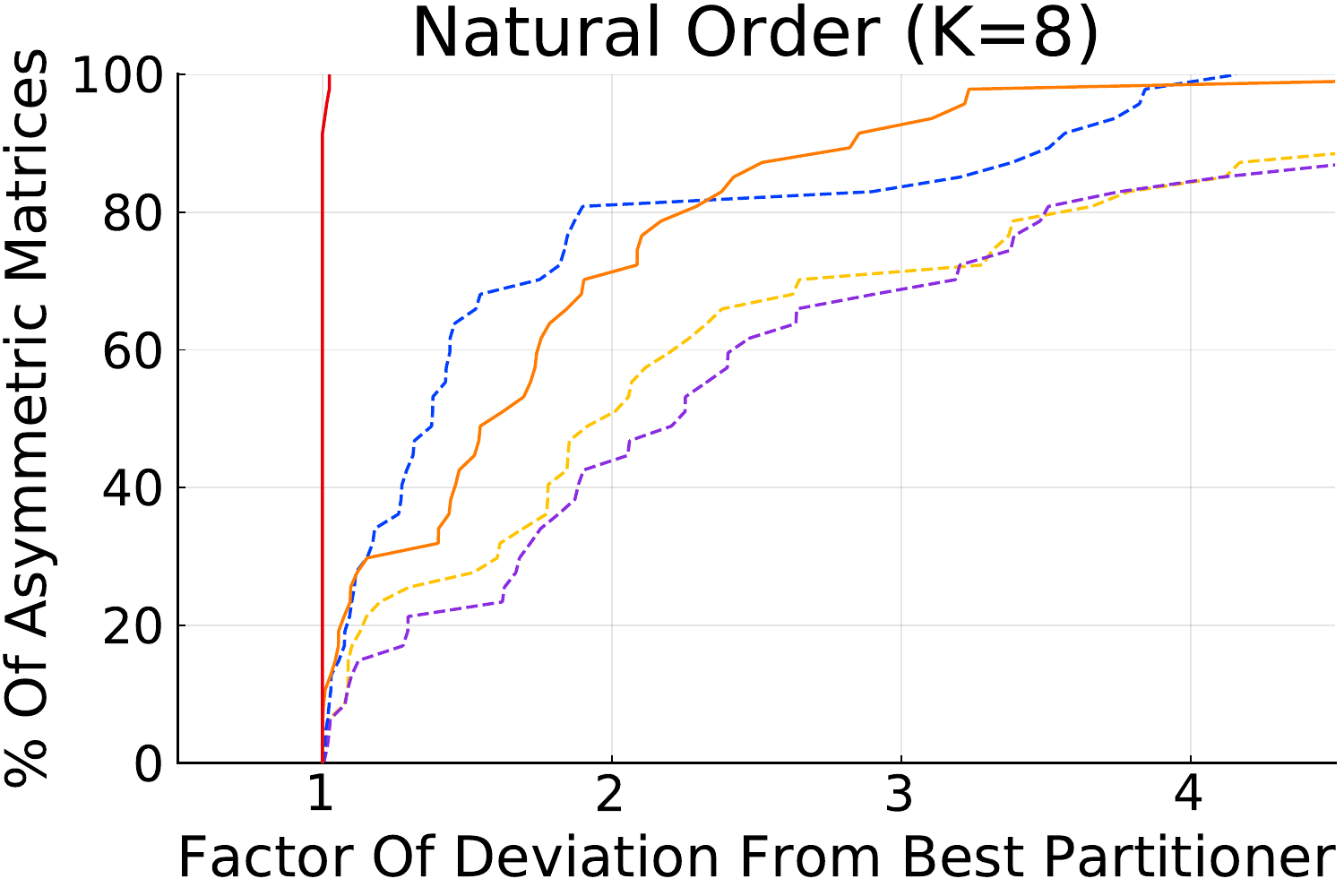}%
    \includegraphics[]{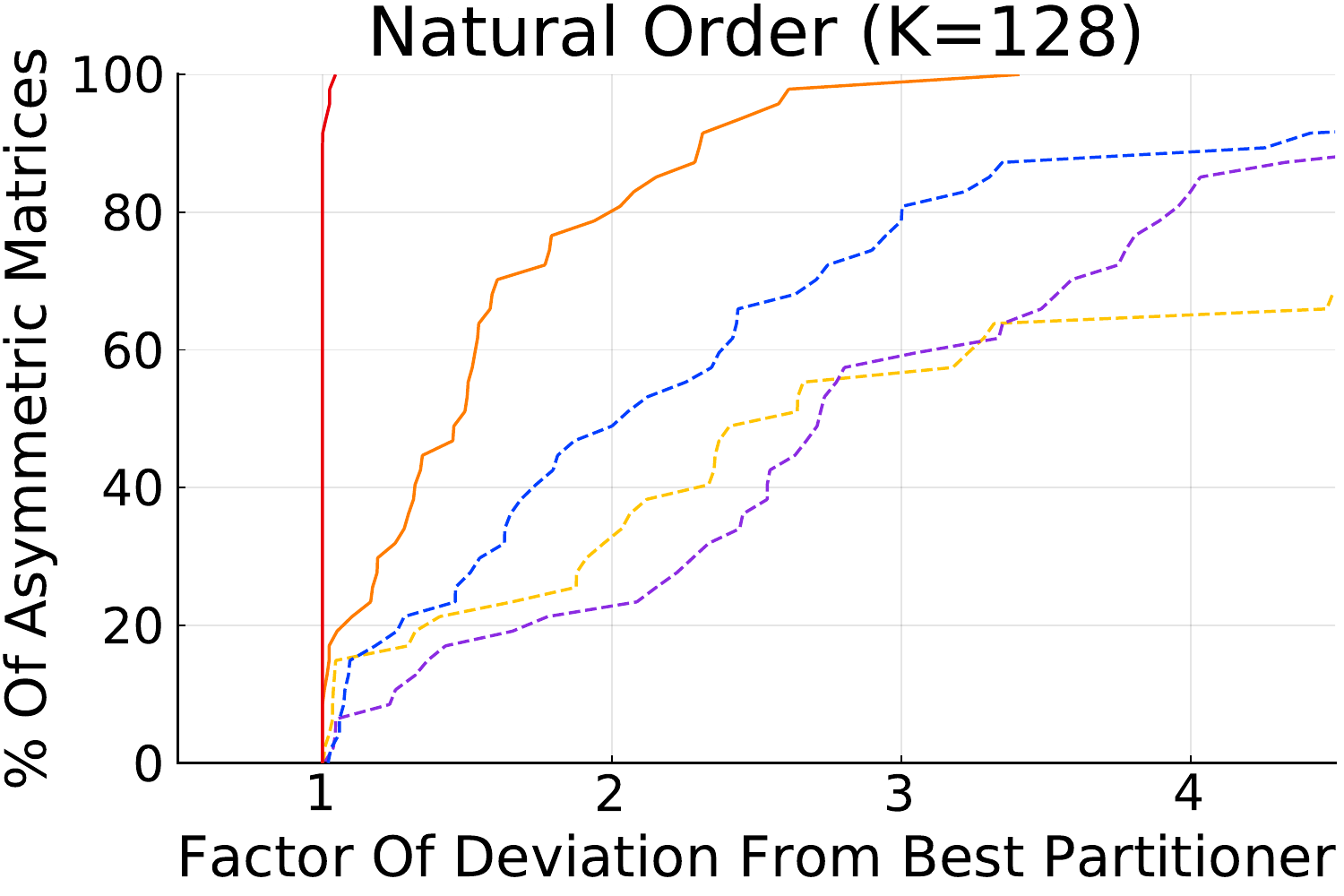}%
    \includegraphics[]{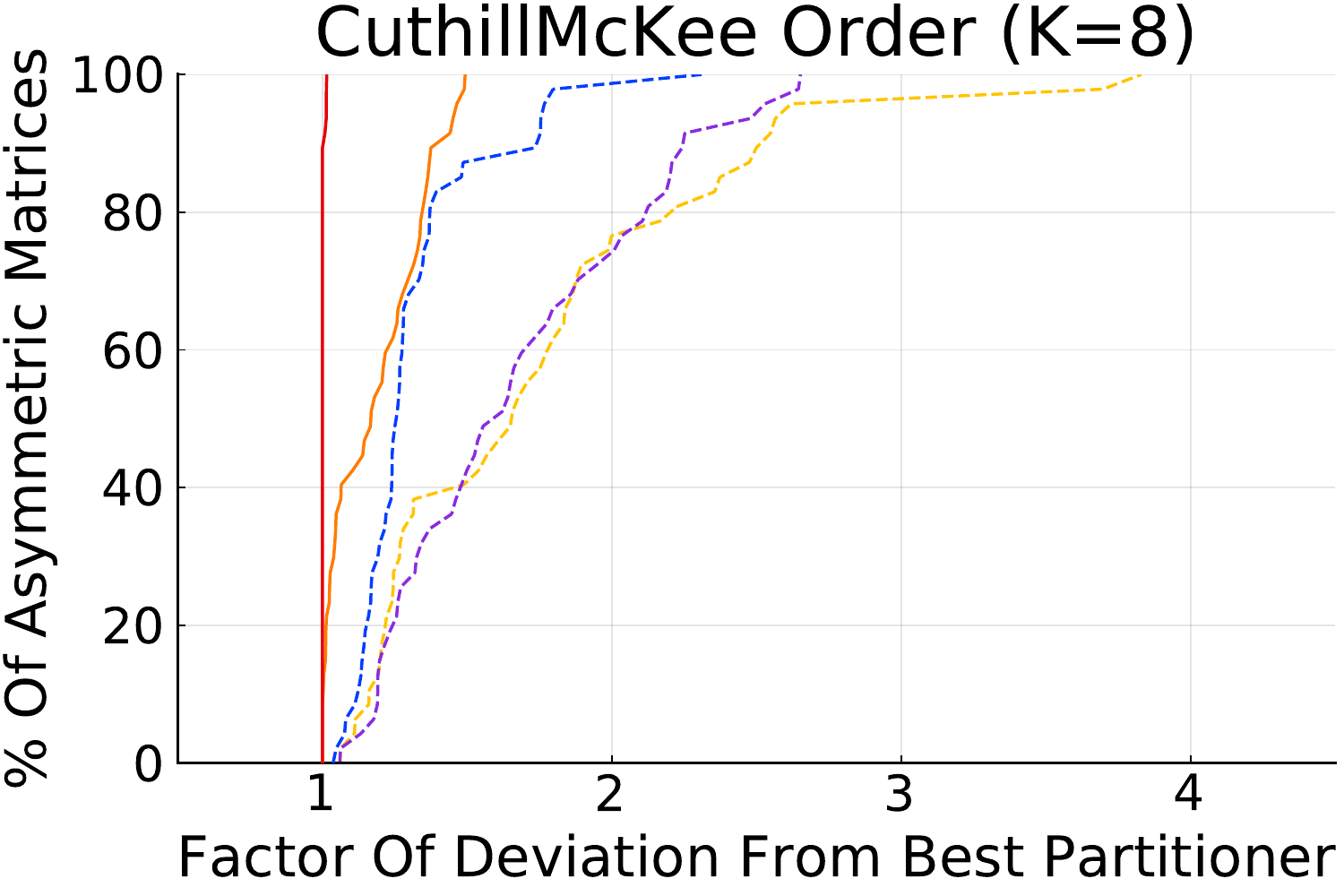}%
    \includegraphics[]{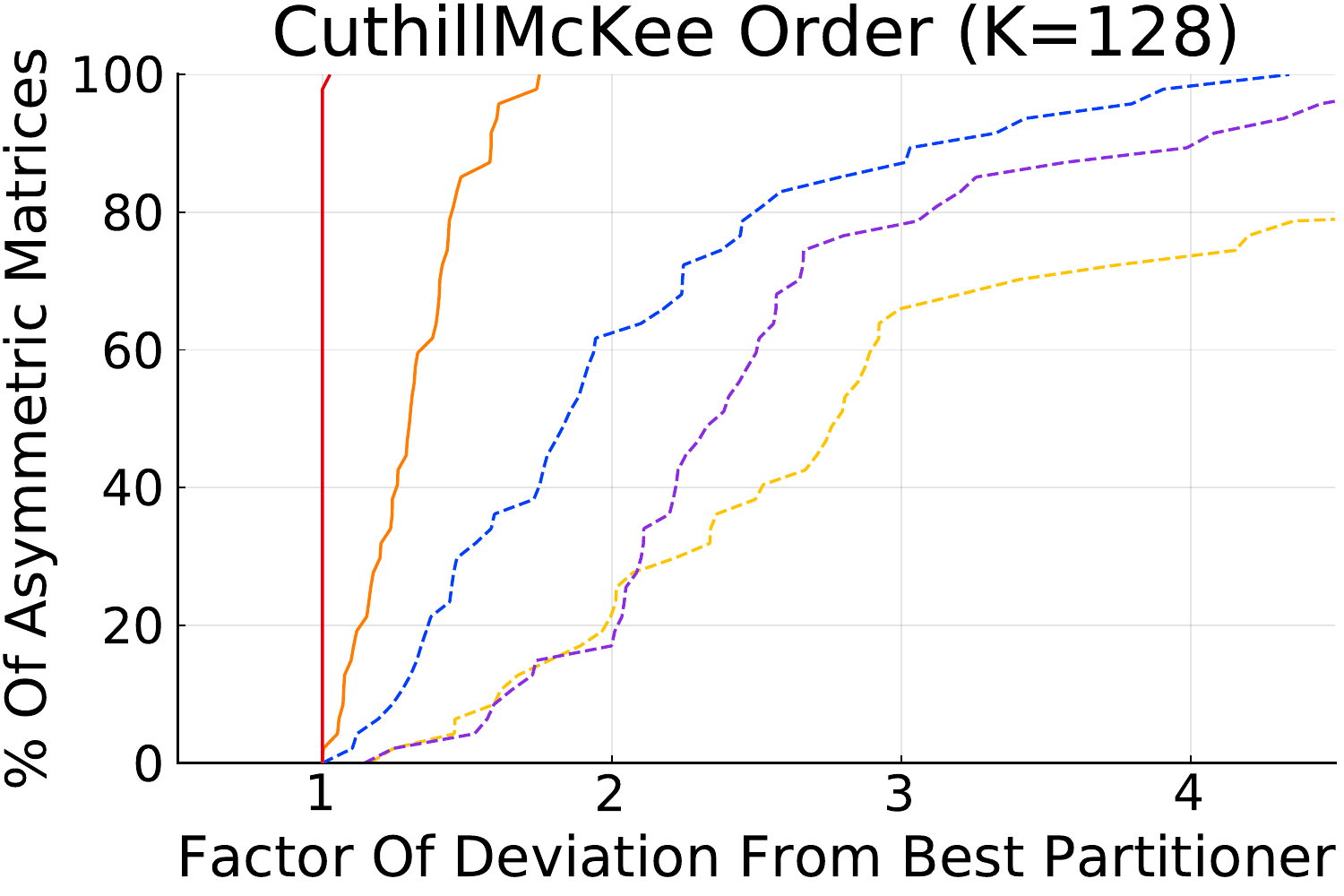}%
    \includegraphics[]{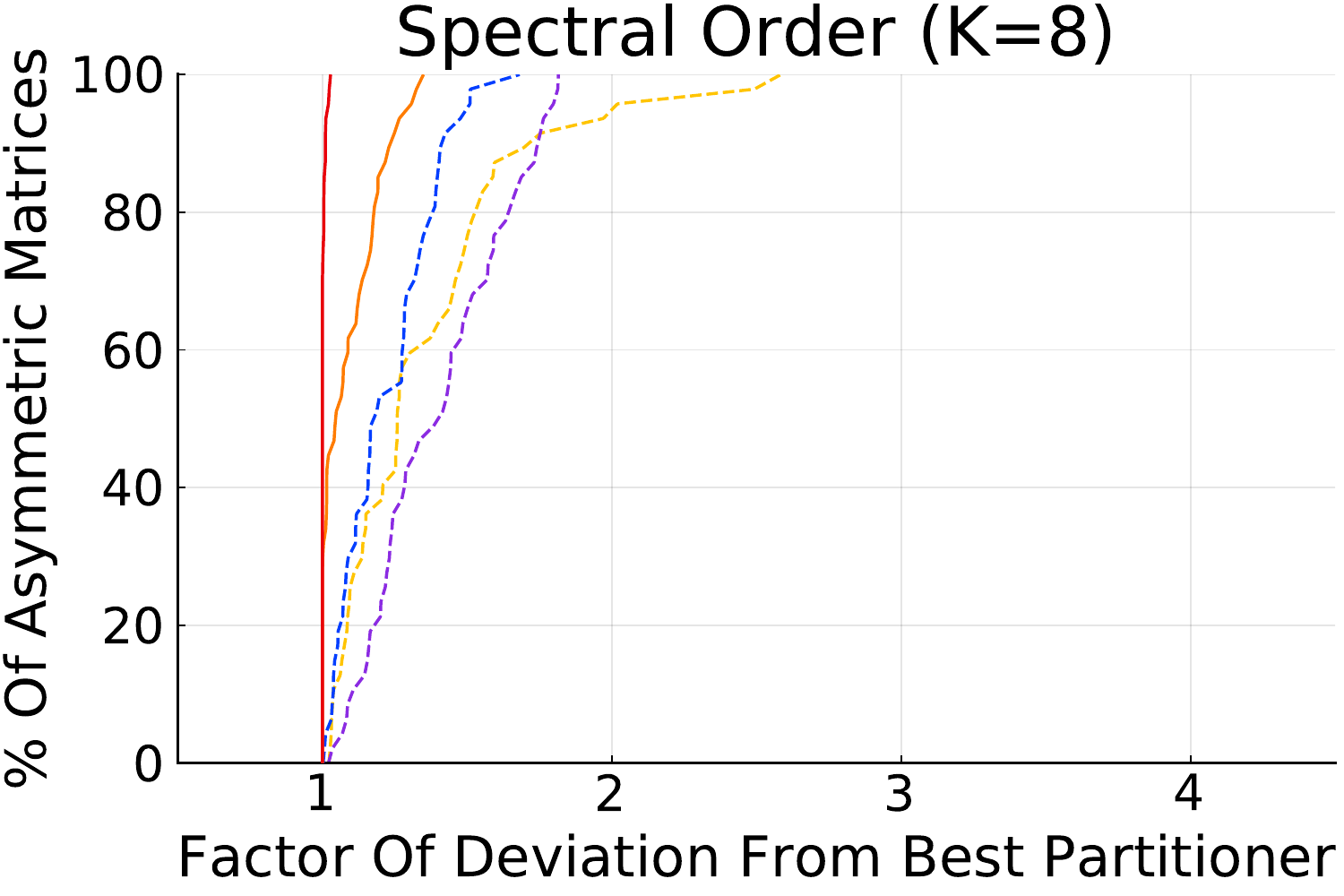}%
    \includegraphics[]{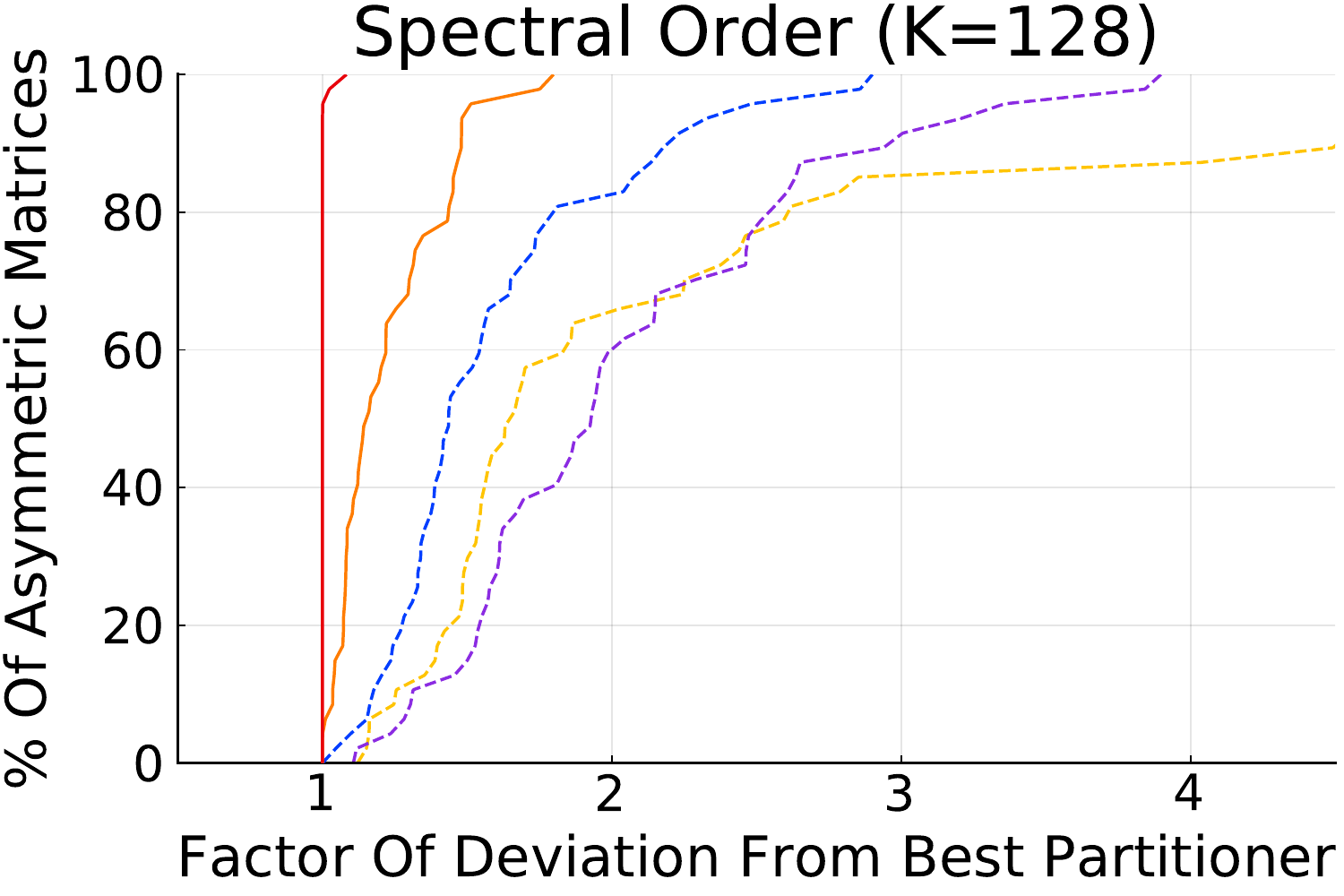}%
    }
    Bottleneck Load $+$ Connectivity On Symmetric Matrices
    \resizebox{\linewidth}{!}{%
    \includegraphics[]{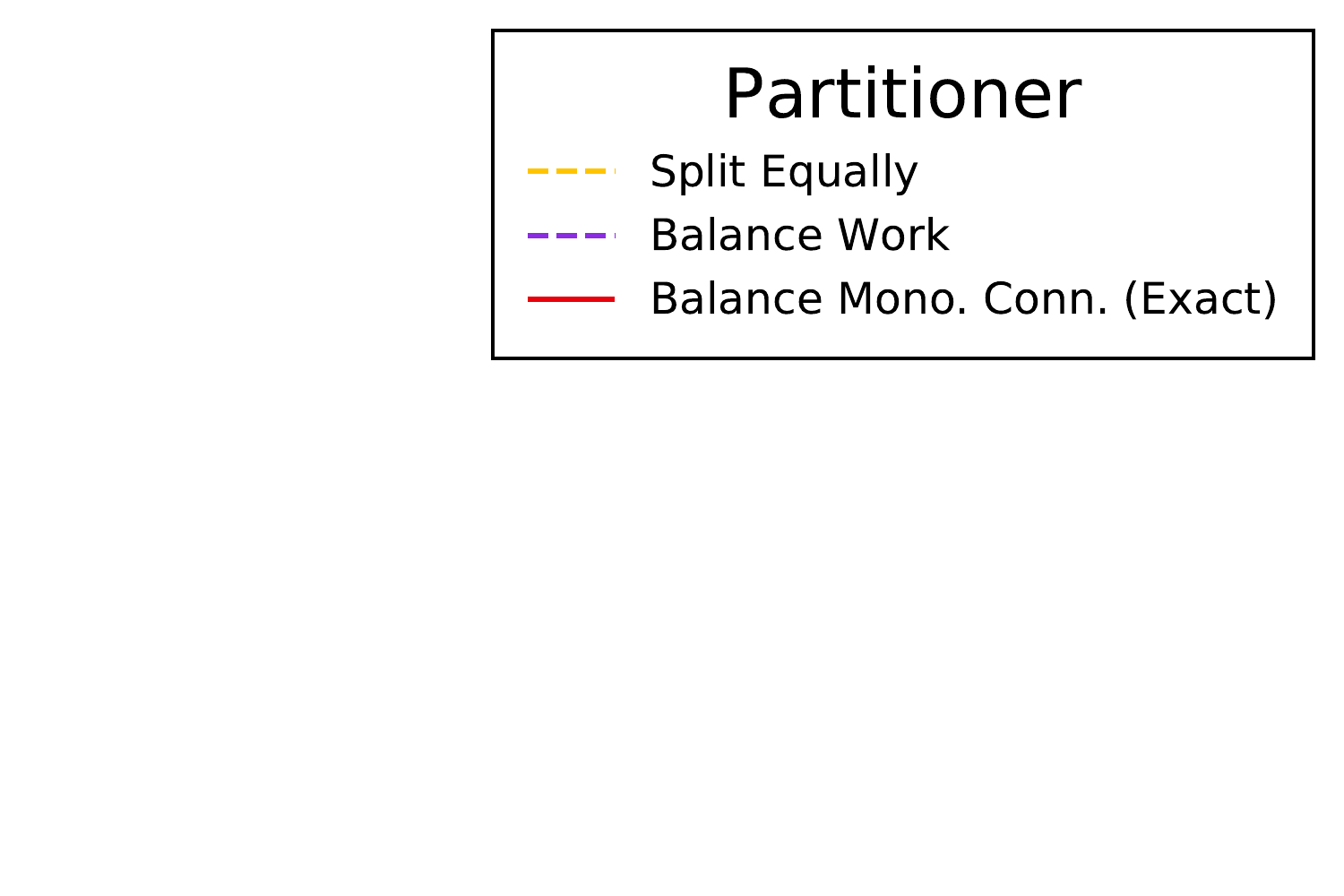}%
    \includegraphics[]{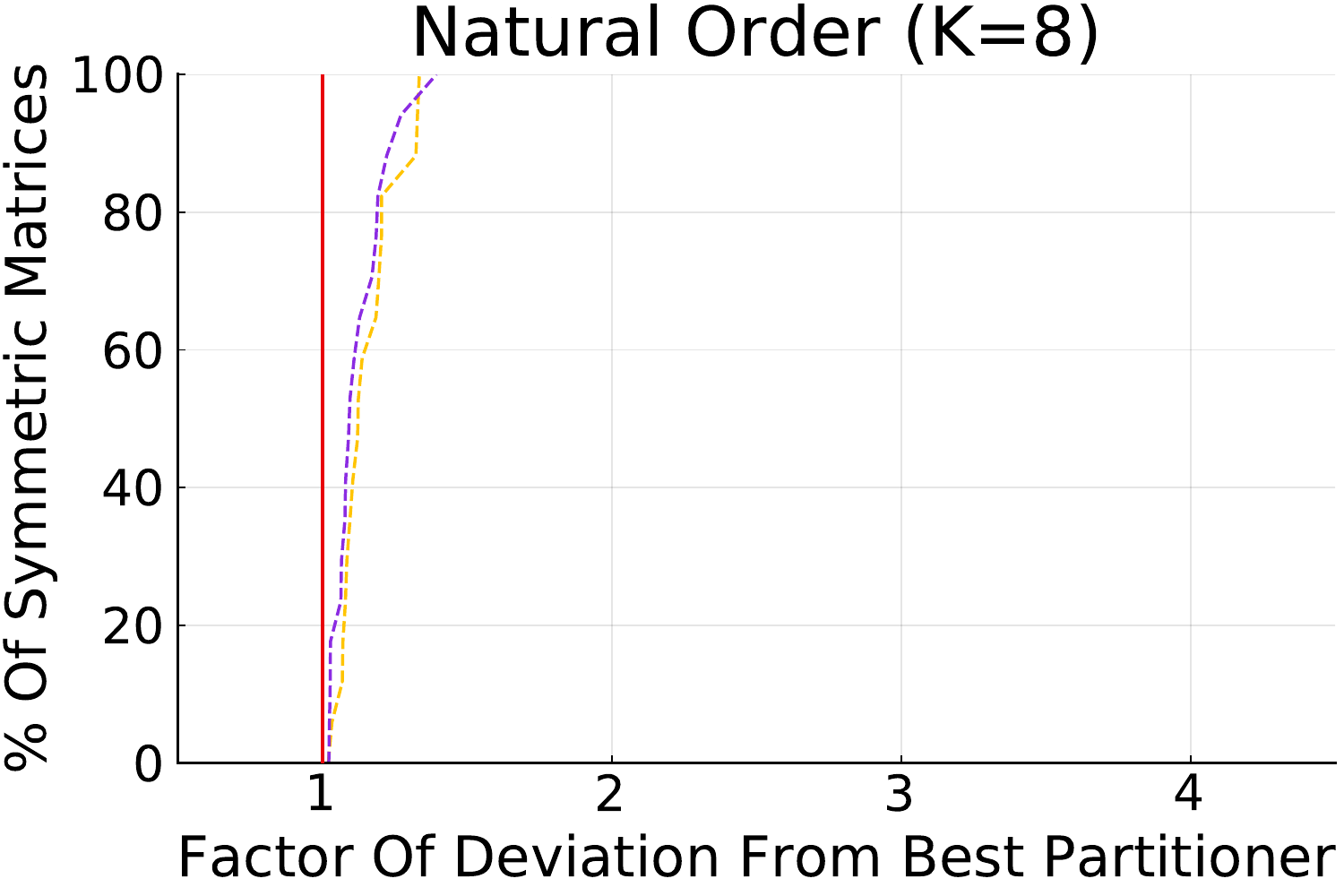}%
    \includegraphics[]{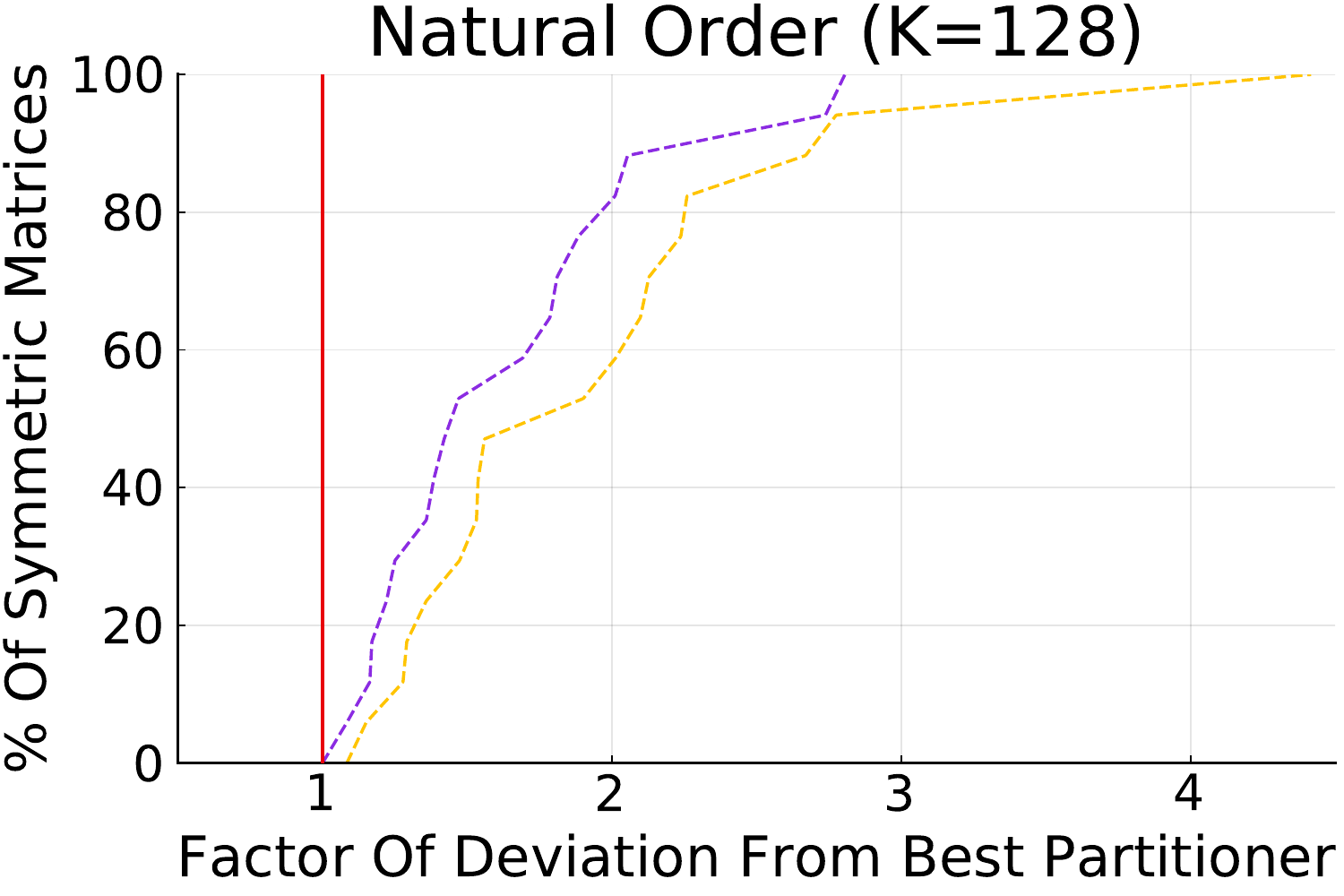}%
    \includegraphics[]{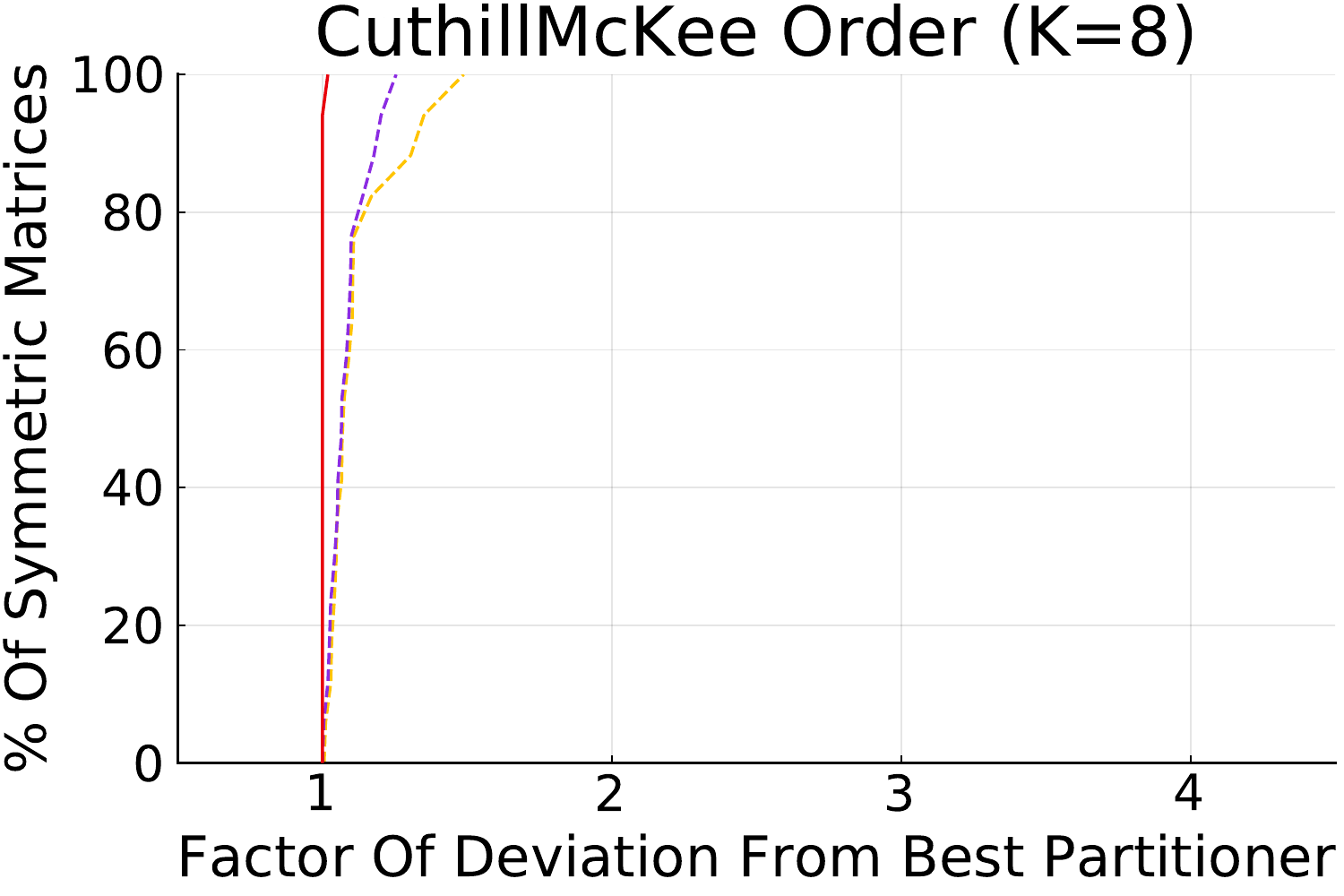}%
    \includegraphics[]{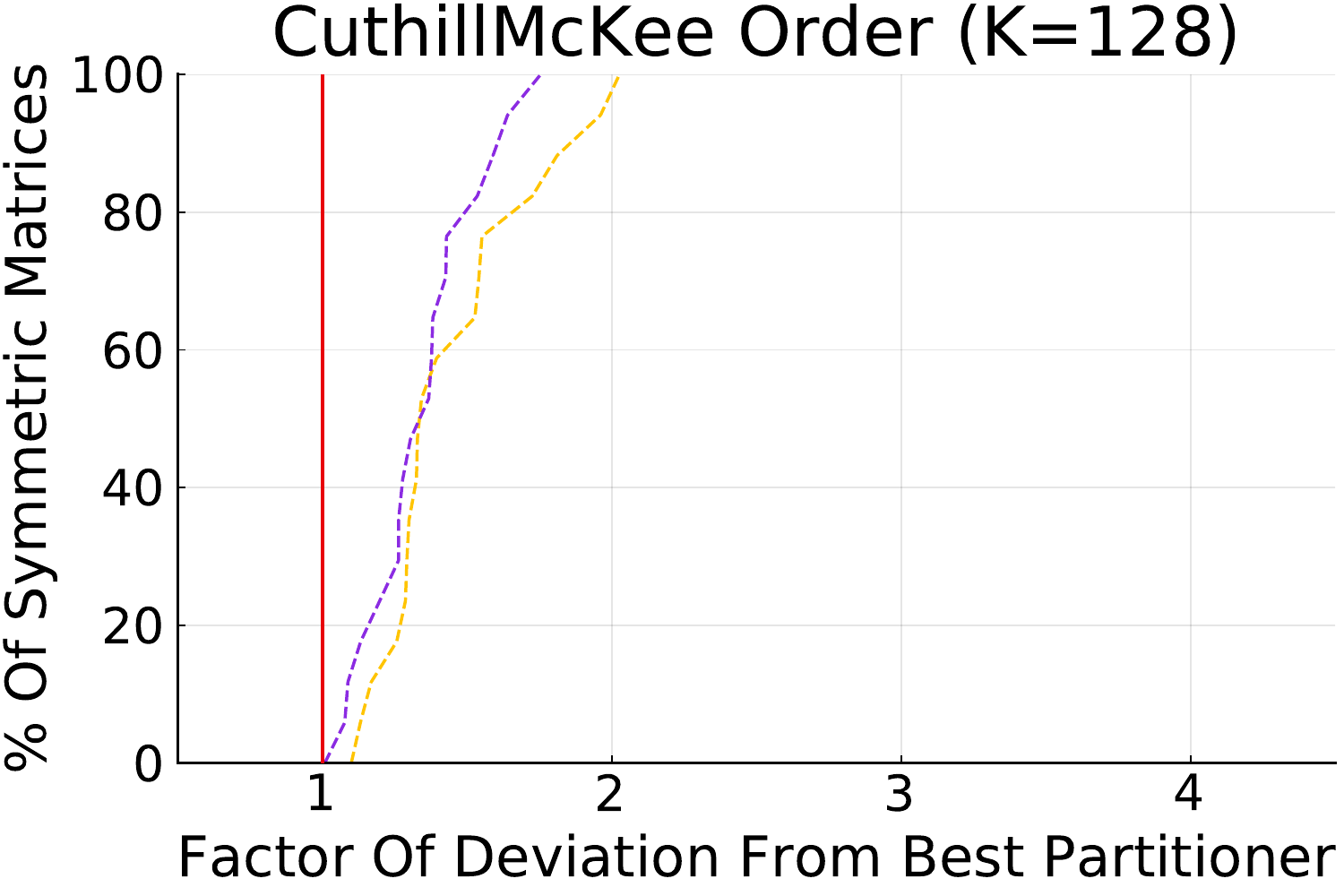}%
    \includegraphics[]{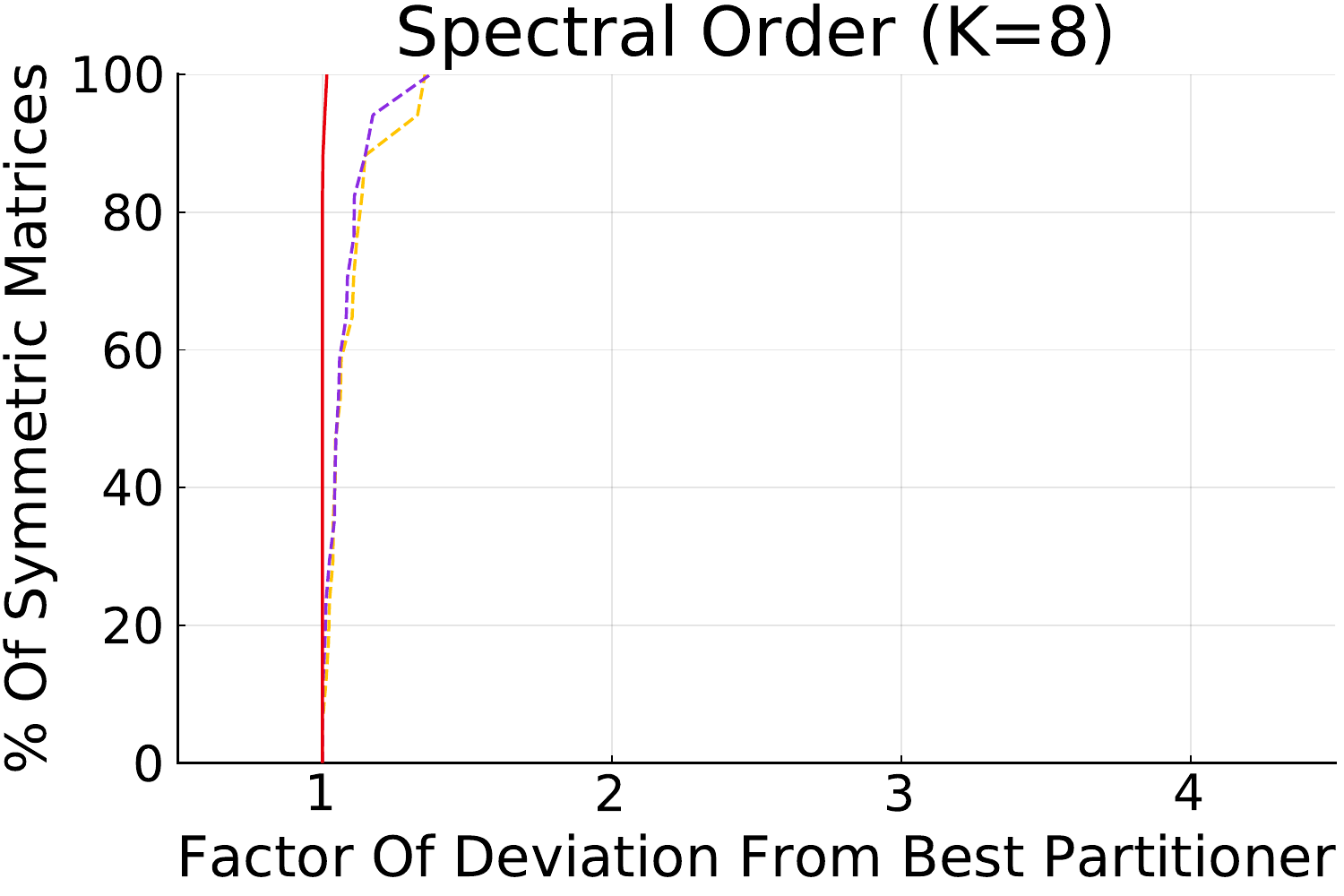}%
    \includegraphics[]{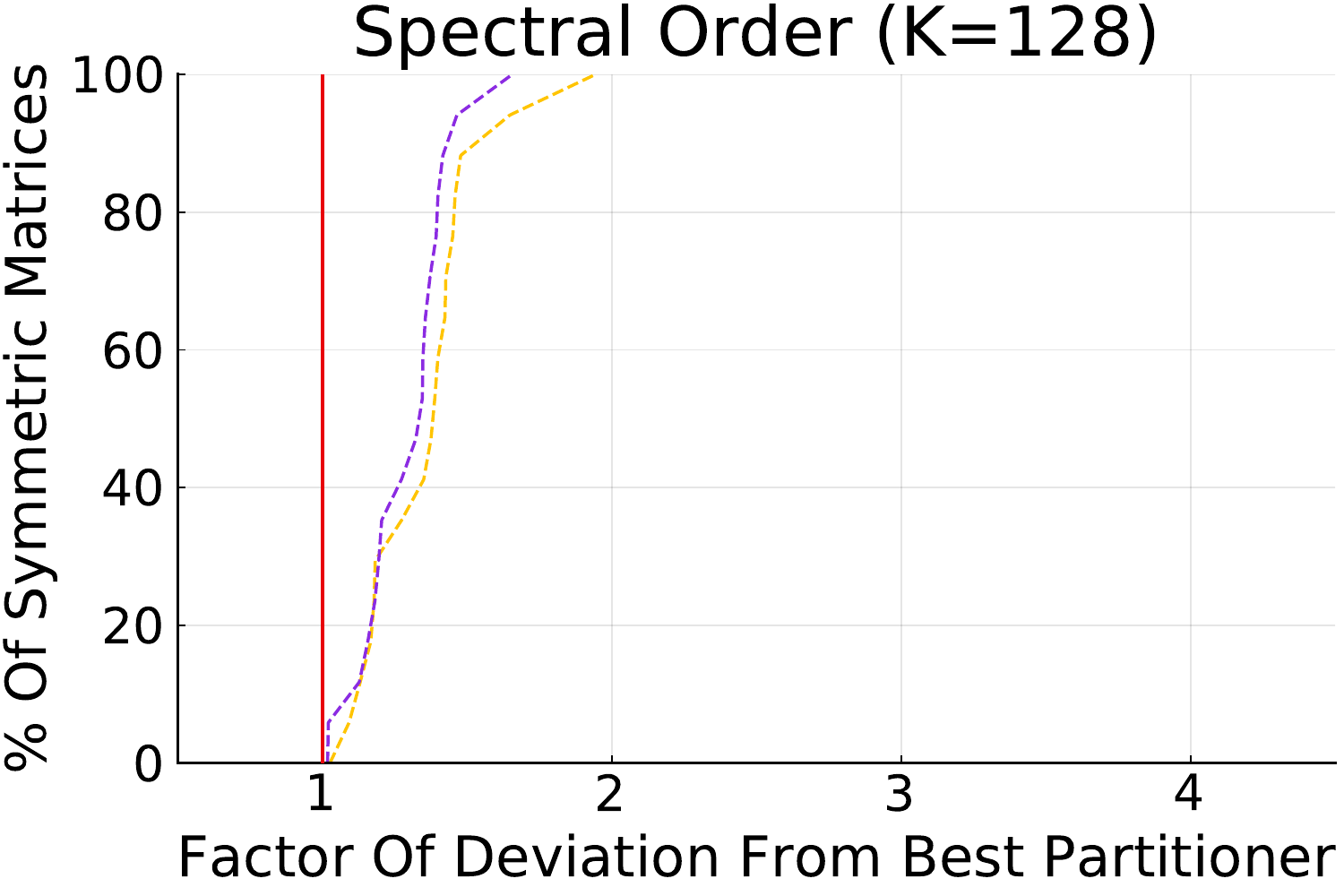}%
    }
    Load Balanced Total Connectivity On All Matrices
    \resizebox{\linewidth}{!}{%
    \includegraphics[]{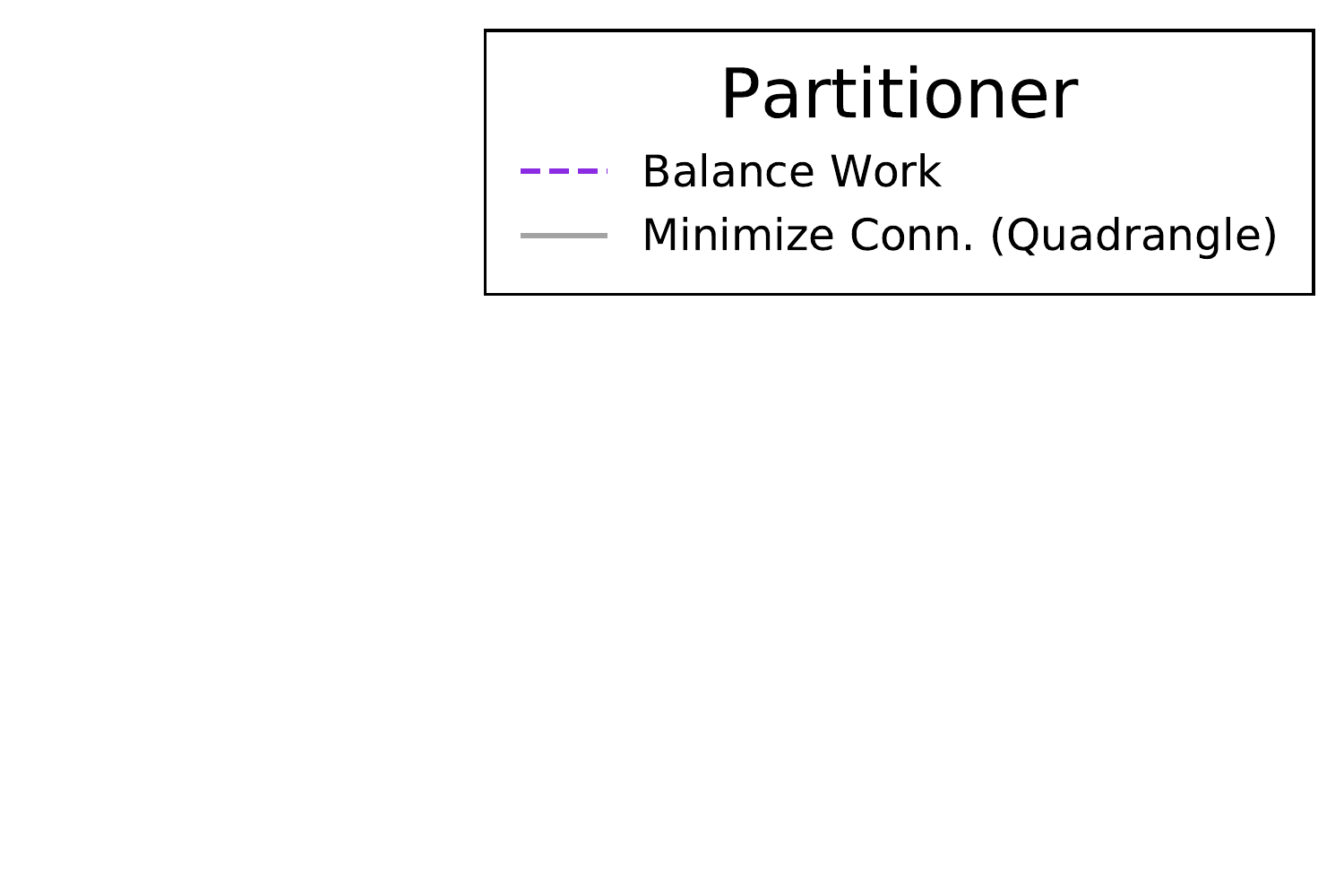}%
    \includegraphics[]{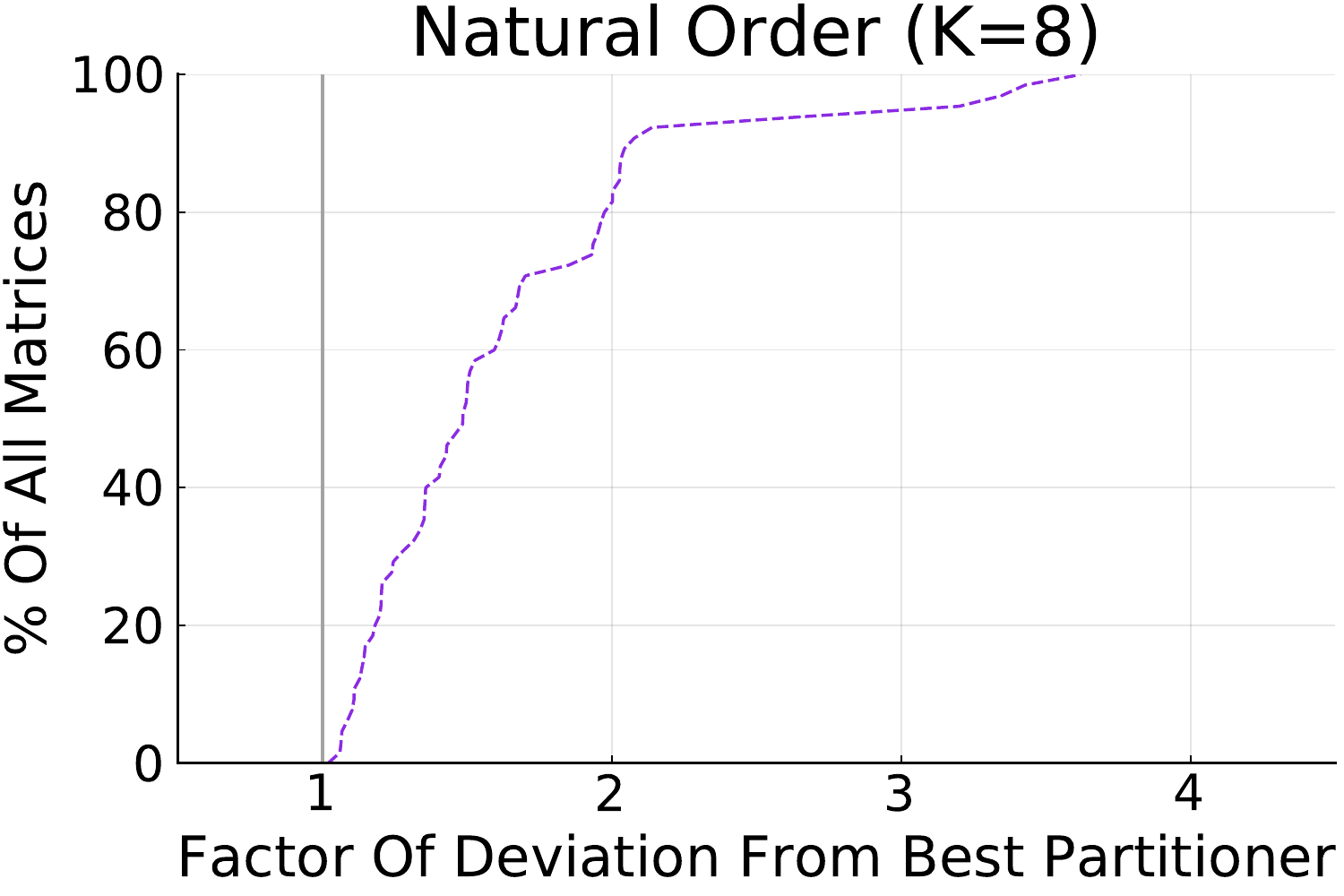}%
    \includegraphics[]{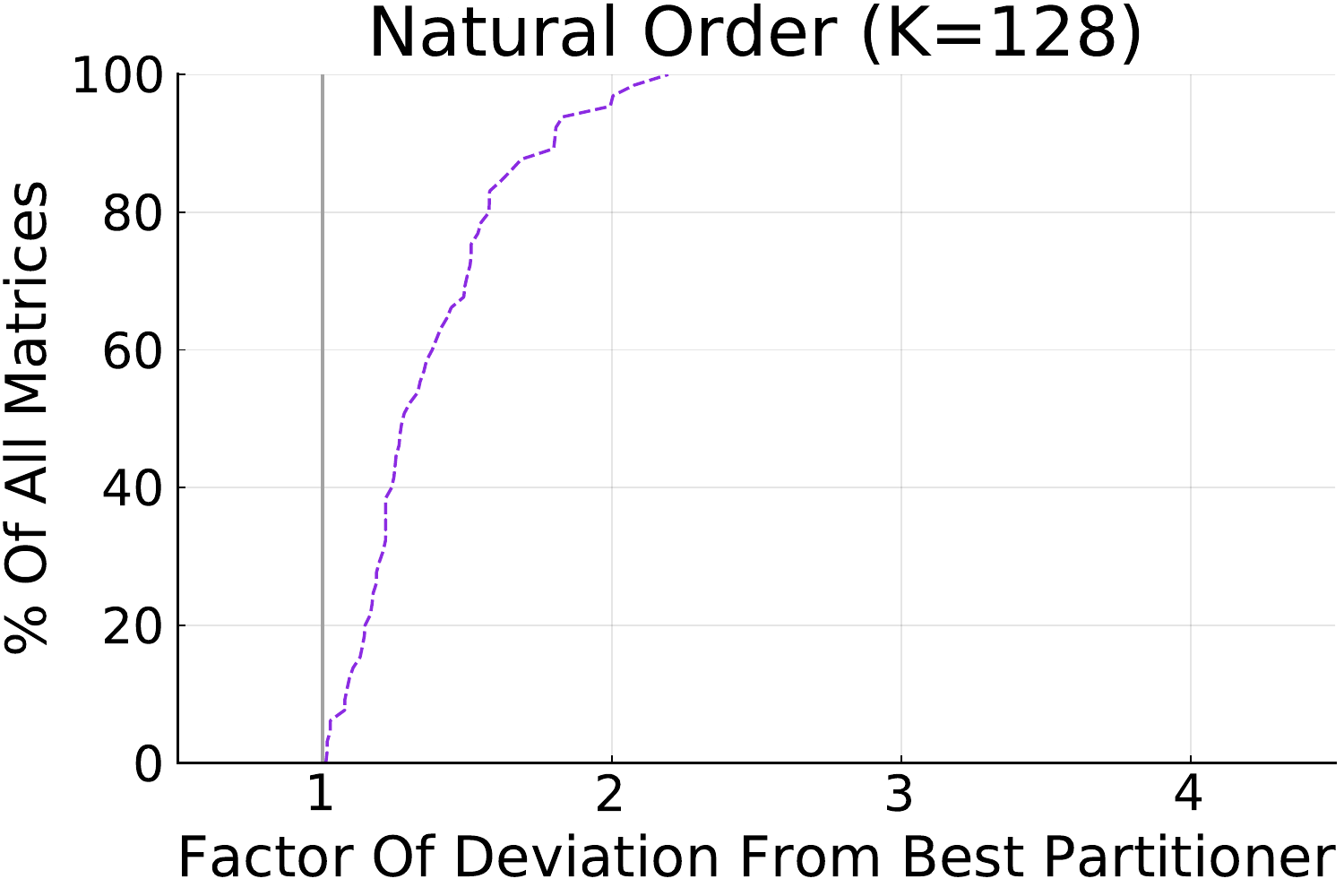}%
    \includegraphics[]{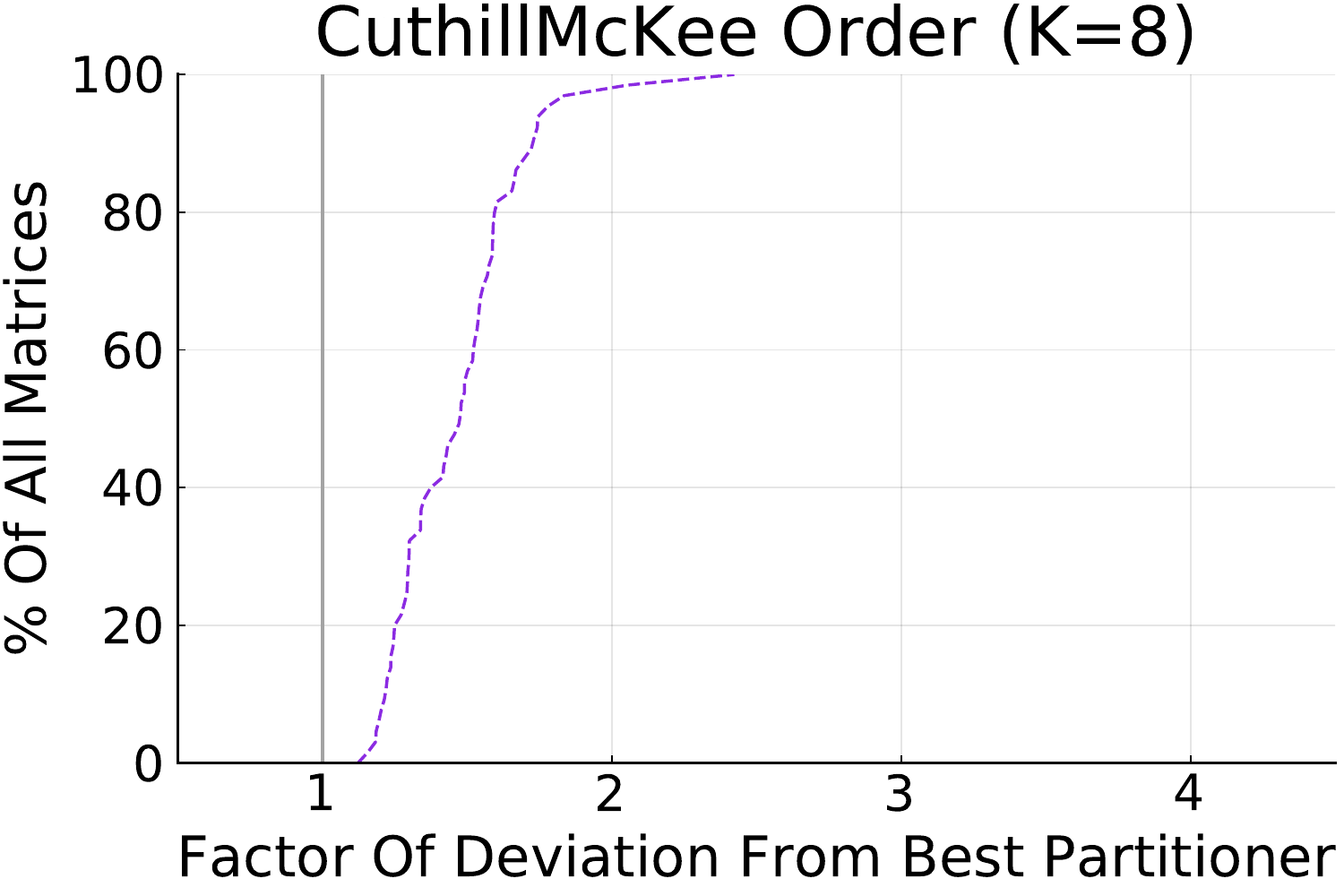}%
    \includegraphics[]{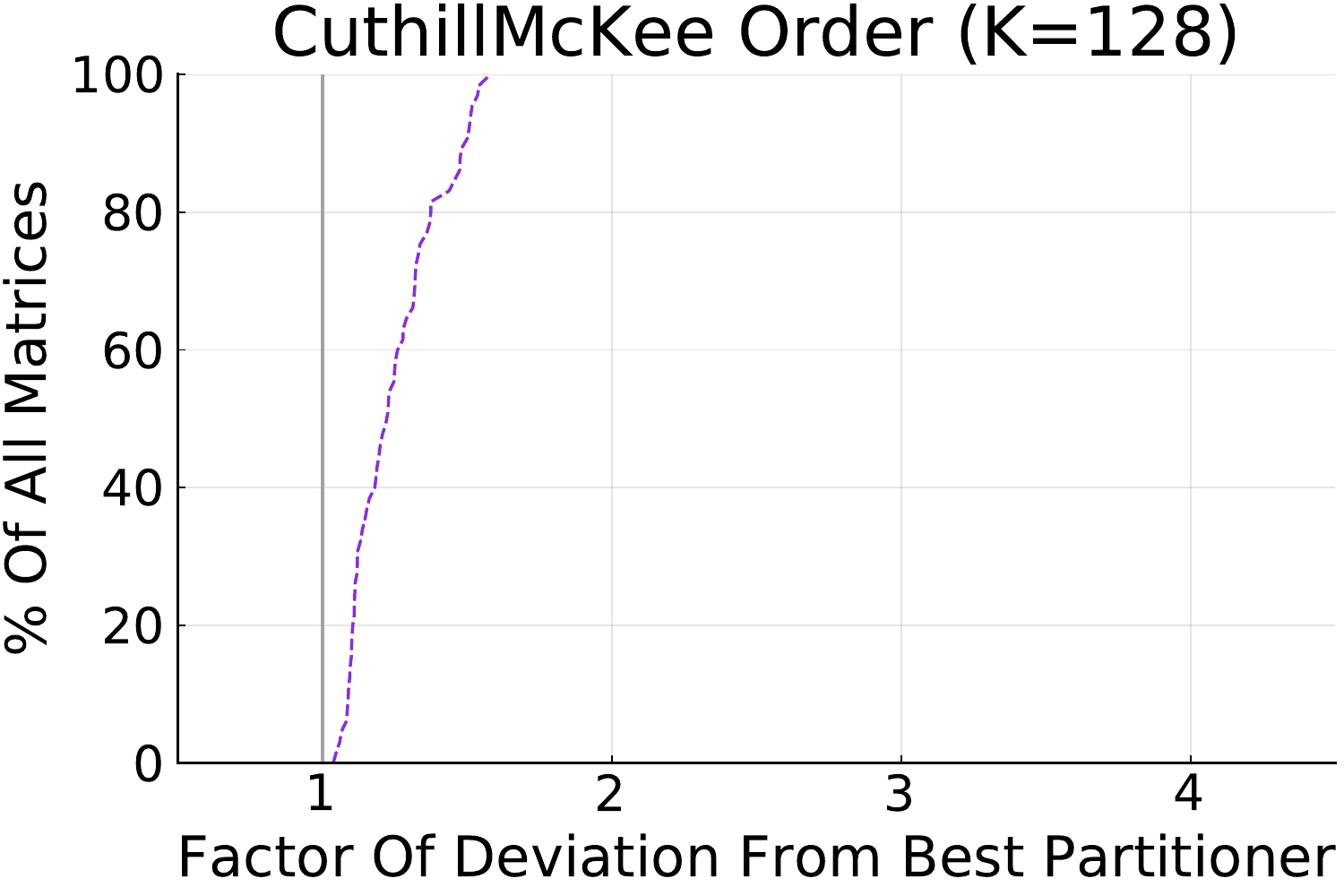}%
    \includegraphics[]{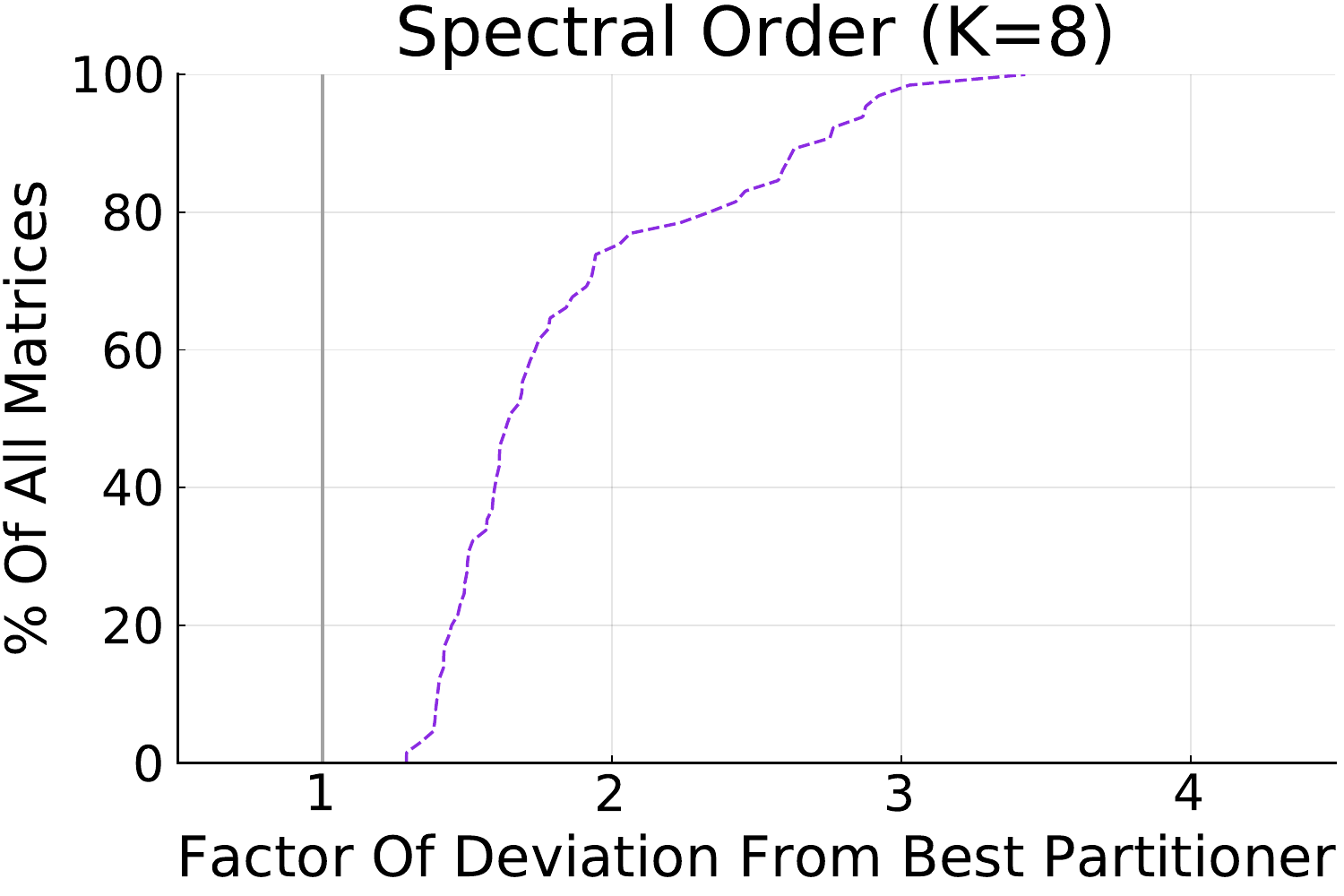}%
    \includegraphics[]{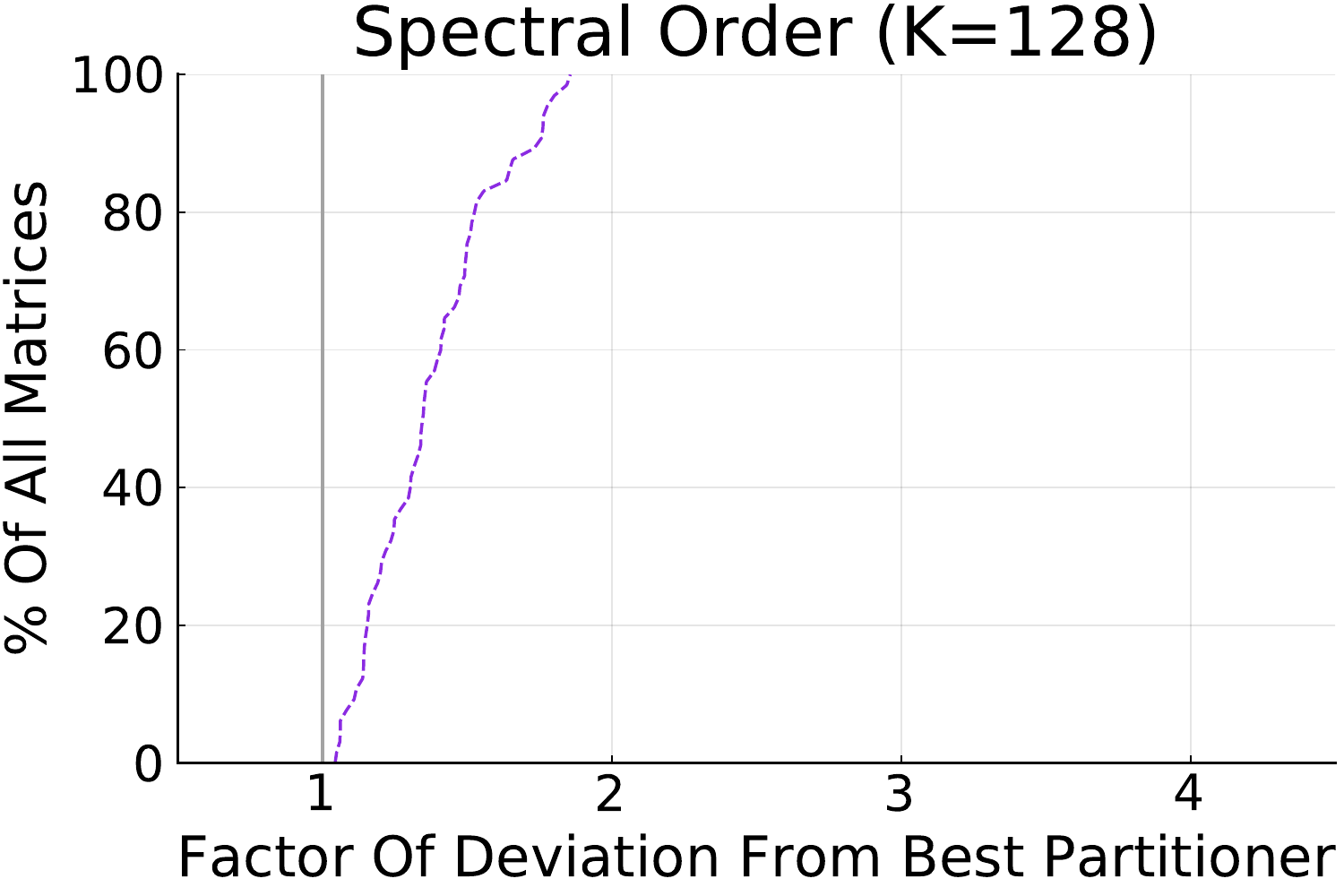}%
    }
    Load Balanced Hyperedge Cut On All Matrices
    \resizebox{\linewidth}{!}{%
    \includegraphics[]{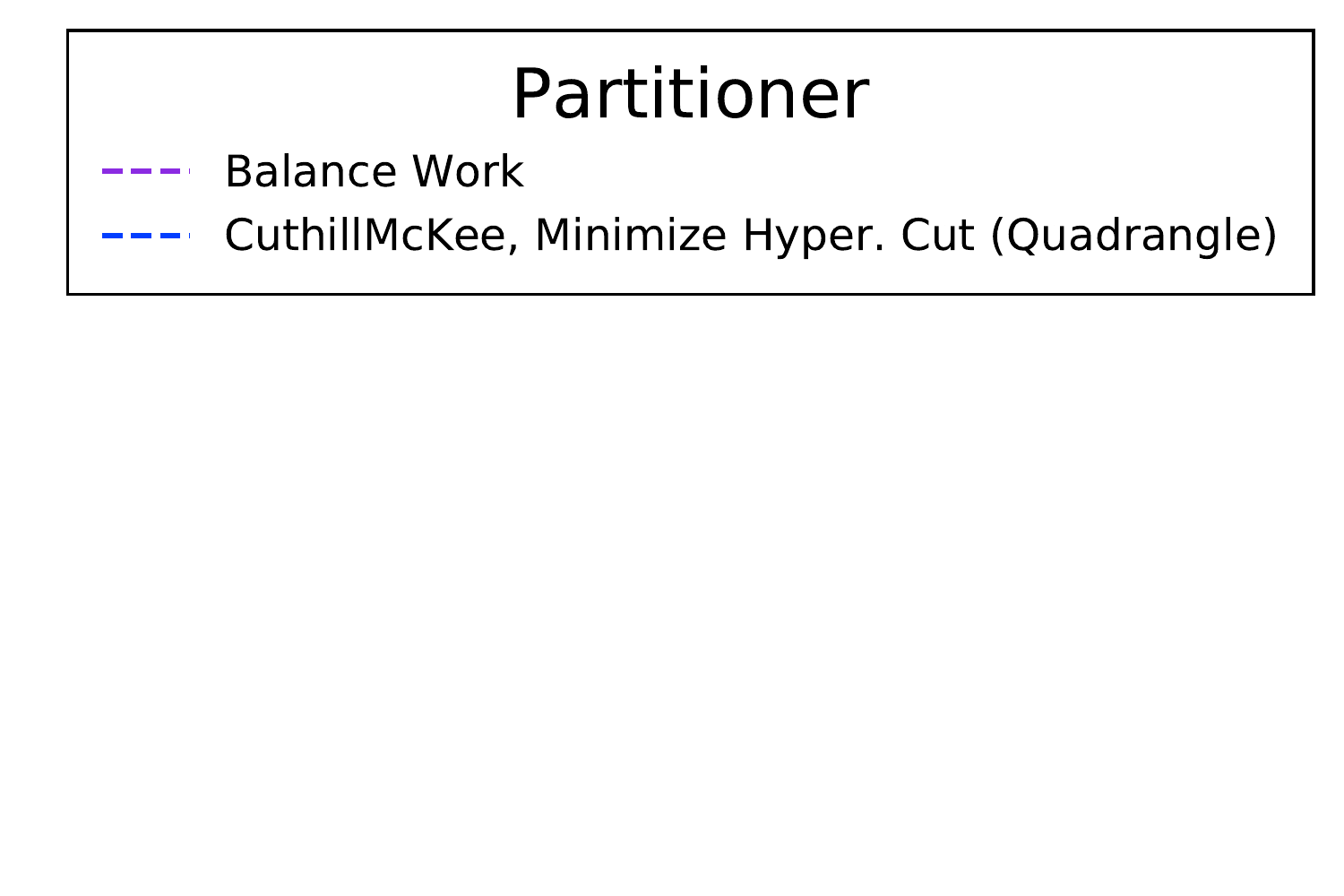}%
    \includegraphics[]{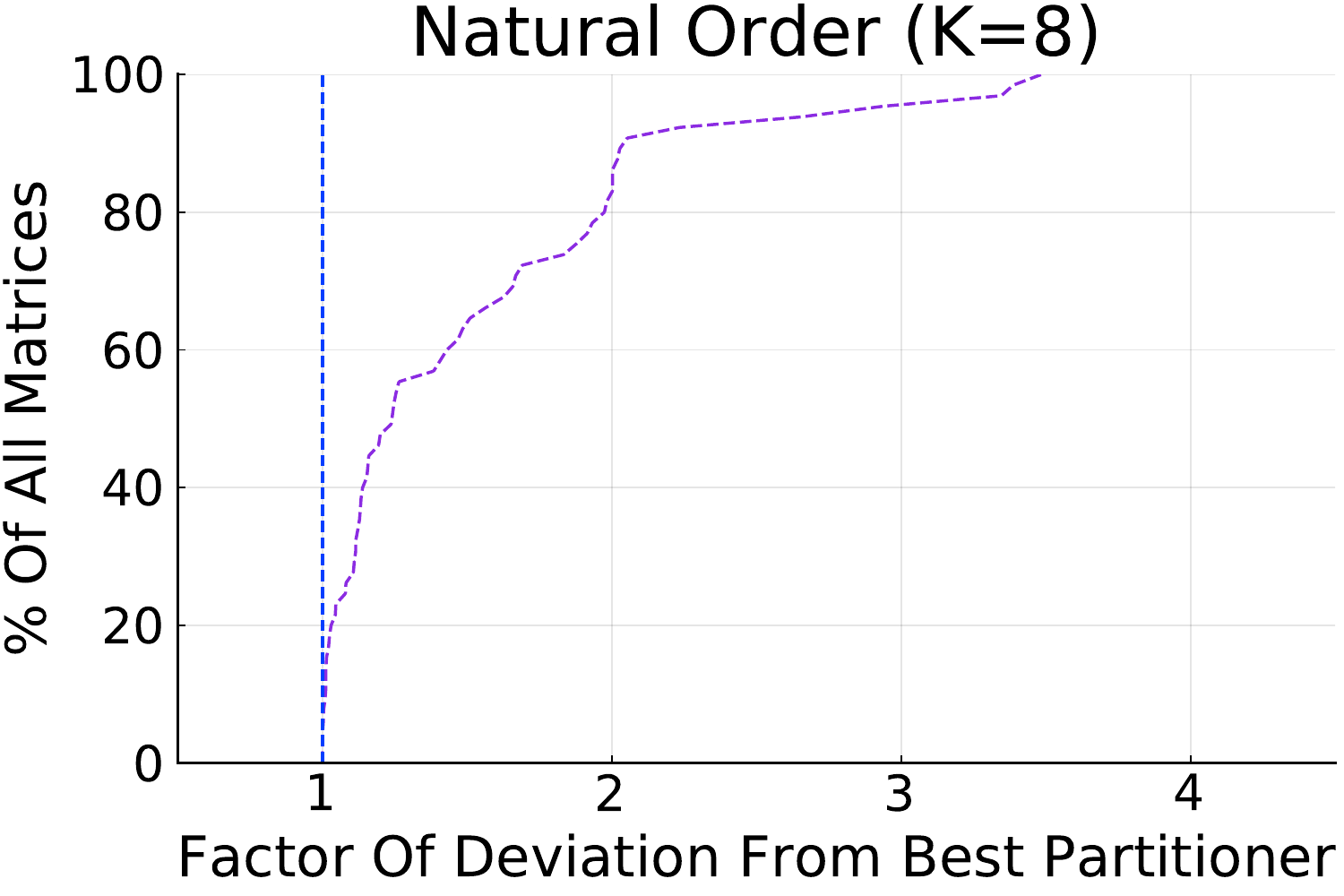}%
    \includegraphics[]{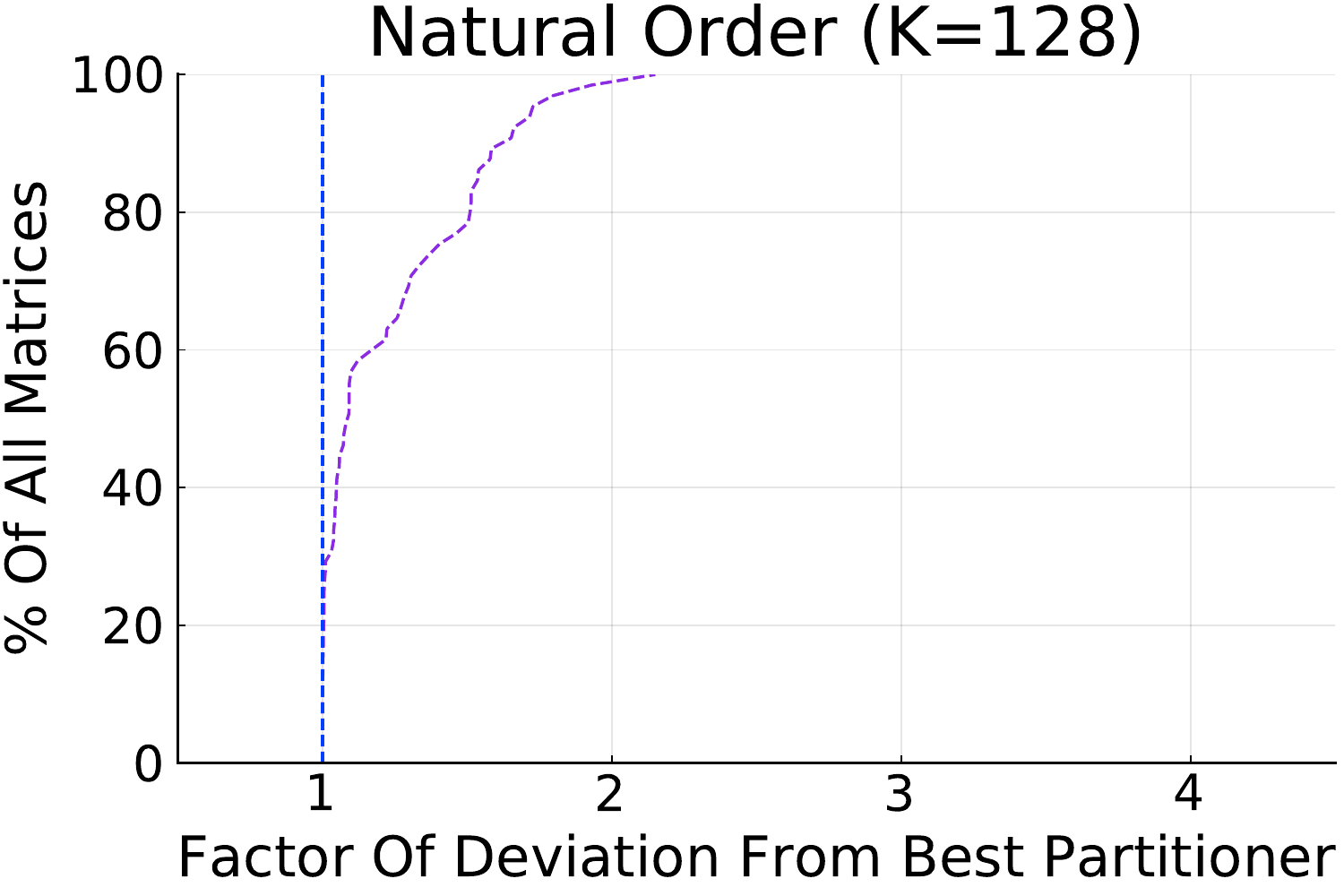}%
    \includegraphics[]{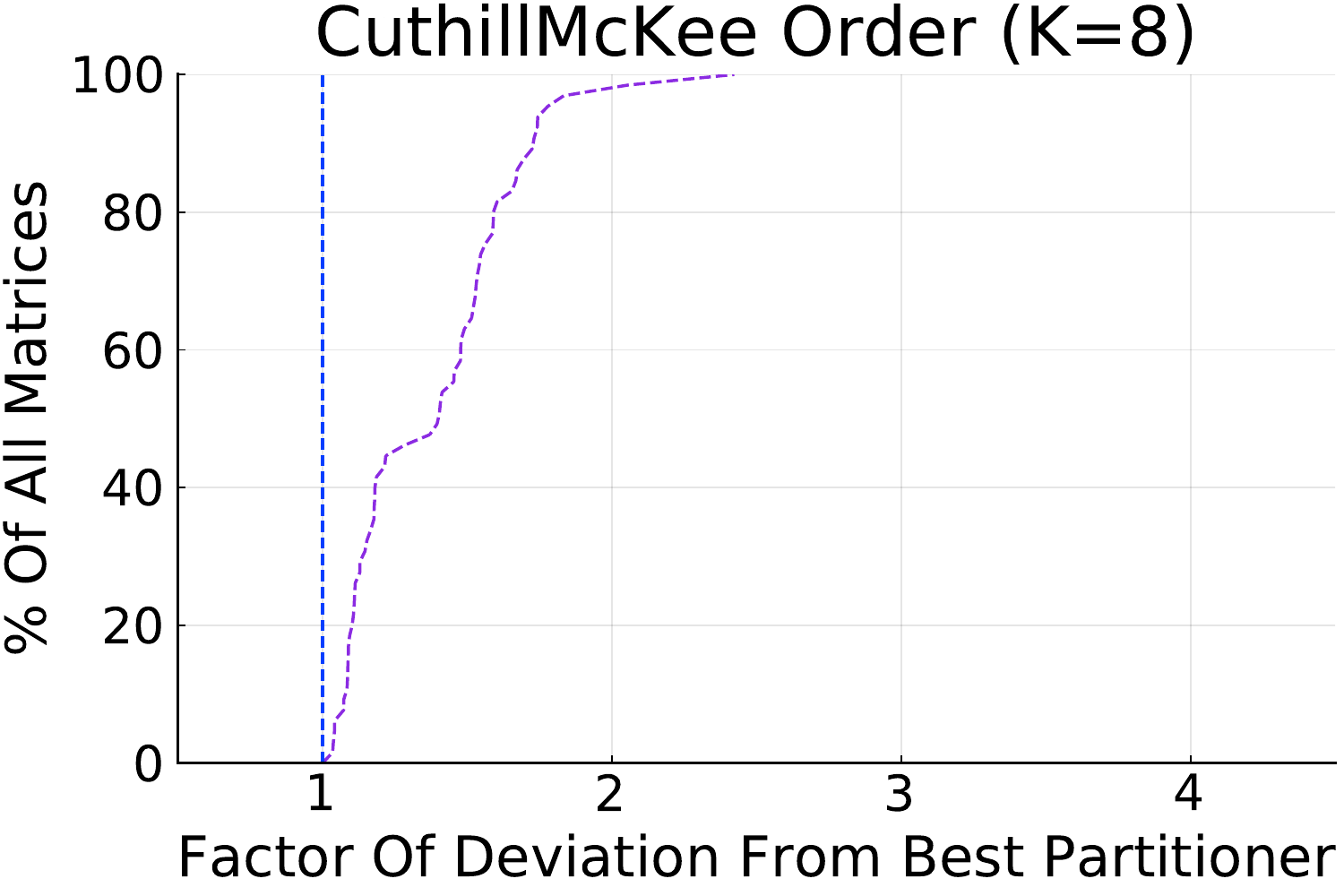}%
    \includegraphics[]{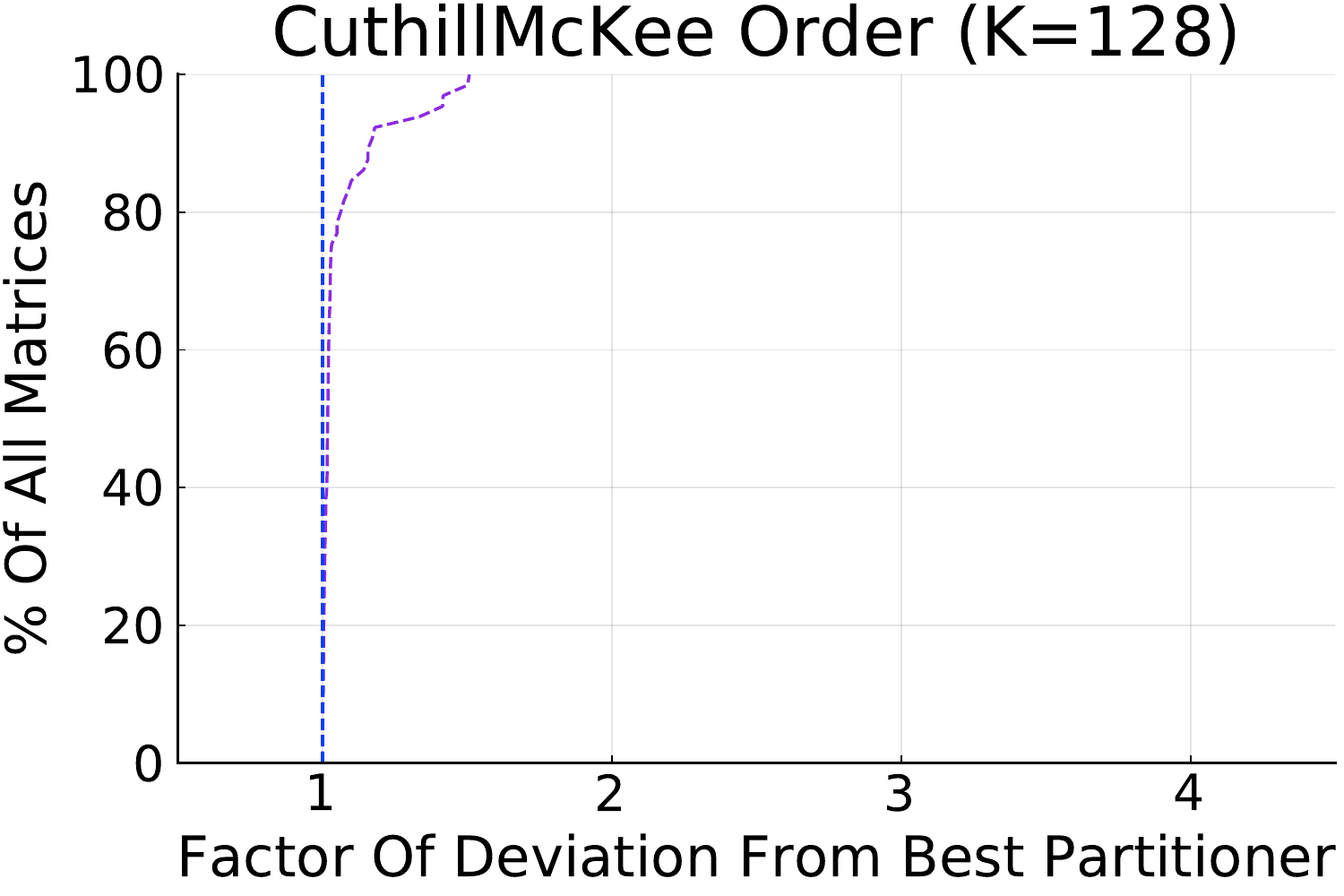}%
    \includegraphics[]{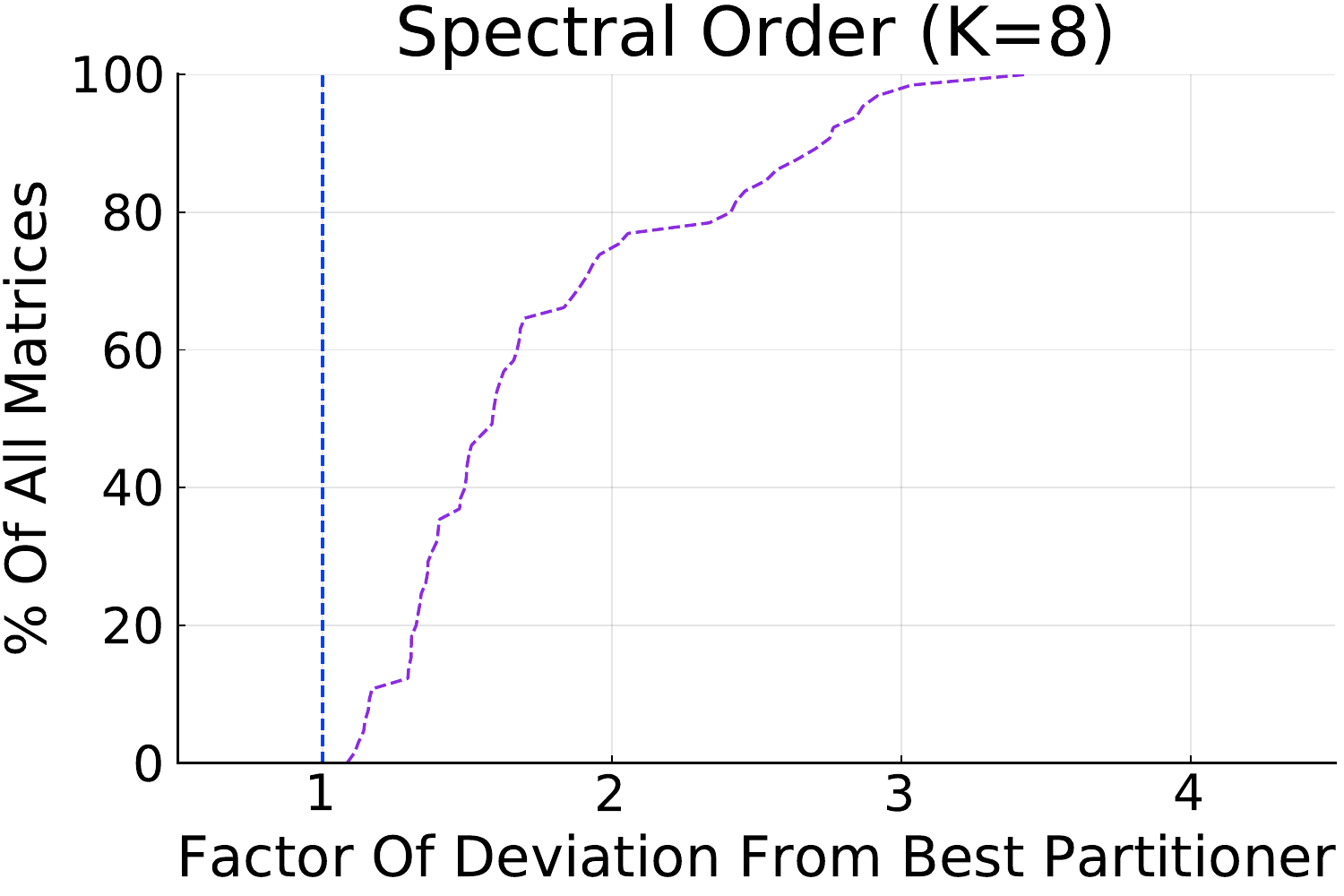}%
    \includegraphics[]{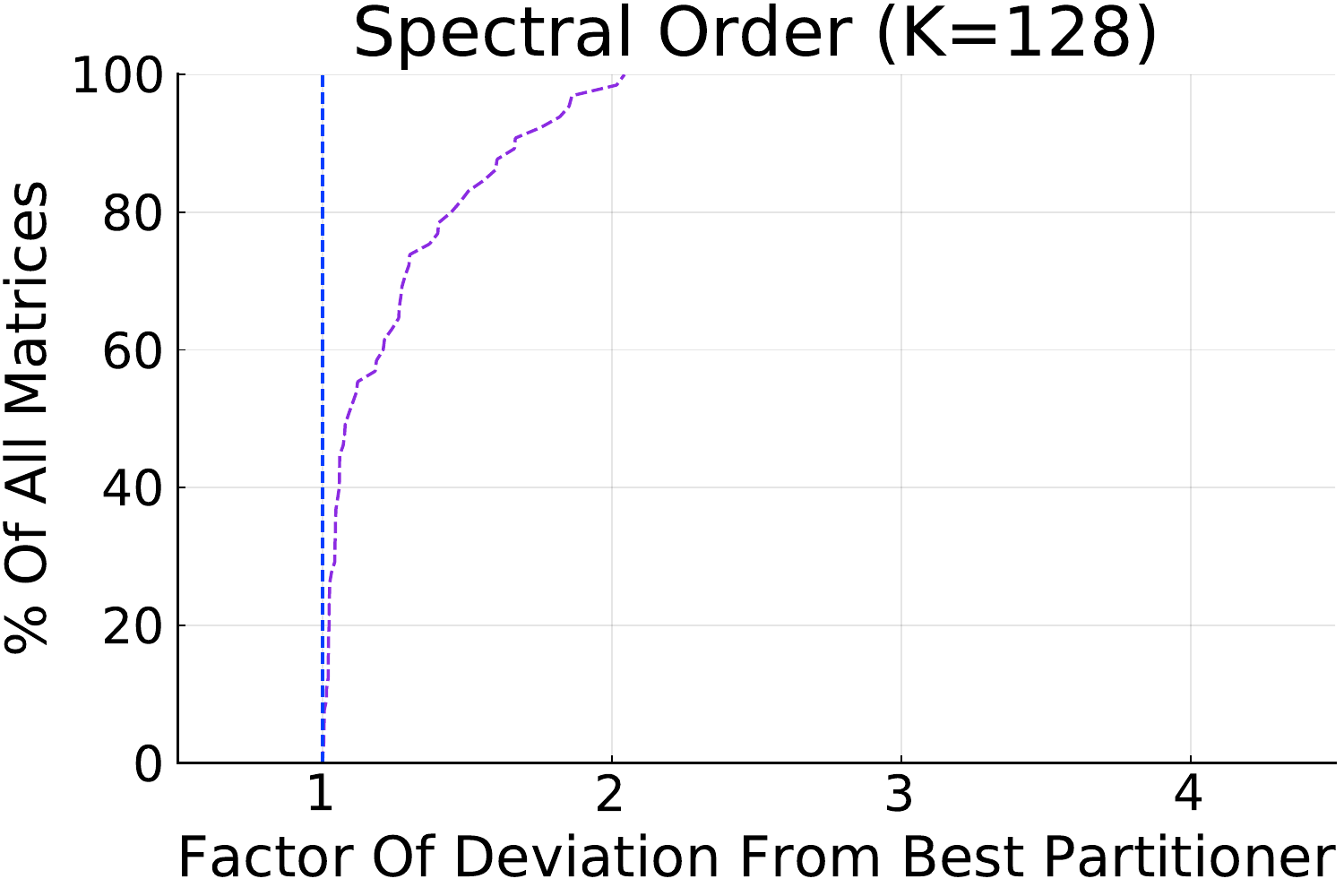}%
    }
    Load Balanced Edge Cut On Symmetric Matrices
    \resizebox{\linewidth}{!}{%
    \includegraphics[]{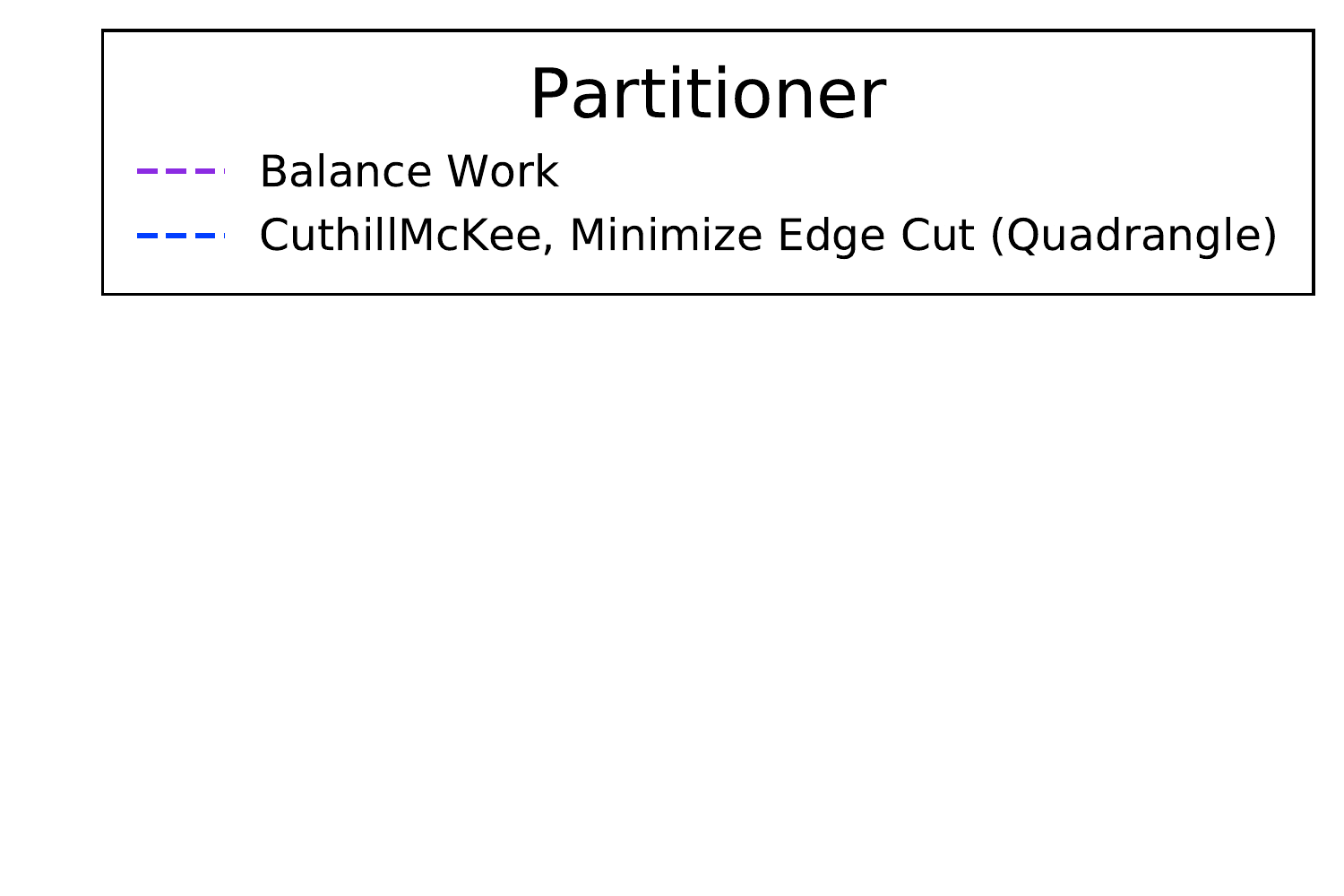}%
    \includegraphics[]{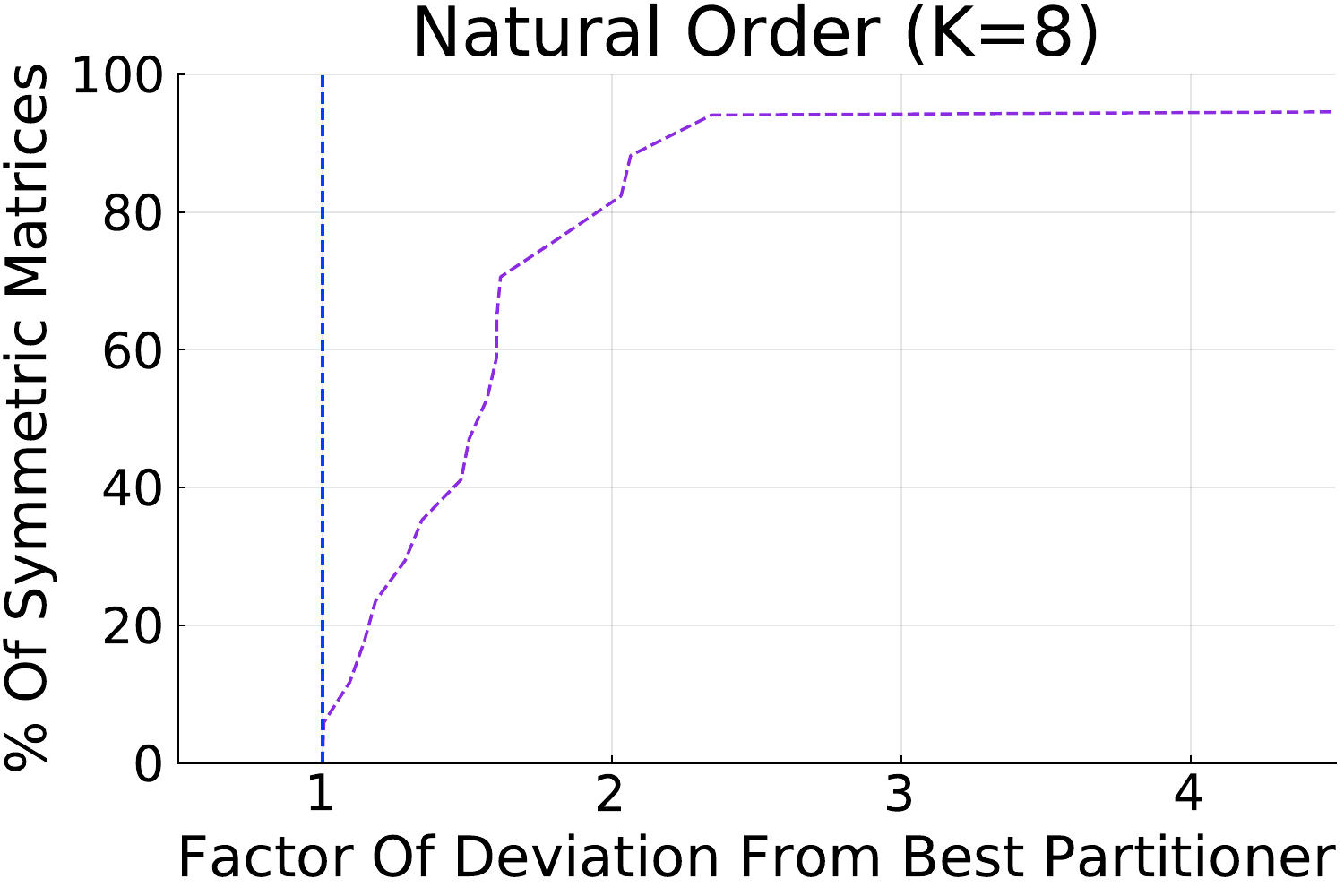}%
    \includegraphics[]{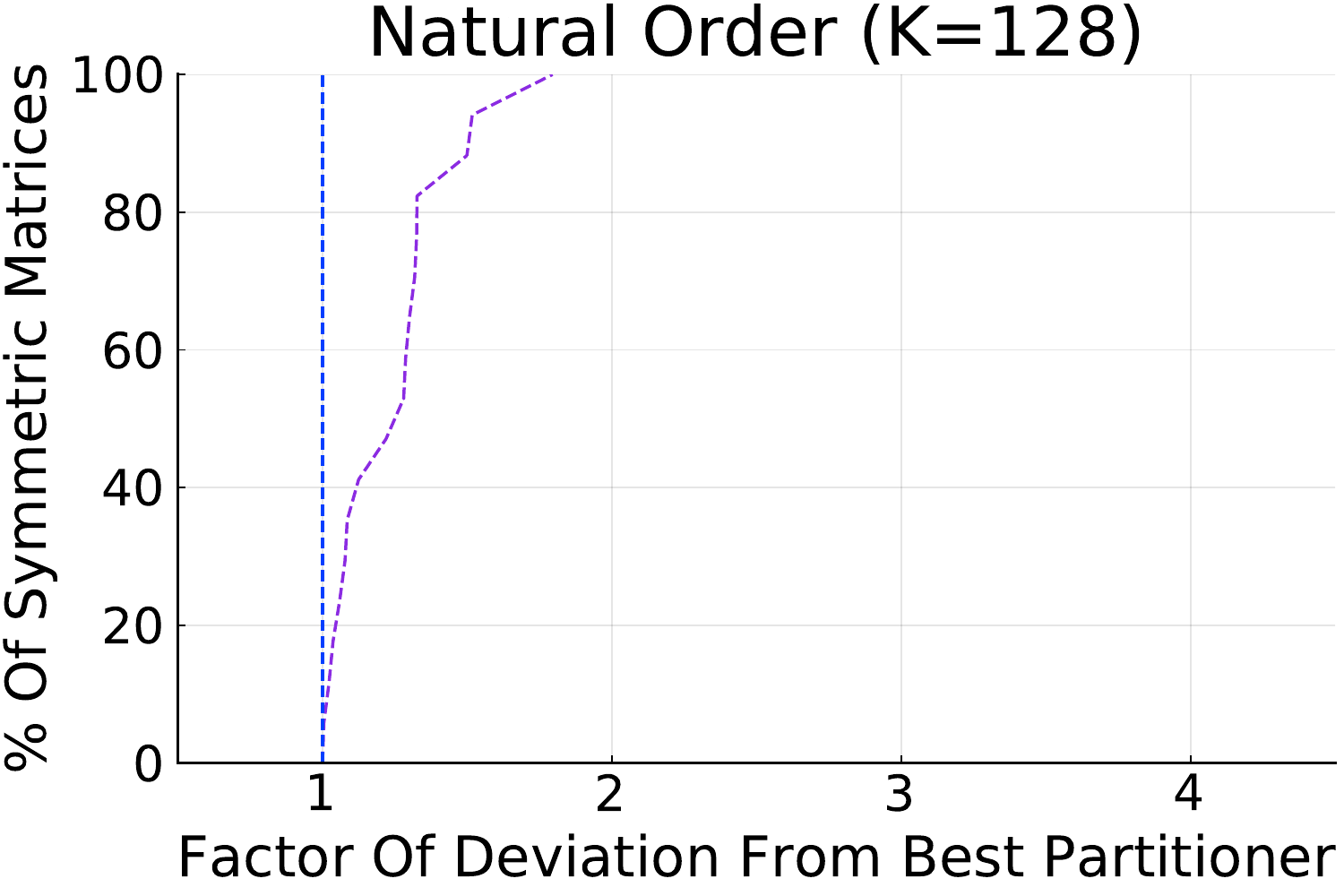}%
    \includegraphics[]{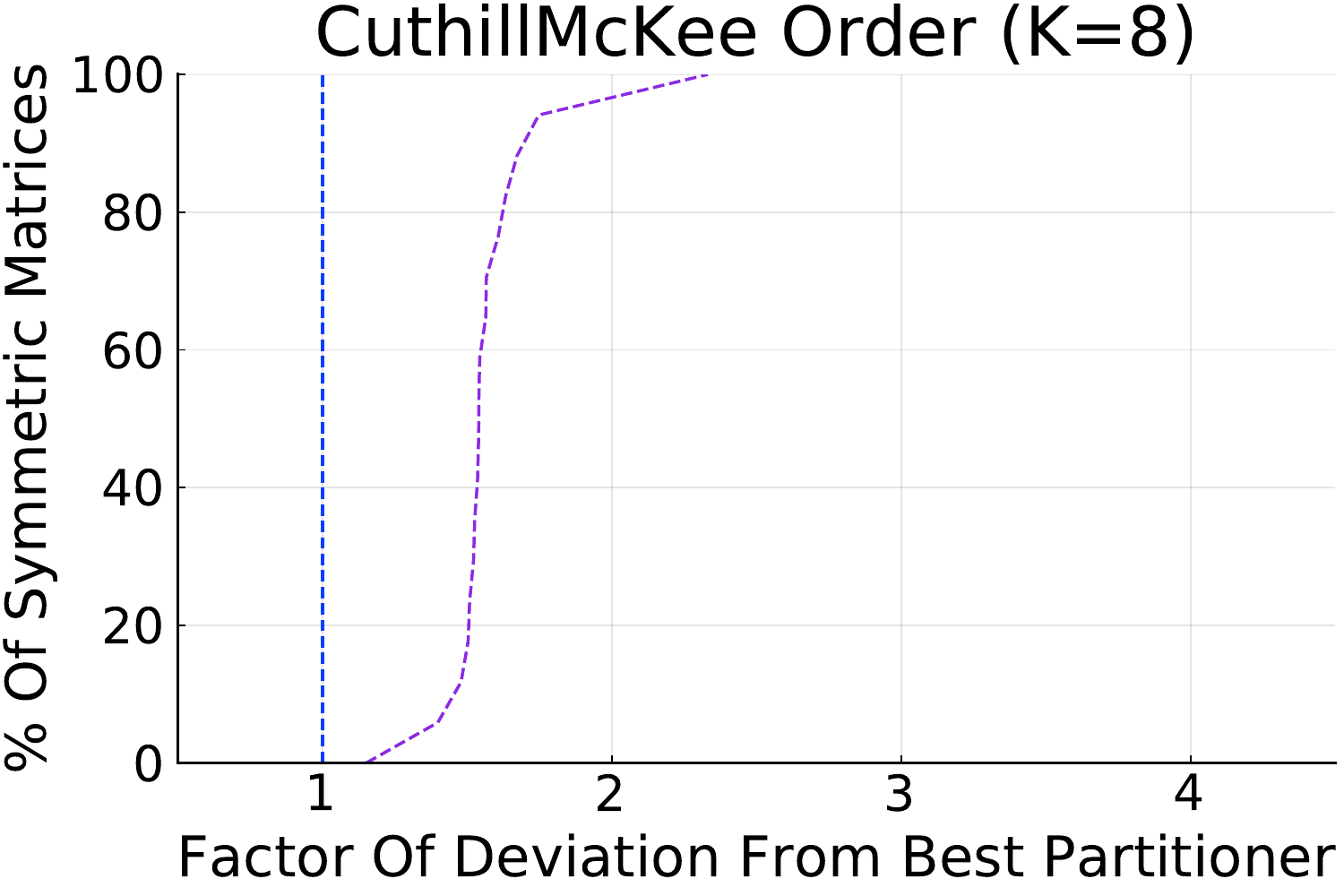}%
    \includegraphics[]{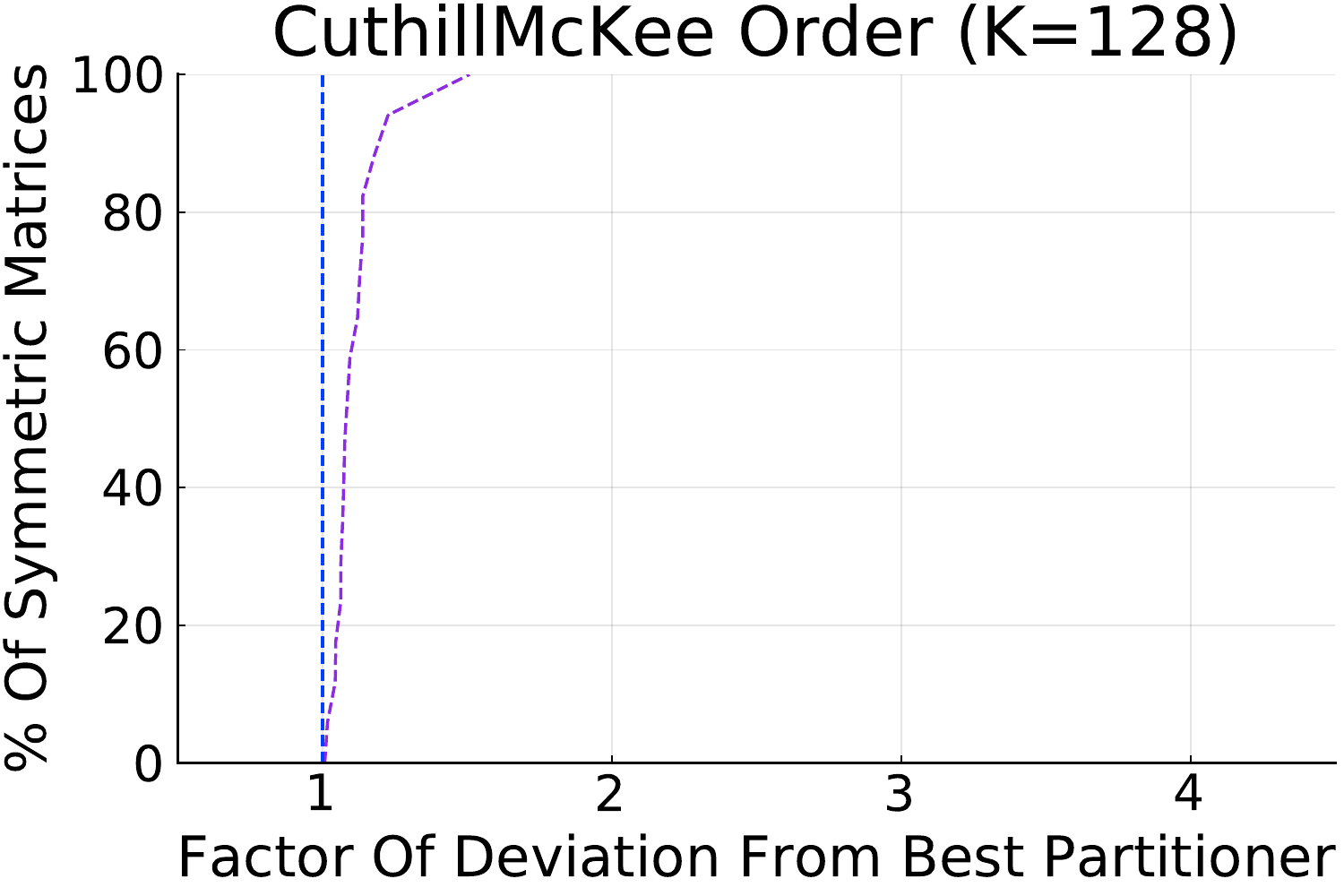}%
    \includegraphics[]{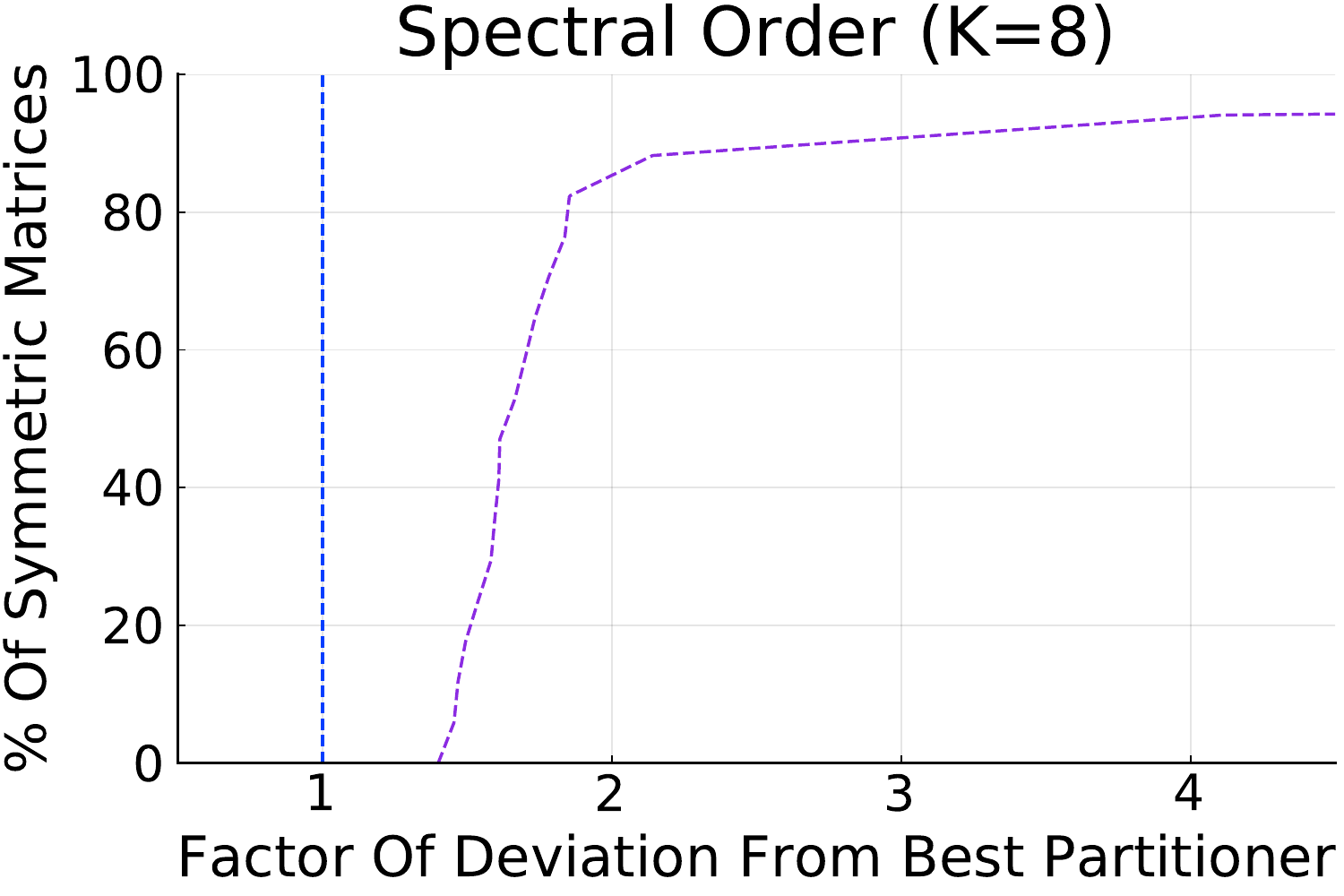}%
    \includegraphics[]{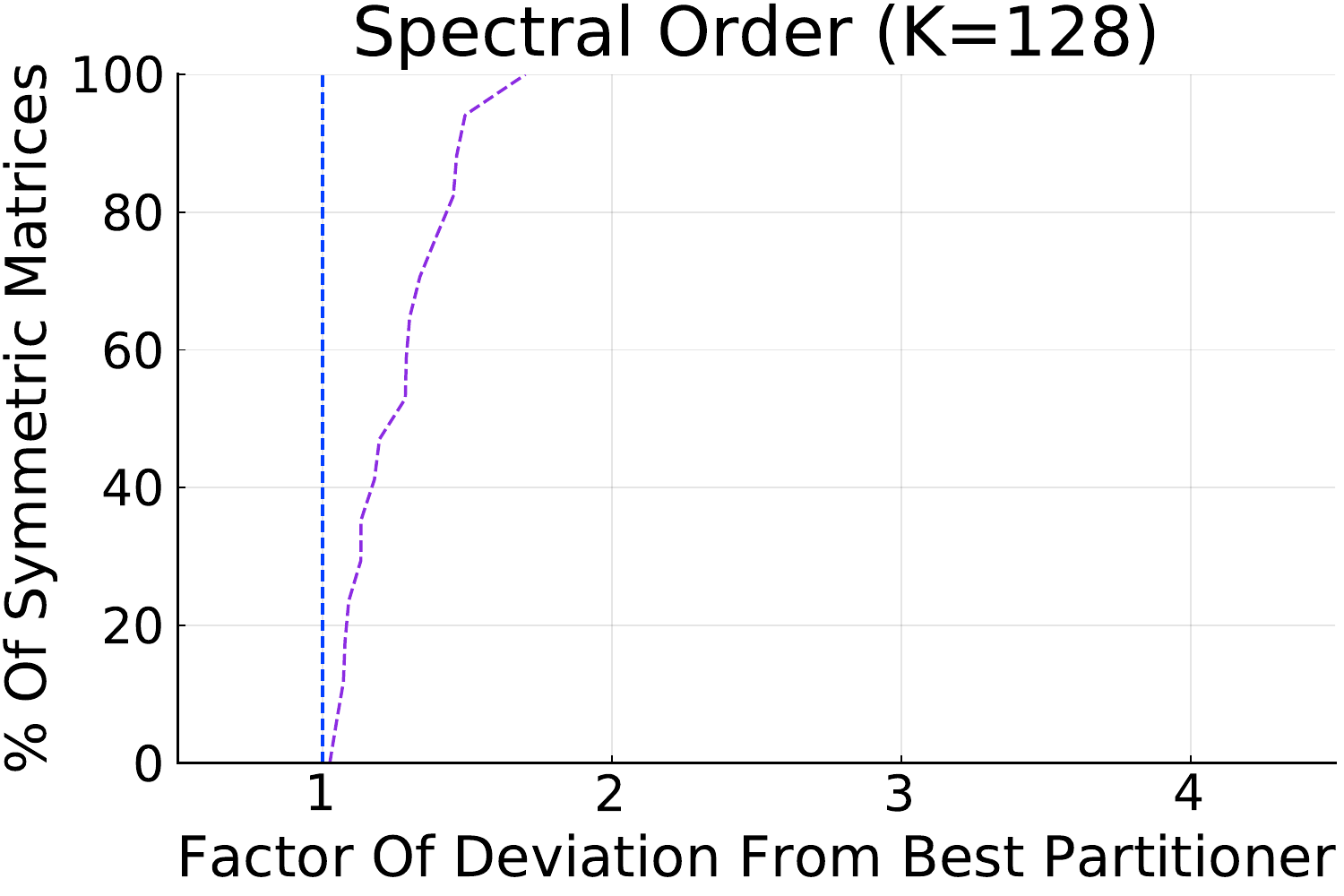}%
    }
    Cache Blocked Connectivity On All Matrices
    \resizebox{\linewidth}{!}{%
    \includegraphics[]{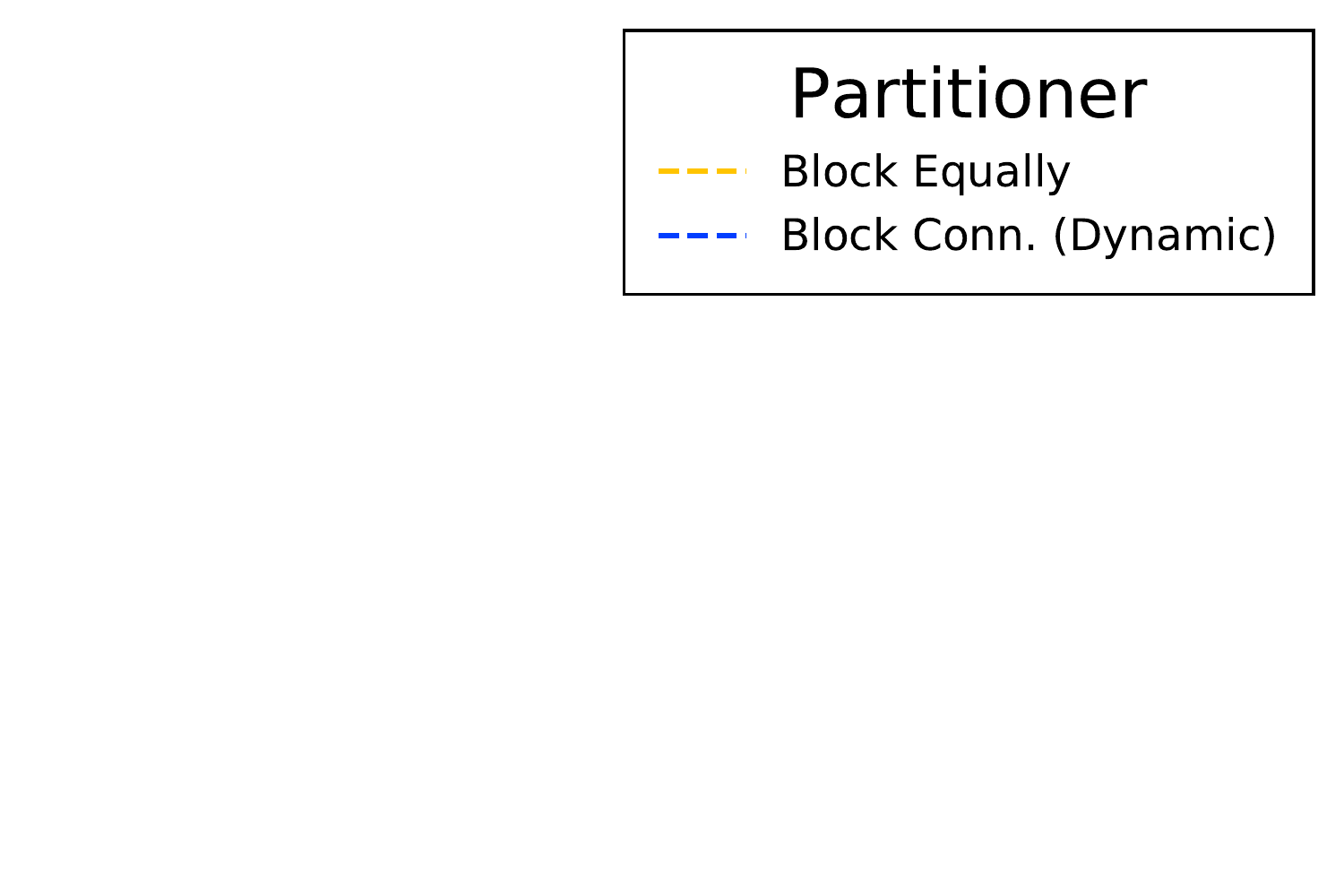}%
    \includegraphics[]{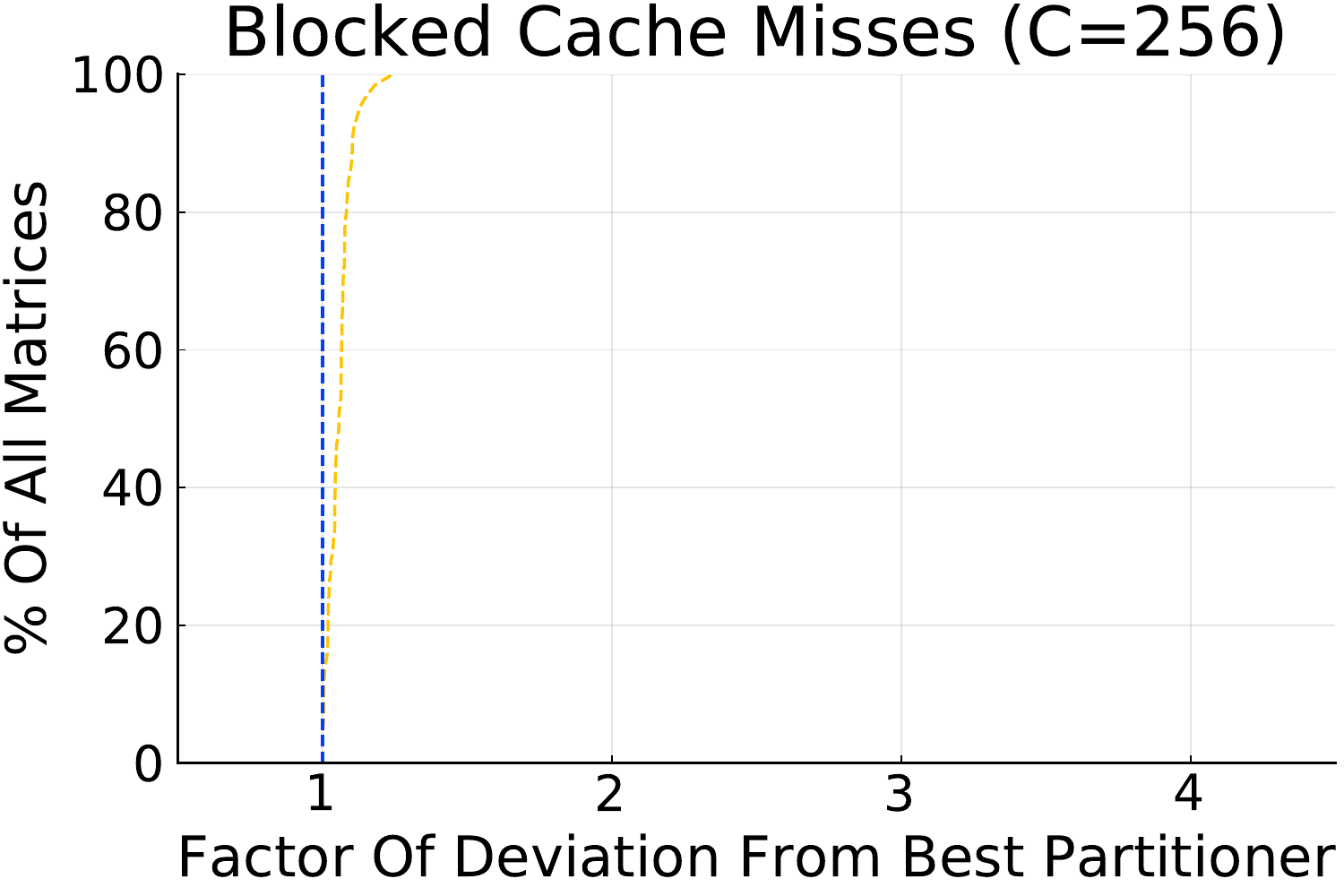}%
    \includegraphics[]{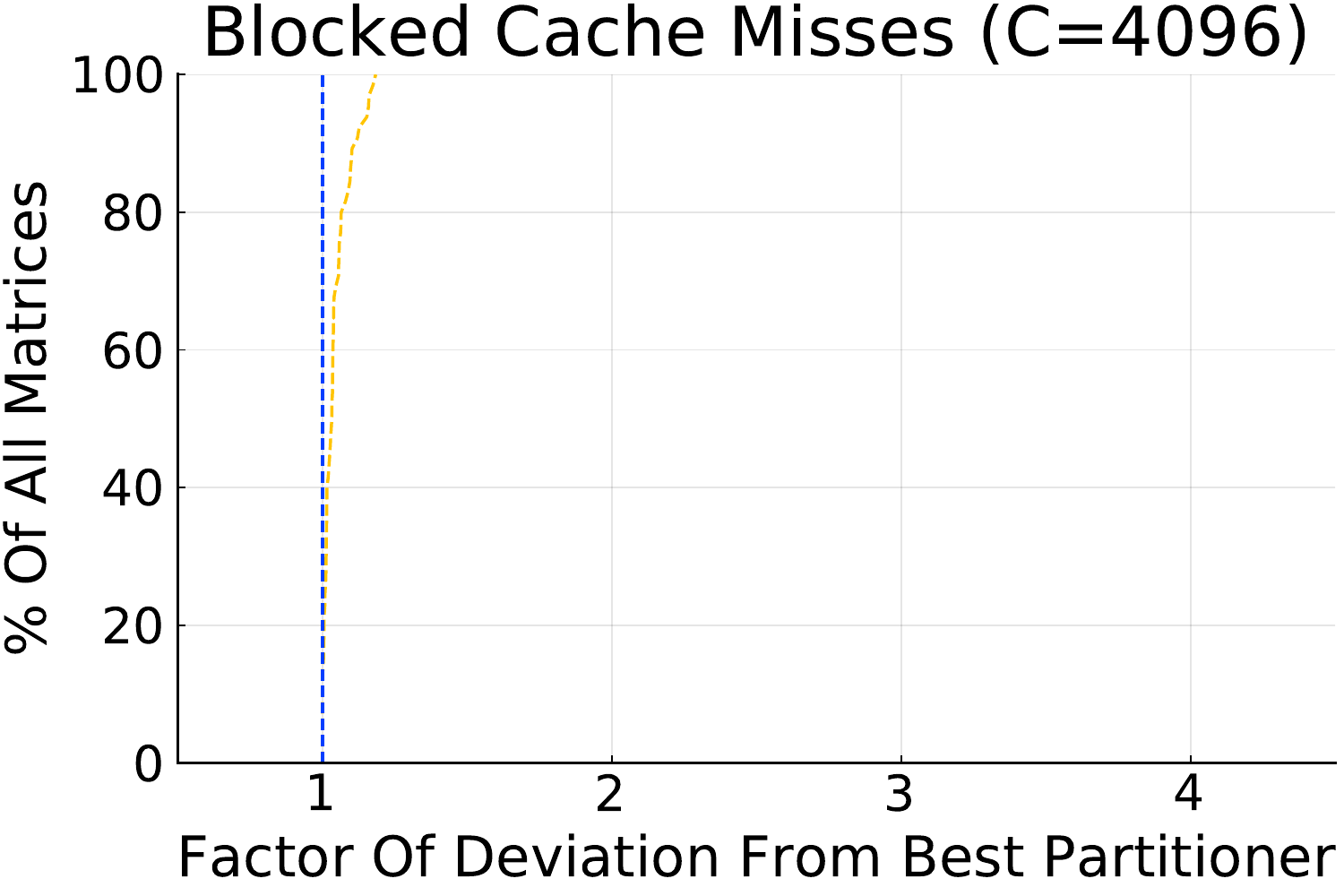}%
    \includegraphics[]{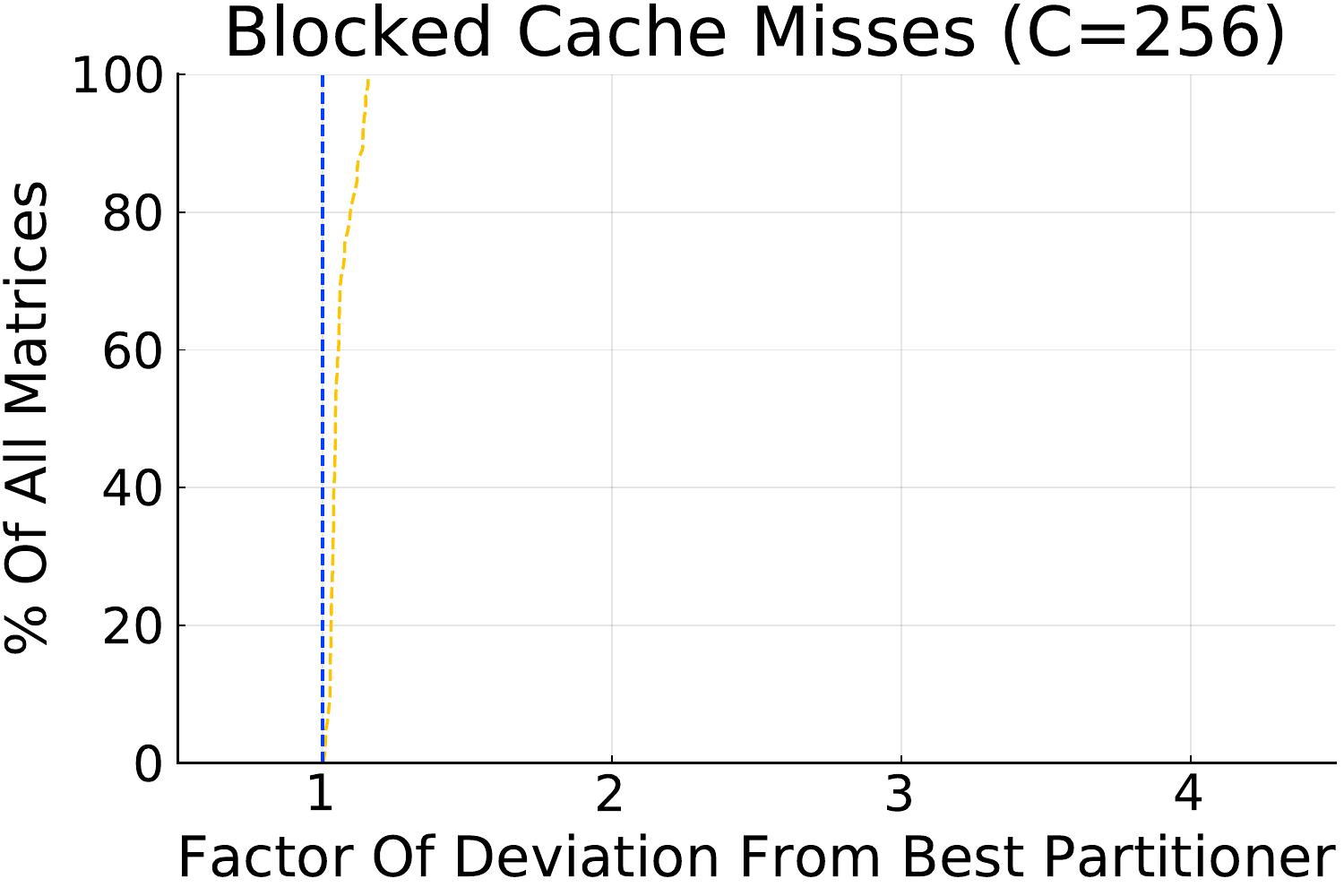}%
    \includegraphics[]{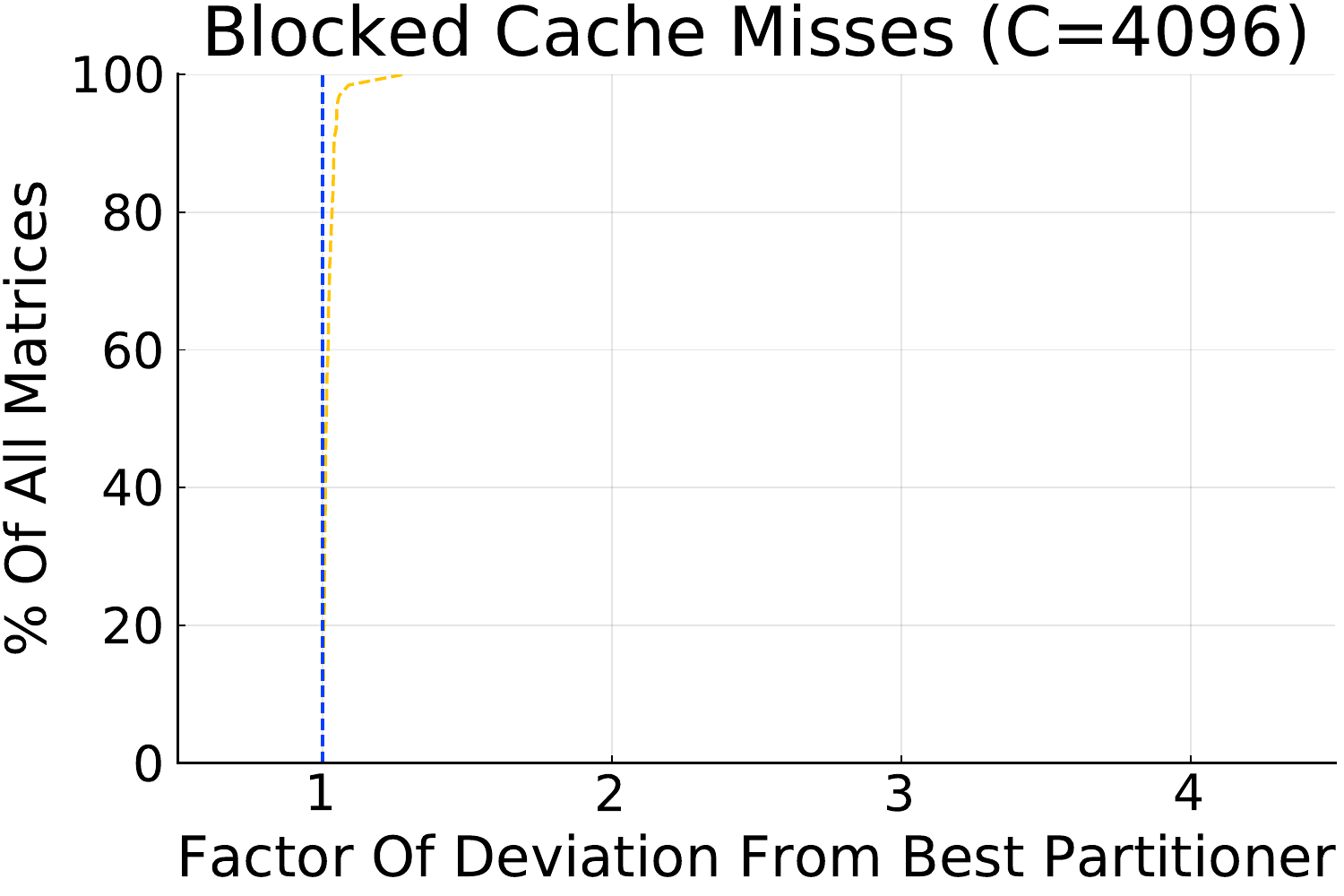}%
    \includegraphics[]{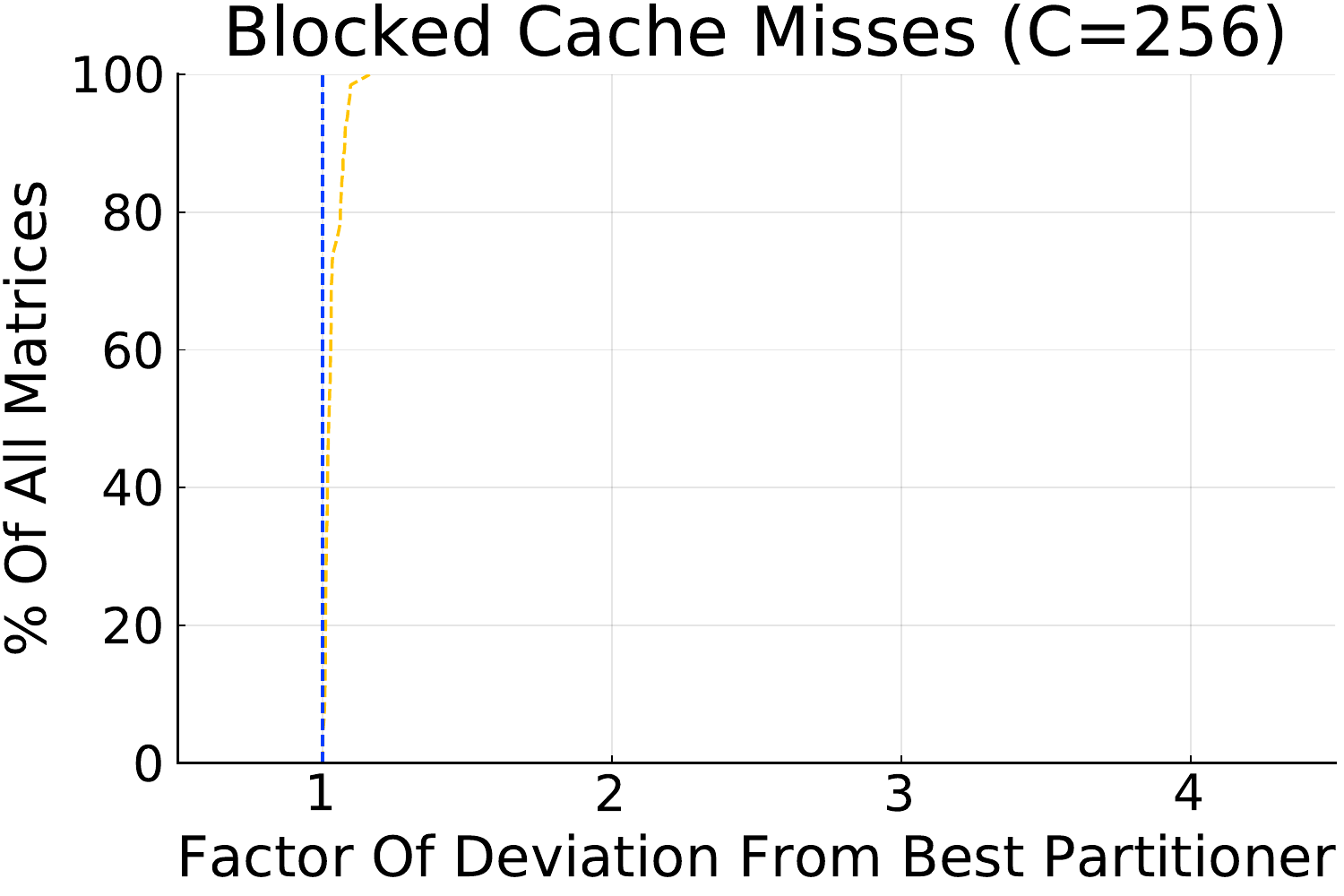}%
    \includegraphics[]{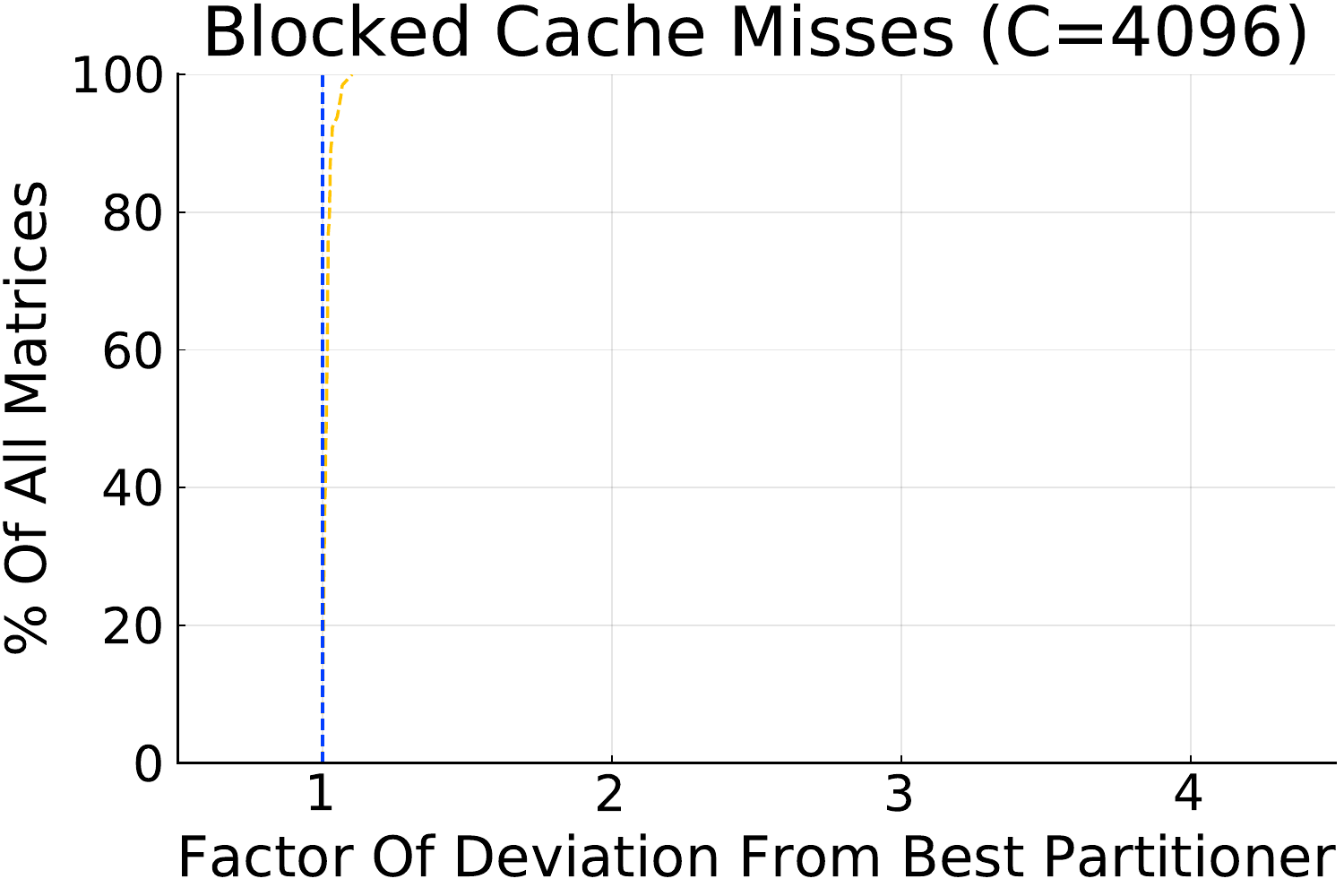}%
    }

    \caption{Performance profiles comparing normalized modeled quality of our
    general (possibly noncontiguous) partitioners (Table
    \ref{tbl:partitioners}) on symmetric and asymmetric test matrices (Table
    \ref{tbl:matrices}) in realistic and infinite reuse situations. Quality
    is measured with cost \eqref{eq:nonsymmetriccost}, using the coefficients
    $c_{\textbf{entry}} = 1$, $c_{\textbf{row}} = 10$, and
    $c_{\textbf{message}} = 100$. For symmetric matrices, we require that the
    associated partitions be symmetric (we use the same partition for rows
    and columns). Our asymmetric test matrices also include their transposes.
    Some of the partitioners may reorder the matrix; setup time includes
    reordering operations.}\label{fig:quality}
\end{figure*}

Figure \ref{fig:quality} compares the quality of contiguous partitions under the
several cost metrics we have addressed in this work. This figure shows that even
in the contiguous regime, cost-aware partitioners can improve significantly over
simpler strategies such as equal splitting or work balancing.  The benefits were
especially pronounced for bottleneck partitioning, often improving by more than
a factor of $3\times$.  Similar benefits were observed when partitioning
Cuthill-McKee or spectrally reordered symmetric matrices, but were less
pronounced when $K$ was smaller, since these algorithms cluster nonzeros more
evenly along the diagonal. The asymmetric bottleneck partitioner ``Balance Comm,
Balance Comm'' constrains the second partition to be contiguous, and is
therefore more competitive on these bandwidth-reduced reordered matrices where
nonzeros occur in a roughly diagonal pattern.  The asymmetric ``Balance Comm,
Assign Greedy'' and ``Balance Work, Assign Local'' partitioners allow the column
partition to be noncontiguous, and are therefore more robust to natural
orderings.  While our partitioners significantly improved on the quality of the
equal splitting strategy, no such improvements were found when blocking to
minimize cache misses.

\begin{figure}
    \centering
    {\large Contiguous Partitioner Runtimes}

    \centering
    Symmetric Matrices
    \resizebox{\linewidth}{!}{%
    \includegraphics[]{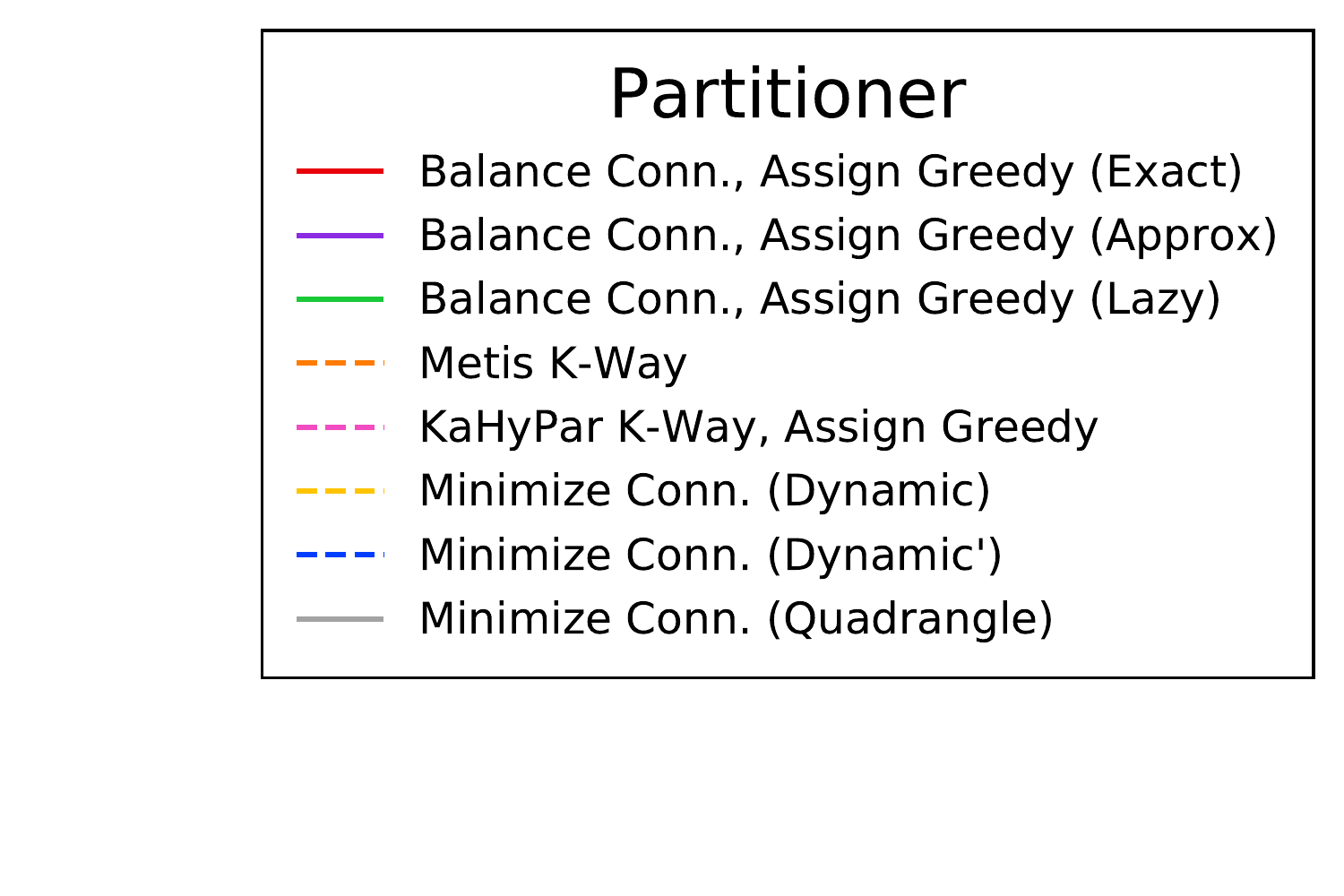}%
    \includegraphics[]{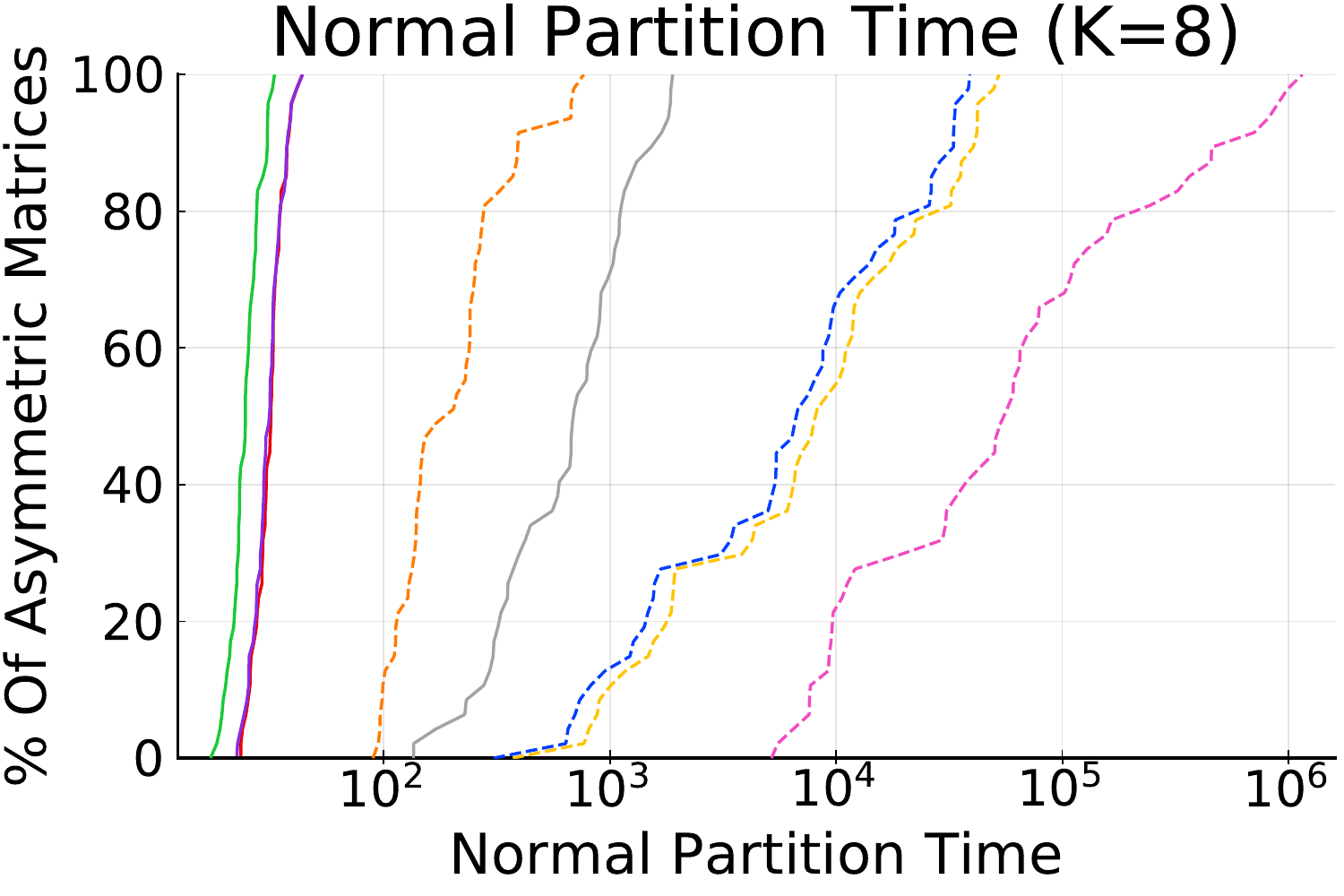}%
    \includegraphics[]{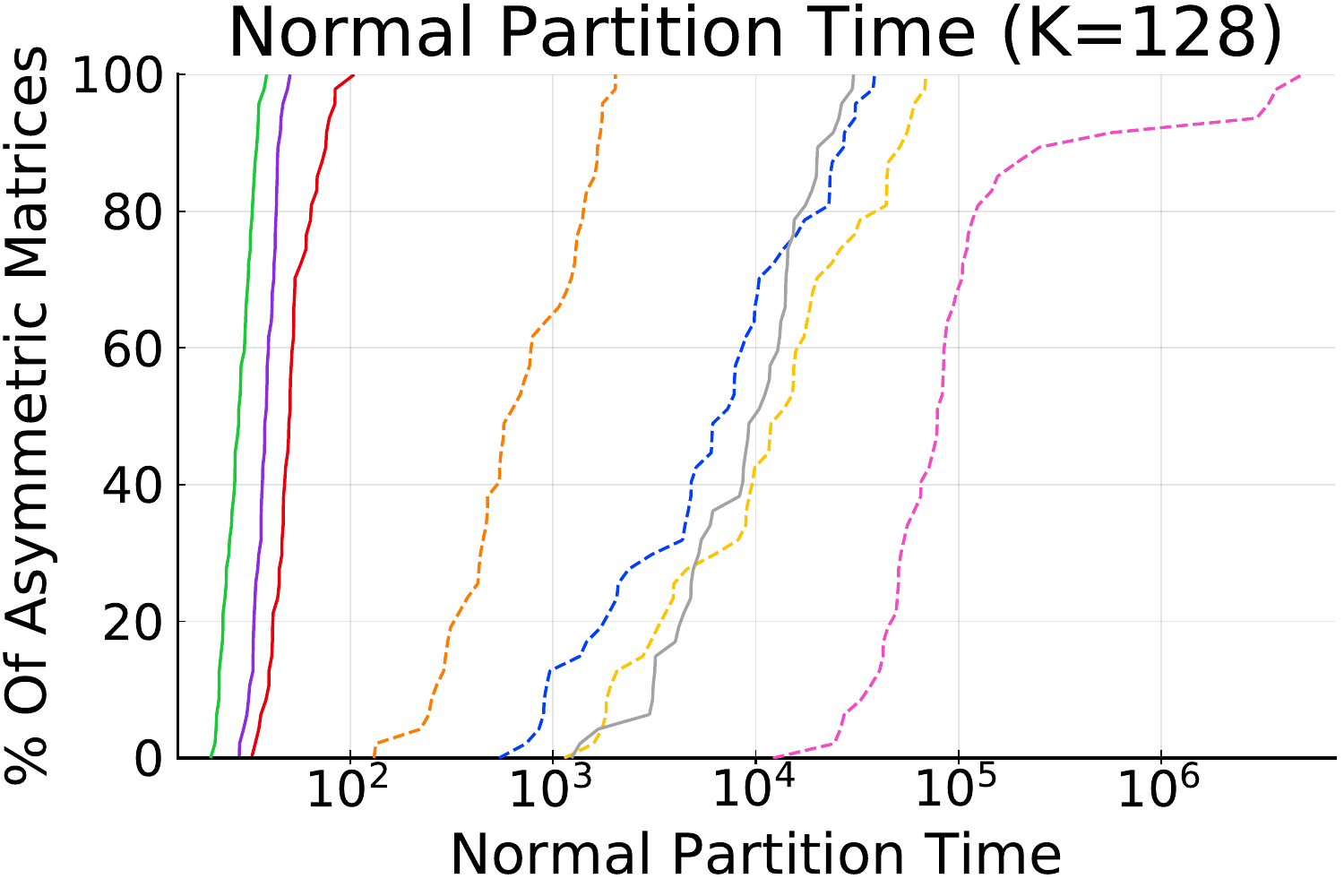}%
    }
    Asymmetric Matrices
    \resizebox{\linewidth}{!}{%
    \includegraphics[]{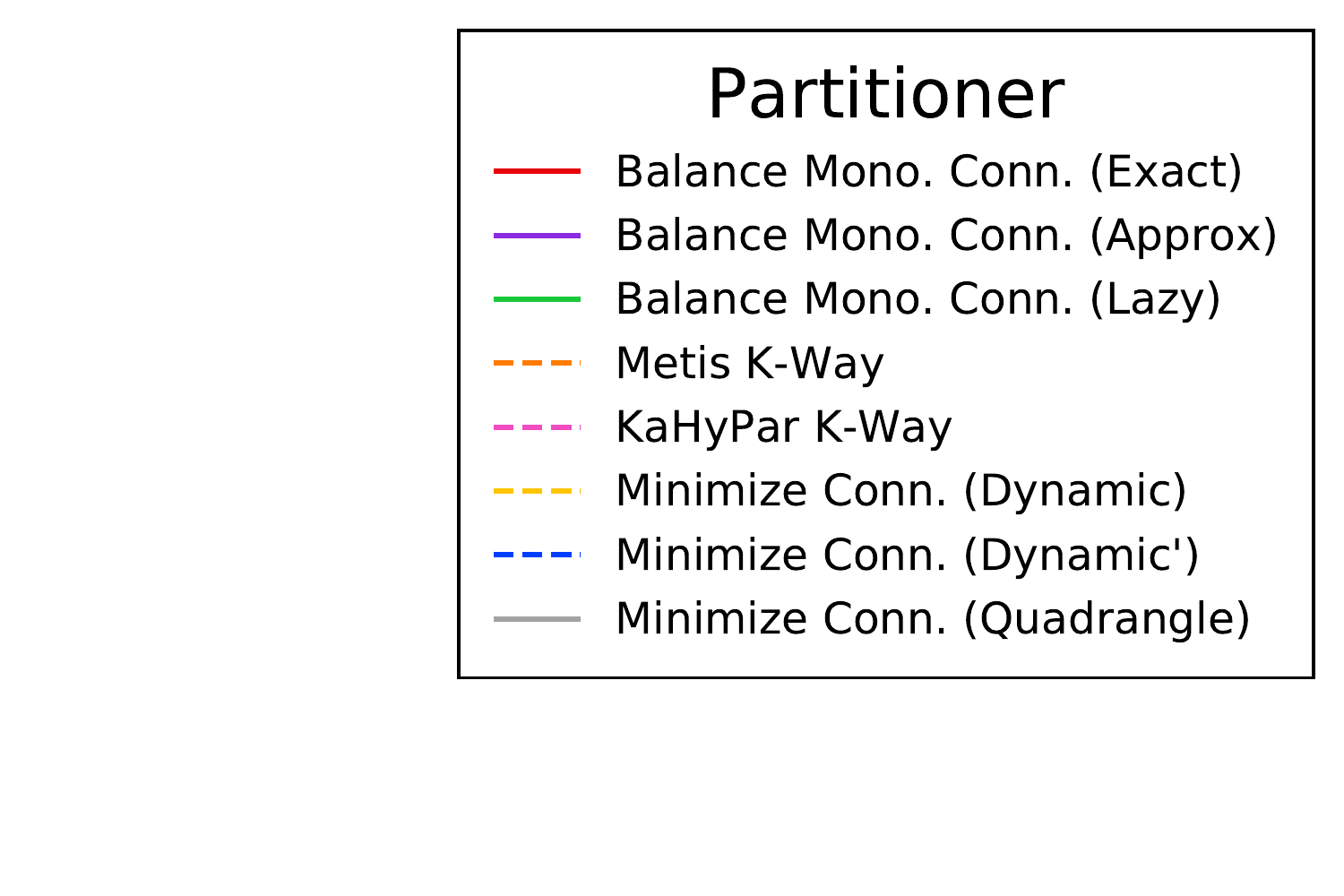}%
    \includegraphics[]{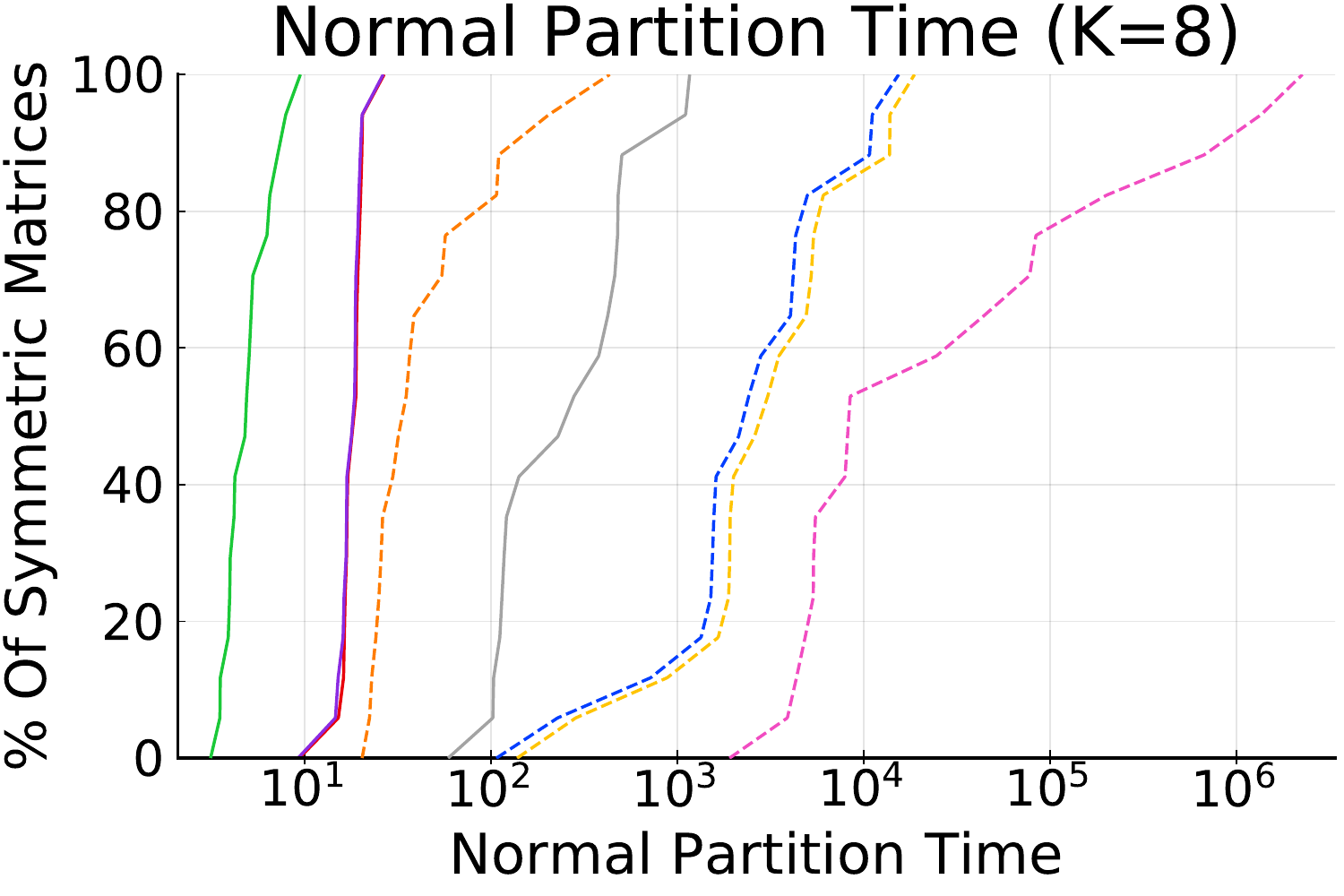}%
    \includegraphics[]{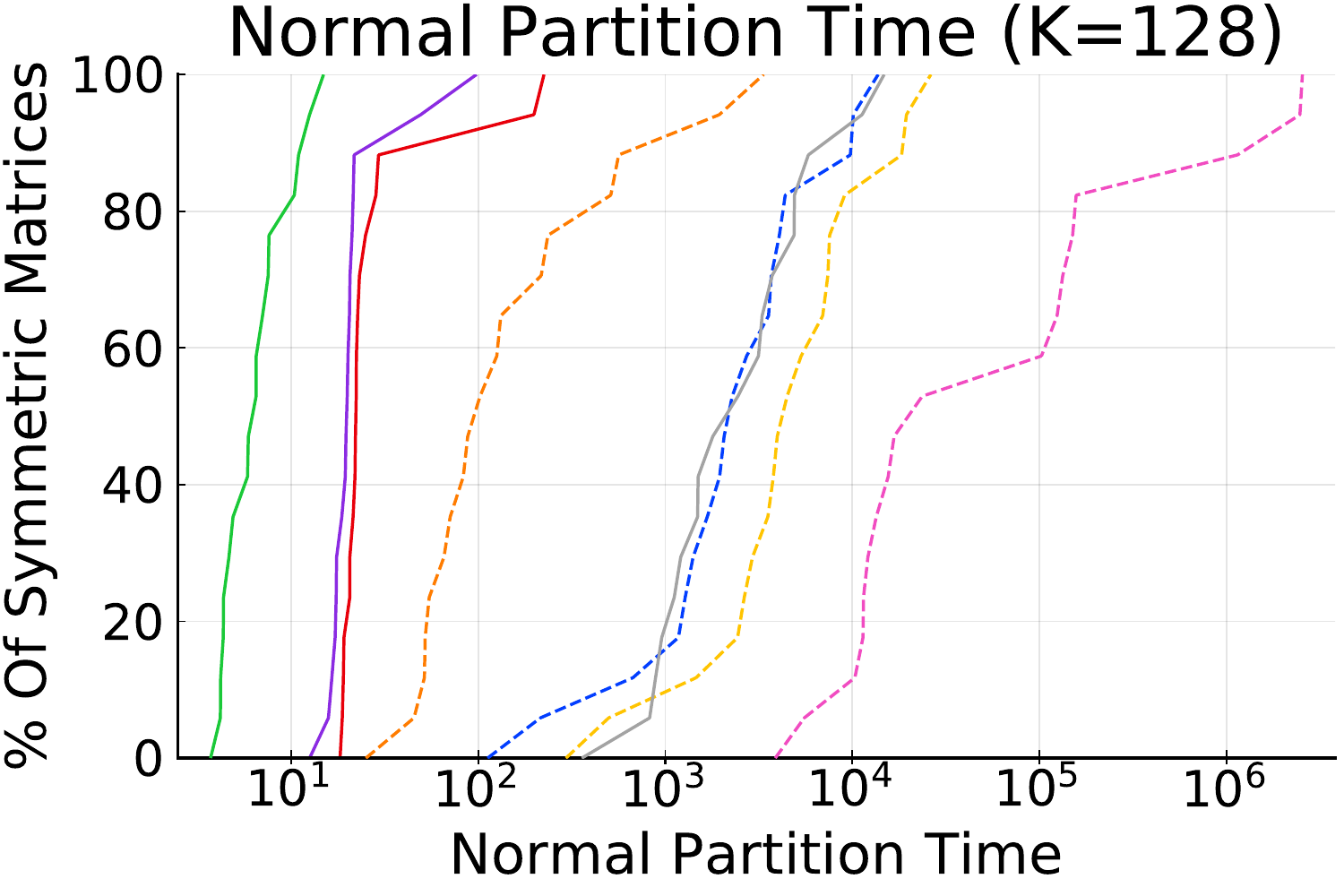}%
    }

    \caption{Performance profiles comparing normalized partitioning runtime of
    our partitioners (Table \ref{tbl:partitioners}) on symmetric and asymmetric
    test matrices (Table \ref{tbl:matrices}) across contiguous and
    noncontiguous, total and bottleneck regimes.}\label{fig:runtime}
\end{figure}

Figure \ref{fig:runtime} compares the normalized runtime of all of our
partitioners.  In general, the relative runtimes of these partitioners agrees
with their asymptotic descriptions. Contiguous bottleneck partitioning is by far
the fastest partitioning strategy, and while contiguous total hyperedge cut 
partitioning is much faster than general total hyperedge cut partitioning
(KaHyPar), it is also slower than general simple edge cut partitioning (Metis).

\begin{figure}
    \centering
    {\large Cache Blocking Runtimes}

    \resizebox{0.6\linewidth}{!}{%
    \includegraphics[]{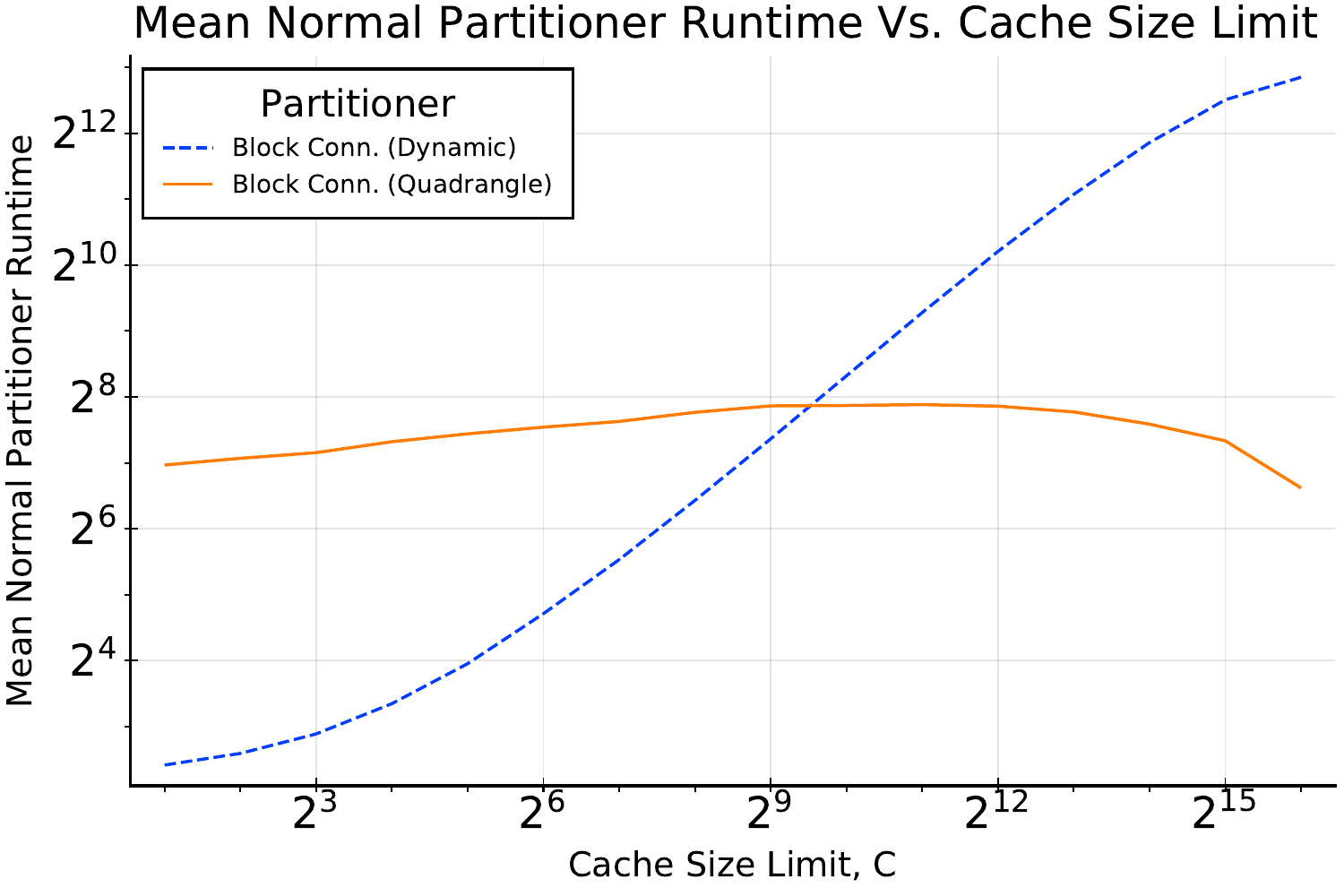}%
    }
    \caption{Mean normalized runtime of our cache blocking algorithm as a
    function of the size constraint on blocks.}\label{fig:cache}
\end{figure}

When total partitioning, using offline dominance counting with $b=2$ and our quadrangle-accelerated
algorithm was significantly faster than the dynamic programming approach with
online dominance counting when the number of parts $K$ was small, but the
opposite was true when $K$ was large. Note that both algorithms reduce the
$K$-partitioning problem to a sequence of $K$ least-weight subsequence problems.
As $K$ becomes smaller, the average weight constraint on each part grows, and
our dynamic programming algorithm endures longer split point searches. When the
average weight constraint is long enough, the quadrangle algorithm's asymptotic
advantage becomes visible. Since the cache blocking problem can be solved as a
single constrained least-weight subsequence problem, Figure \ref{fig:cache}
provides perspective on this effect by comparing the runtime of the dynamic
programming algorithm and the quadrangle algorithm as the weight constraint
grows. We see that the quadrangle algorithm begins to outperform the dynamic
programming algorithm when the weight constraint exceeds about 1000 vertices.
When $K = 8$, the dynamic, simultaneous, and quadrangle total connectivity
partitioners had average runtimes of
$1.16e04$\unskip,
$9.24e03$\unskip,
and
$667$\unskip~SpMVs,
respectively. The quadrangle algorithm resulted in a maximum speedup of
$53\times$\unskip~and
a mean speedup of
$15.1\times$\unskip~
over the dynamic programming algorithm.

When bottleneck partitioning, using offline dominance counting with $H=3$ and our linear time Algorithms
\ref{alg:bisectpartition} or \ref{alg:nicolpartition} was efficient enough to be
practical. However, for approximate partitioning, it was empirically
fastest to use online dominance counters and fuse them directly into our linear
search version of Algorithm \ref{alg:bisectpartition}, even though it is asymptotically slower by a factor of
$\log(c_{\high}/(c_{\low}\epsilon))$. The resulting pseudocode is described in
Appendix \ref{app:lazy}. When $H=3$, the exact algorithm is
expected to run in linear time when $K$ grows slower than $m^{1/3}$. When $K =
8$, on average over our symmetric matrices, the exact, approximate, and lazy
approximate algorithms were able to partition under cost
\eqref{eq:monotonicsymmetriccost} in an average of
$18$\unskip,
$17.7$\unskip,
and
$5.15$\unskip~SpMVs,
respectively. The improved runtime of the linear search algorithm is due to the
simplicity of the implementations, since the cost calculations are fully
inlined.


\begin{figure*}
    \centering
    {\large General Partitioning}

    \centering
    Bottleneck Load $+$ Connectivity On Symmetric Matrices
    \resizebox{\linewidth}{!}{%
    \includegraphics[]{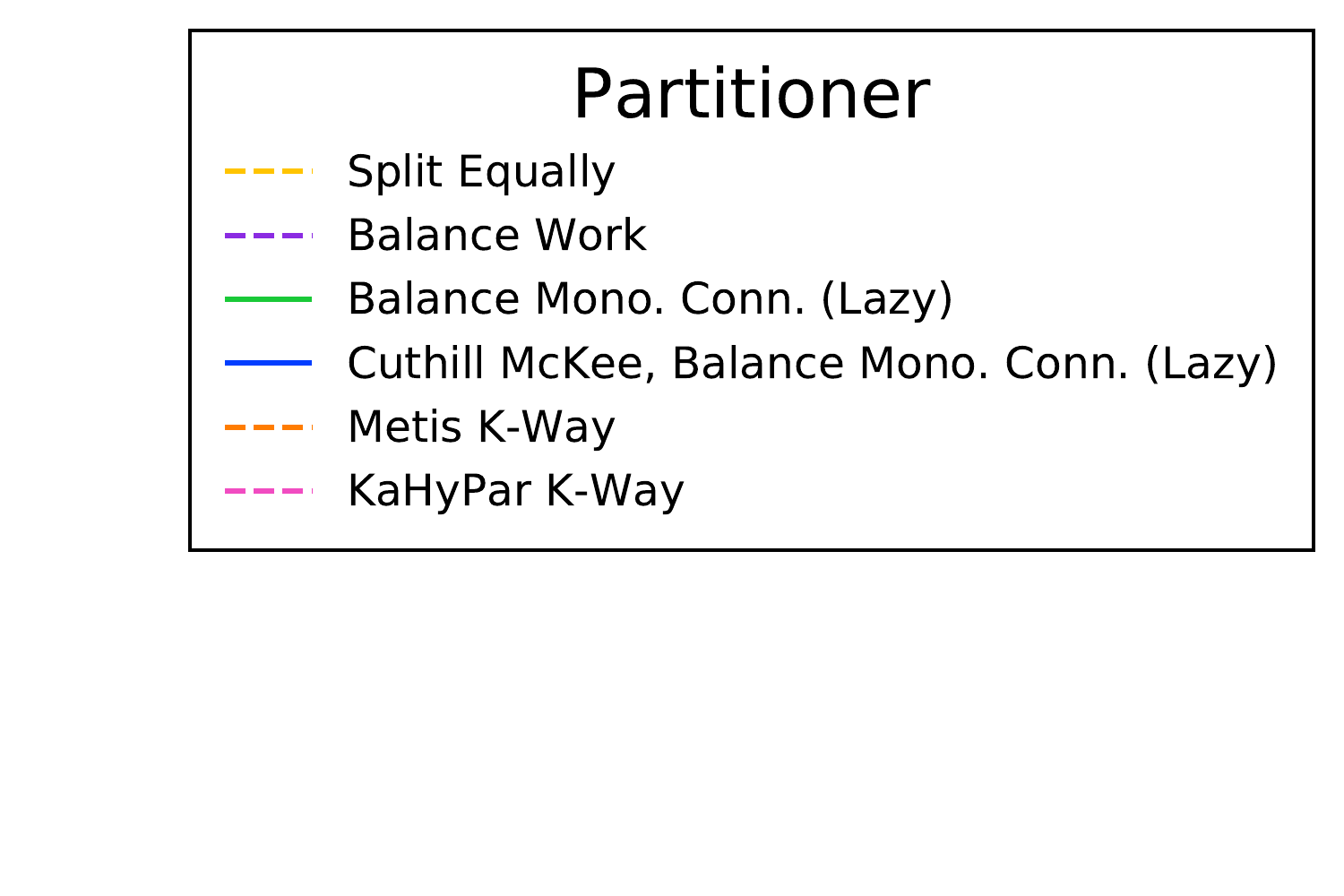}%
    \includegraphics[]{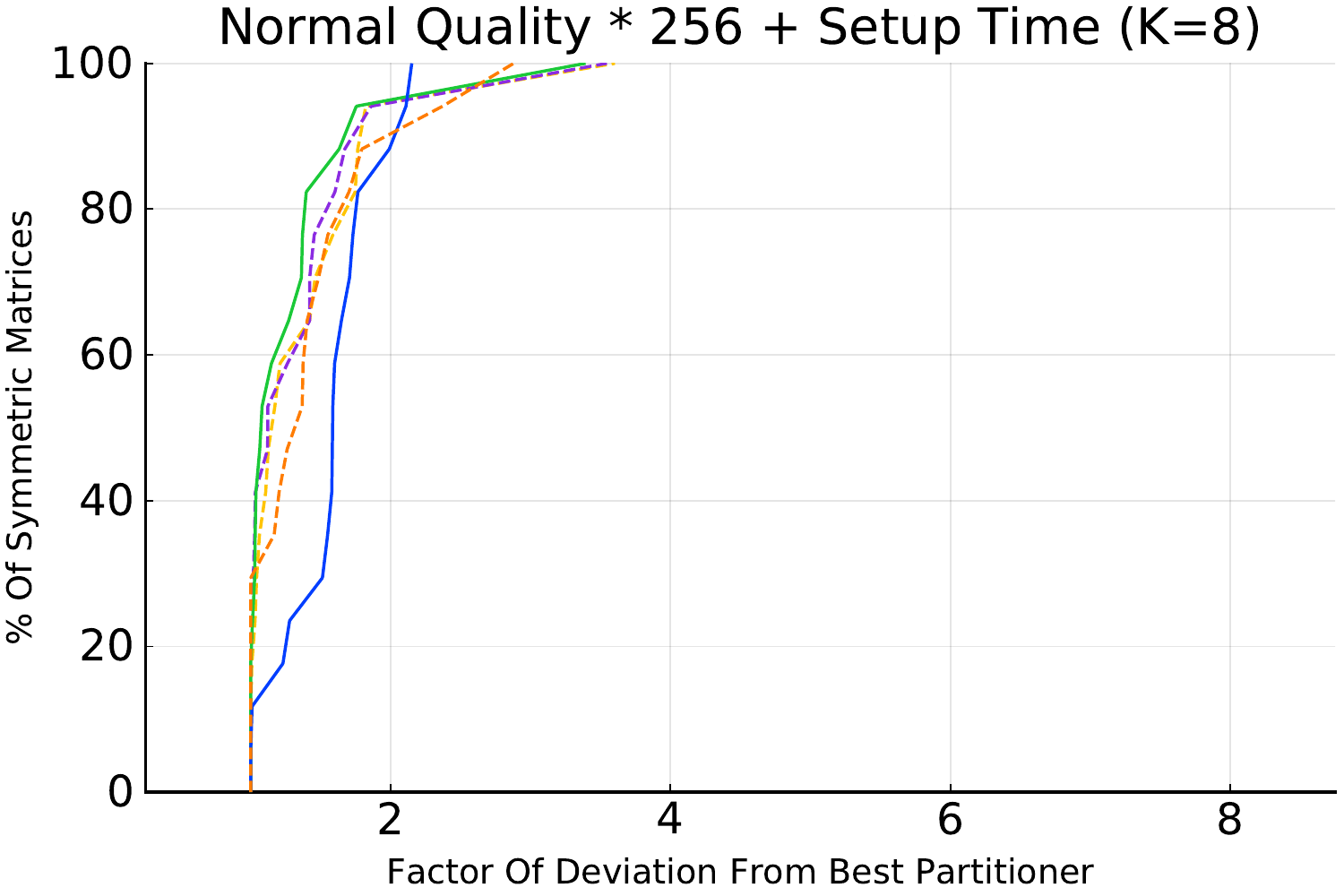}%
    \includegraphics[]{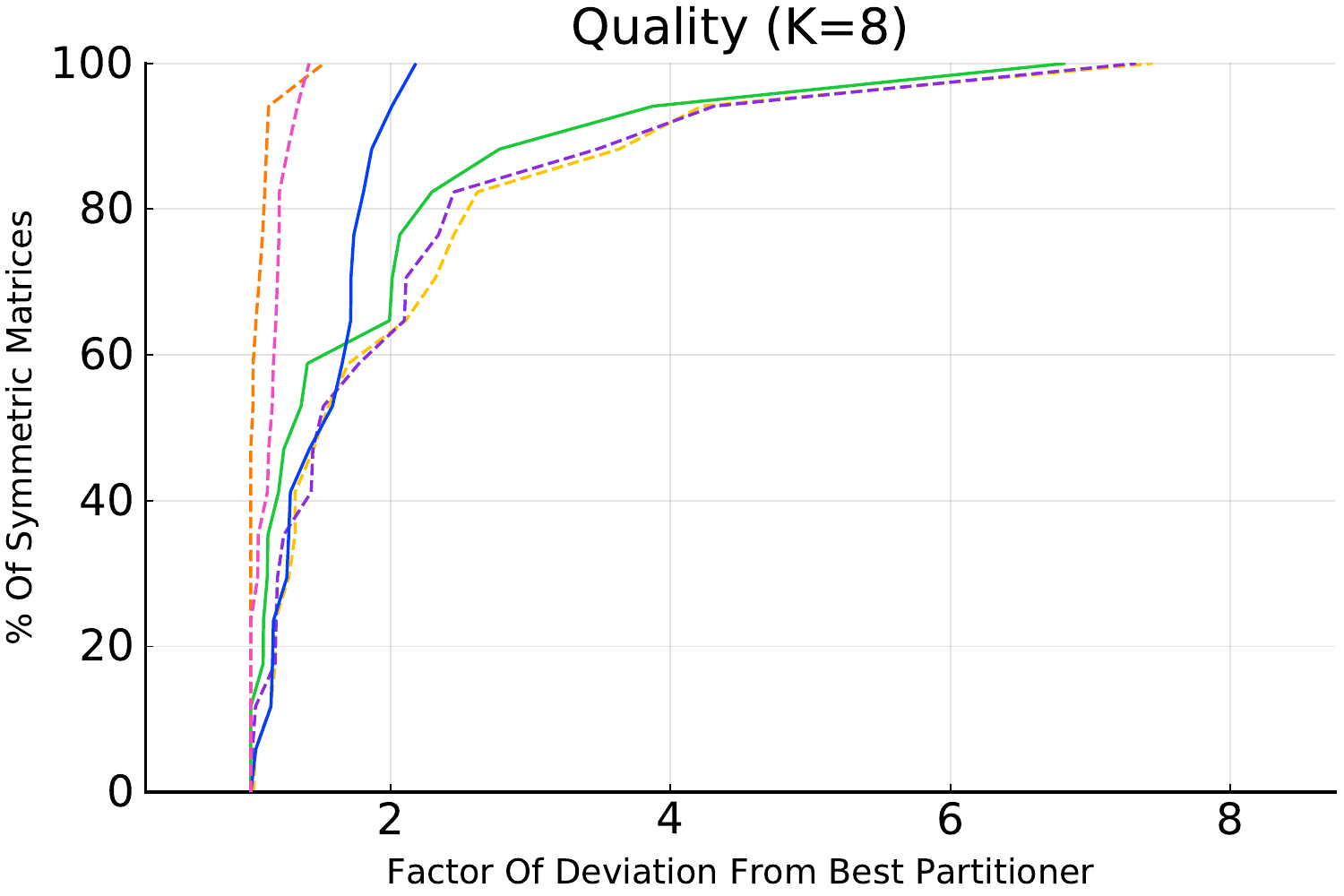}%
    \includegraphics[]{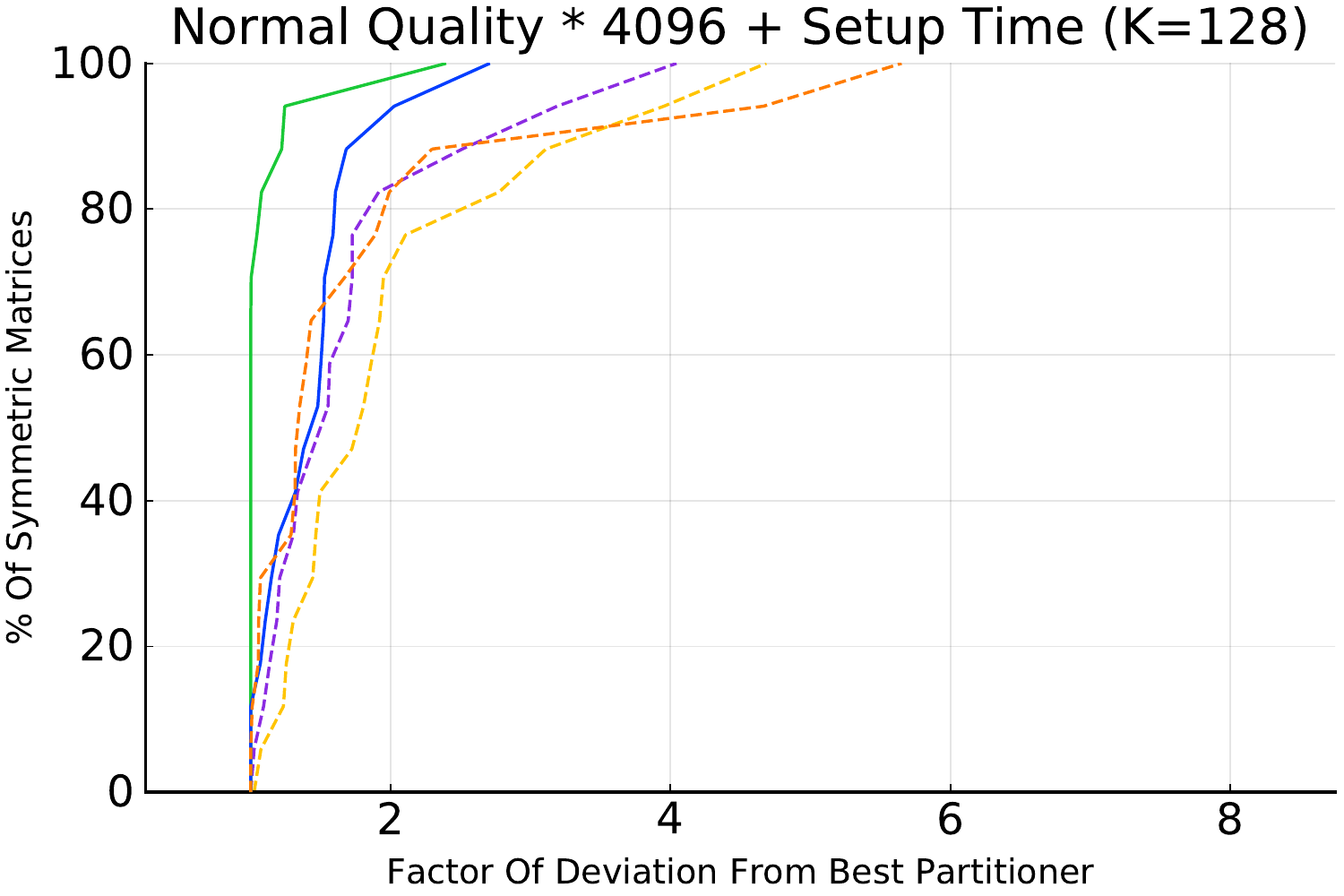}%
    \includegraphics[]{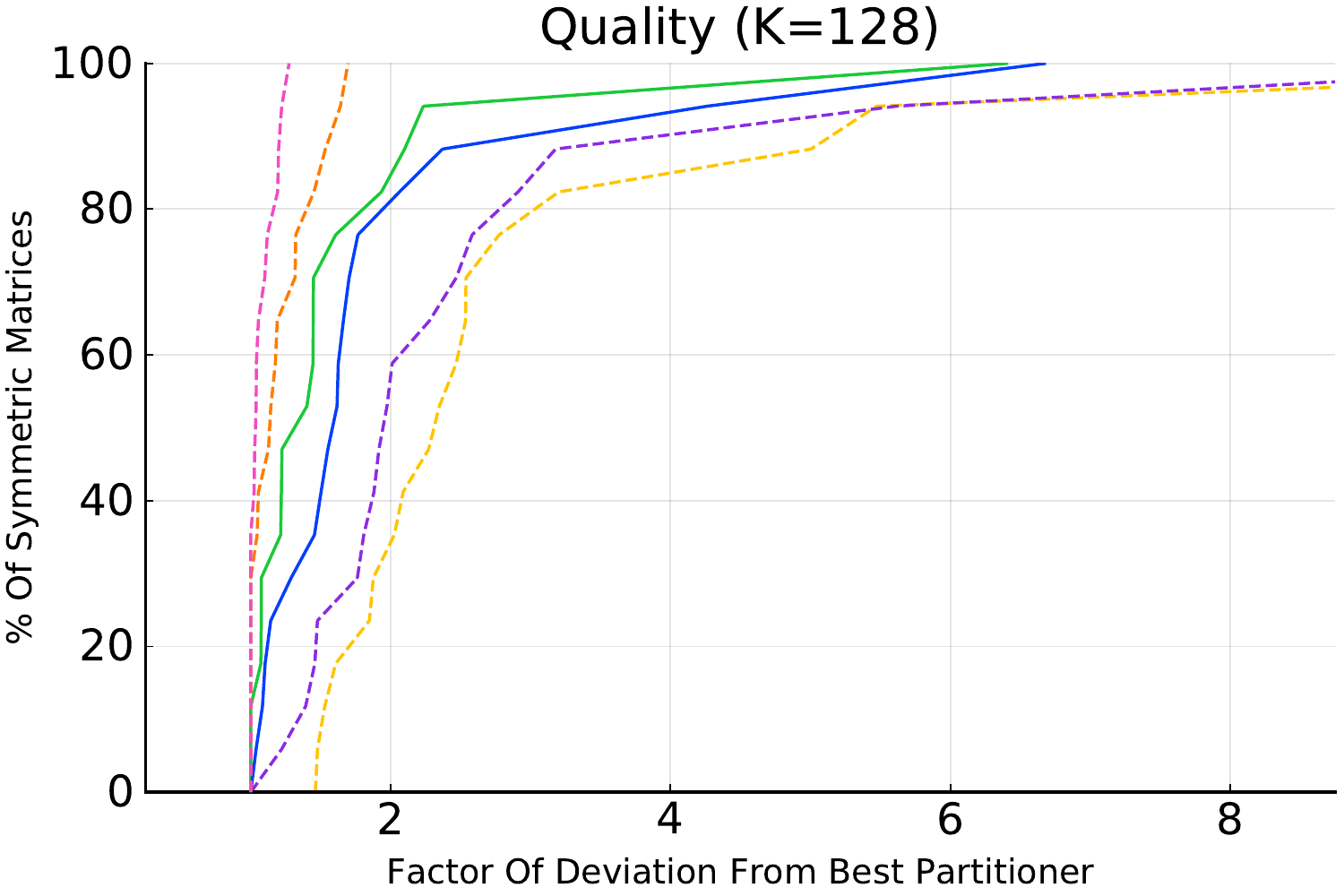}%
    }

    Bottleneck Load $+$ Connectivity On Symmetric Matrices
    \resizebox{\linewidth}{!}{%
    \includegraphics[]{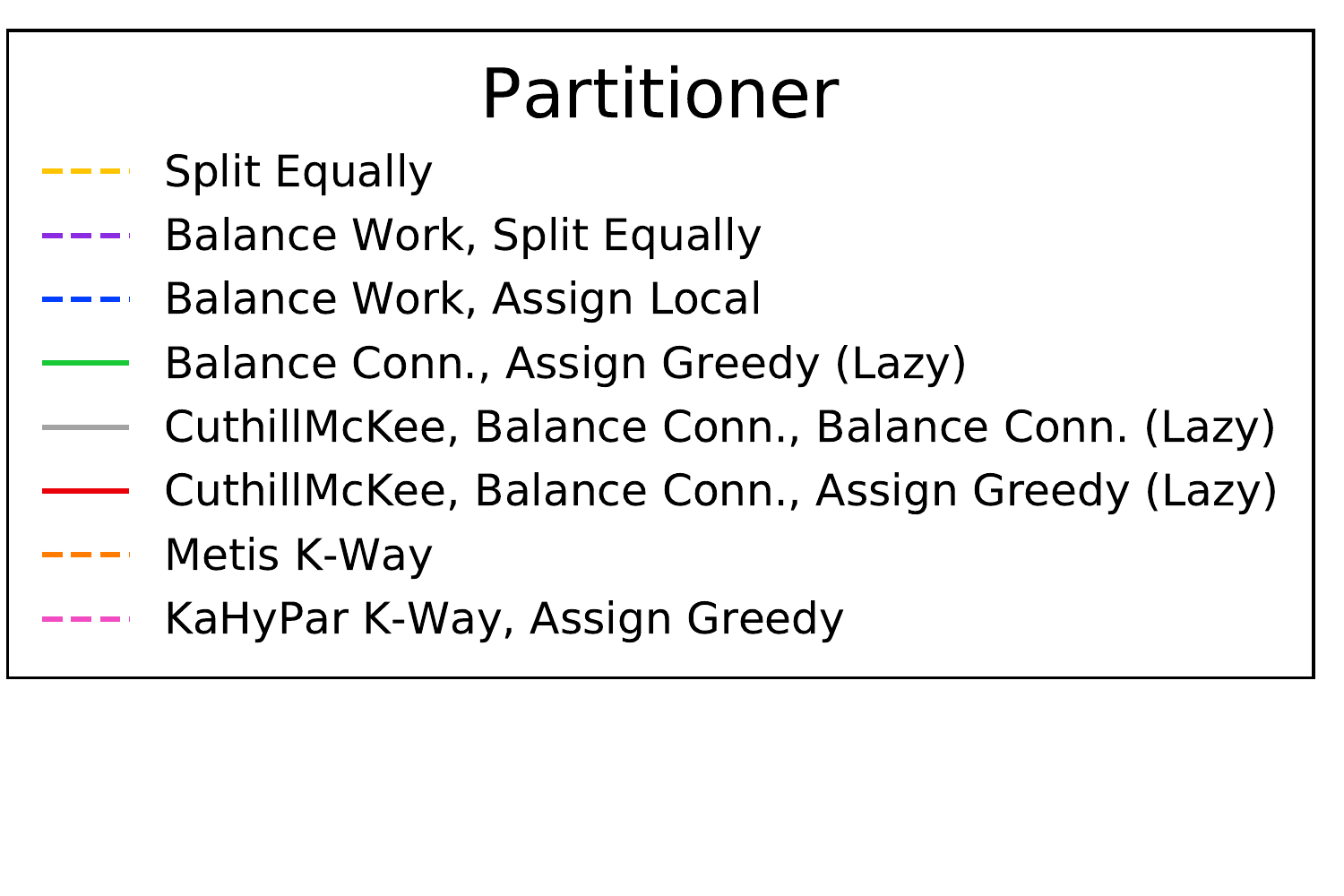}%
    \includegraphics[]{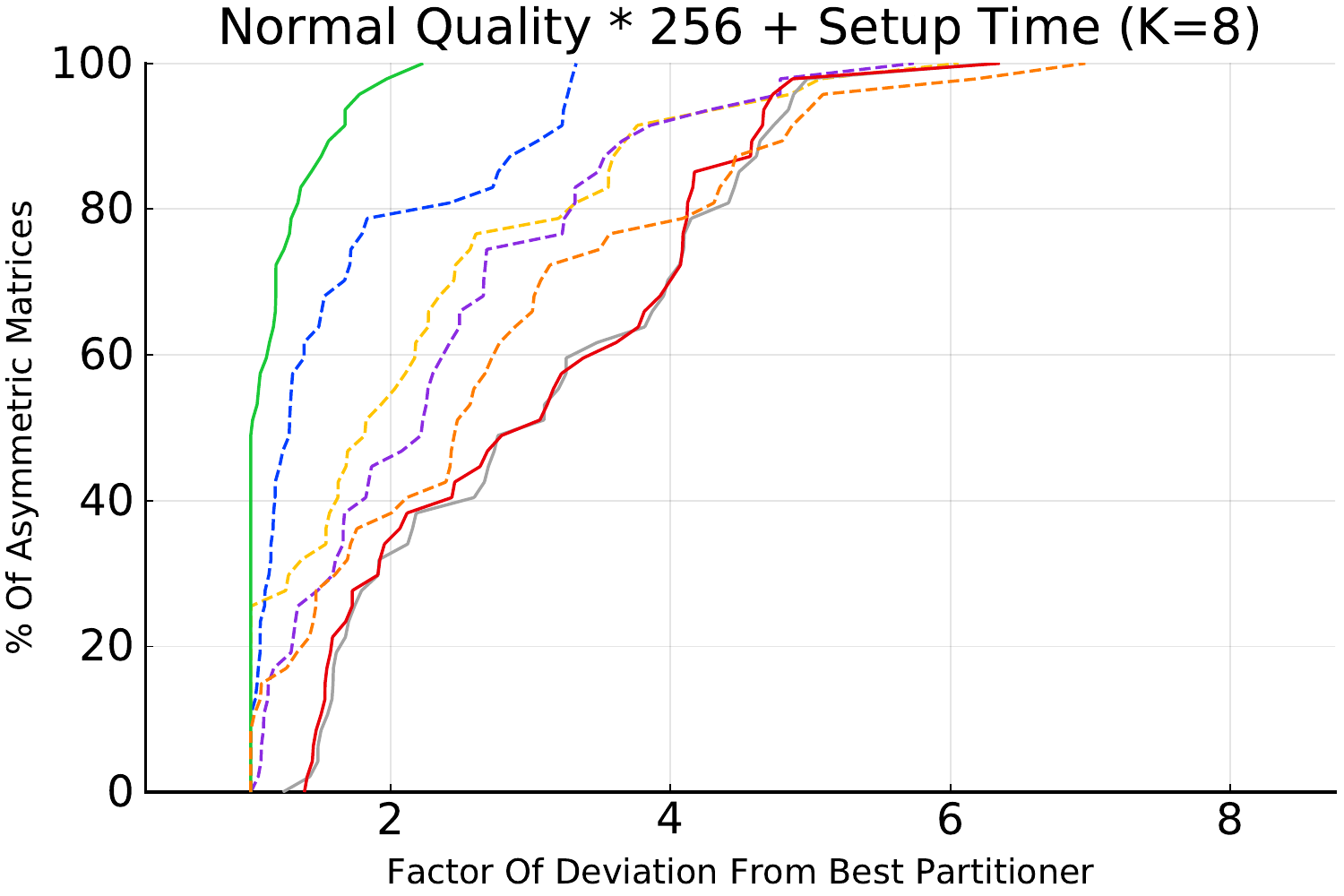}%
    \includegraphics[]{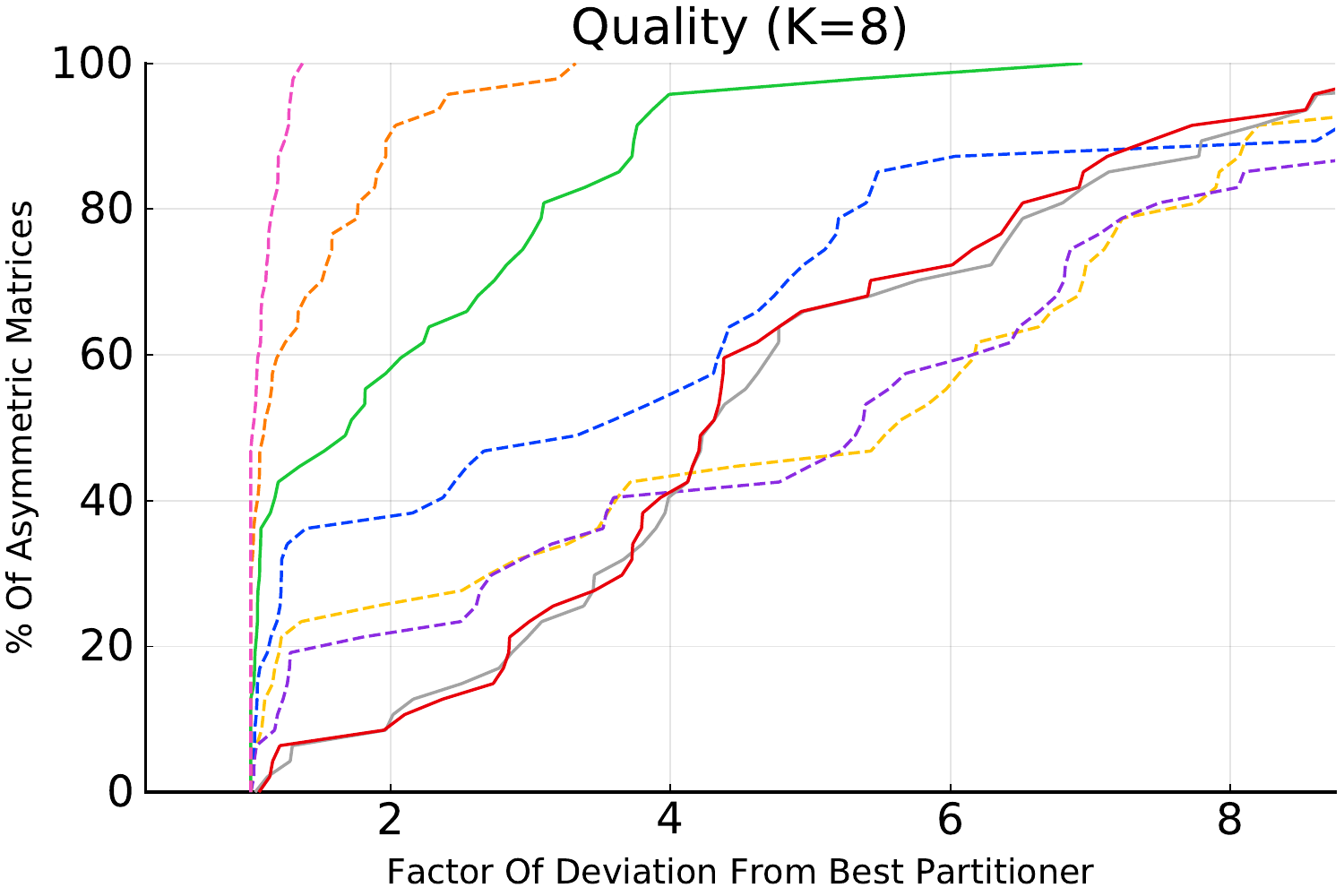}%
    \includegraphics[]{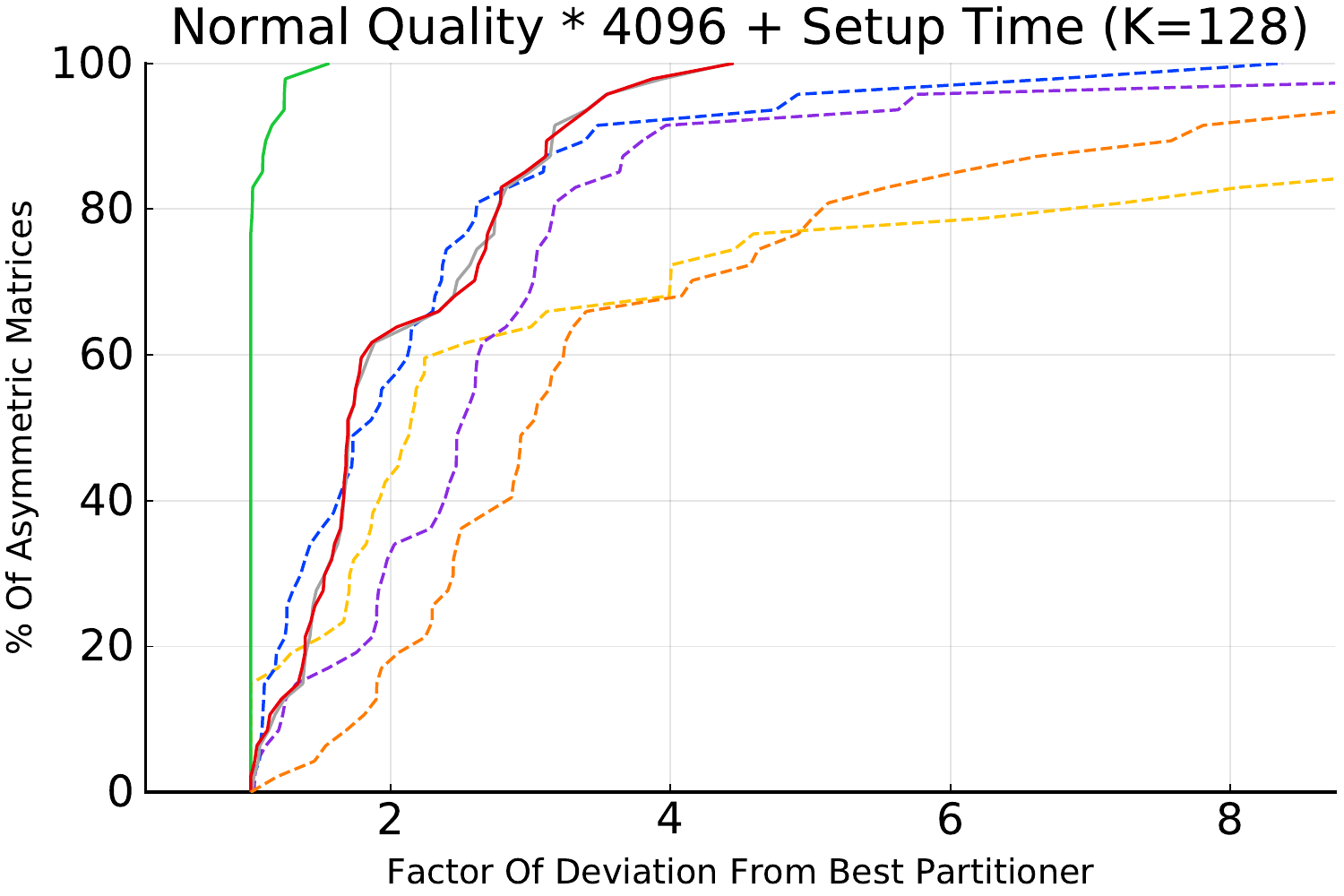}%
    \includegraphics[]{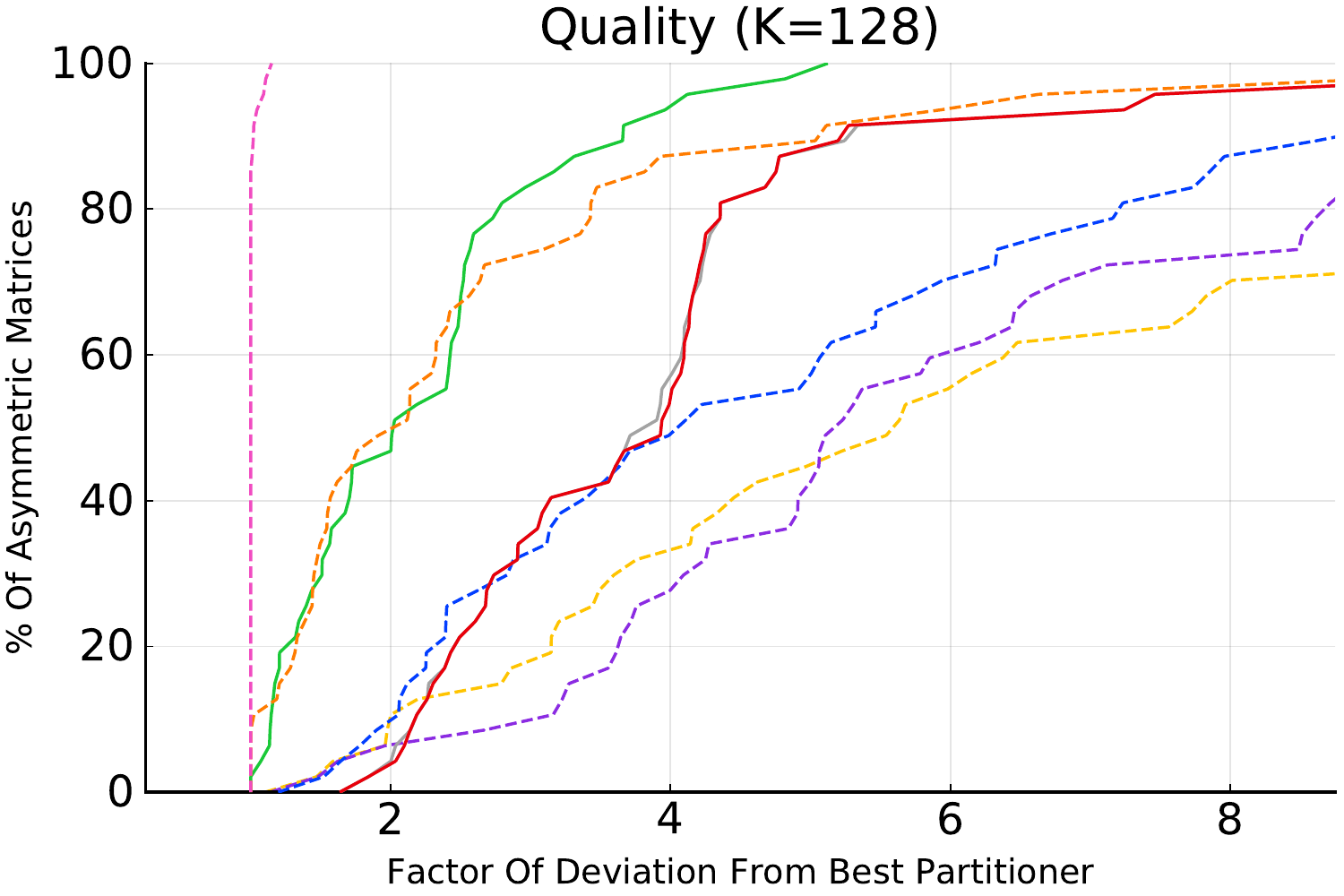}%
    }

    \caption{Performance profiles comparing normalized modeled quality of our
    general (possibly noncontiguous) partitioners (Table
    \ref{tbl:partitioners}) on symmetric and asymmetric test matrices (Table
    \ref{tbl:matrices}) in realistic and infinite reuse situations. Quality
    is measured with cost \eqref{eq:nonsymmetriccost}, using the coefficients
    $c_{\textbf{entry}} = 1$, $c_{\textbf{row}} = 10$, and
    $c_{\textbf{message}} = 100$. For symmetric matrices, we require that the
    associated partitions be symmetric (we use the same partition for rows
    and columns). Our asymmetric test matrices also include their transposes.
    Some of the partitioners may reorder the matrix; setup time includes
    reordering operations.}\label{fig:general_partitioning}
\end{figure*}

Figure \ref{fig:general_partitioning} compares symmetric and asymmetric
partitioners on our bottleneck cost in the case where the partition need not be
contiguous. The figure shows that contiguous communication-aware partitioners
are often competitive with reordering partitioners when we account for setup
time and realistic levels of partition reuse. Furthermore, these results show
that communication-aware bottleneck contiguous partitioning is more effective
when the number of parts is high (and the aspect ratio of each part emphasizes
communication over computation).  Cuthill-McKee reordering before partitioning
is only a good strategy for symmetric matrices when the number of parts is low.
The KaHyPar hypergraph partitioner produced the best quality solutions, but took
far too long to partition, and fell off the drawable region of the graph.

\begin{figure}
    \centering
    PARSEC/H2O (Symmetric Partition)

    \resizebox{\linewidth}{!}{\includegraphics[height=10cm]{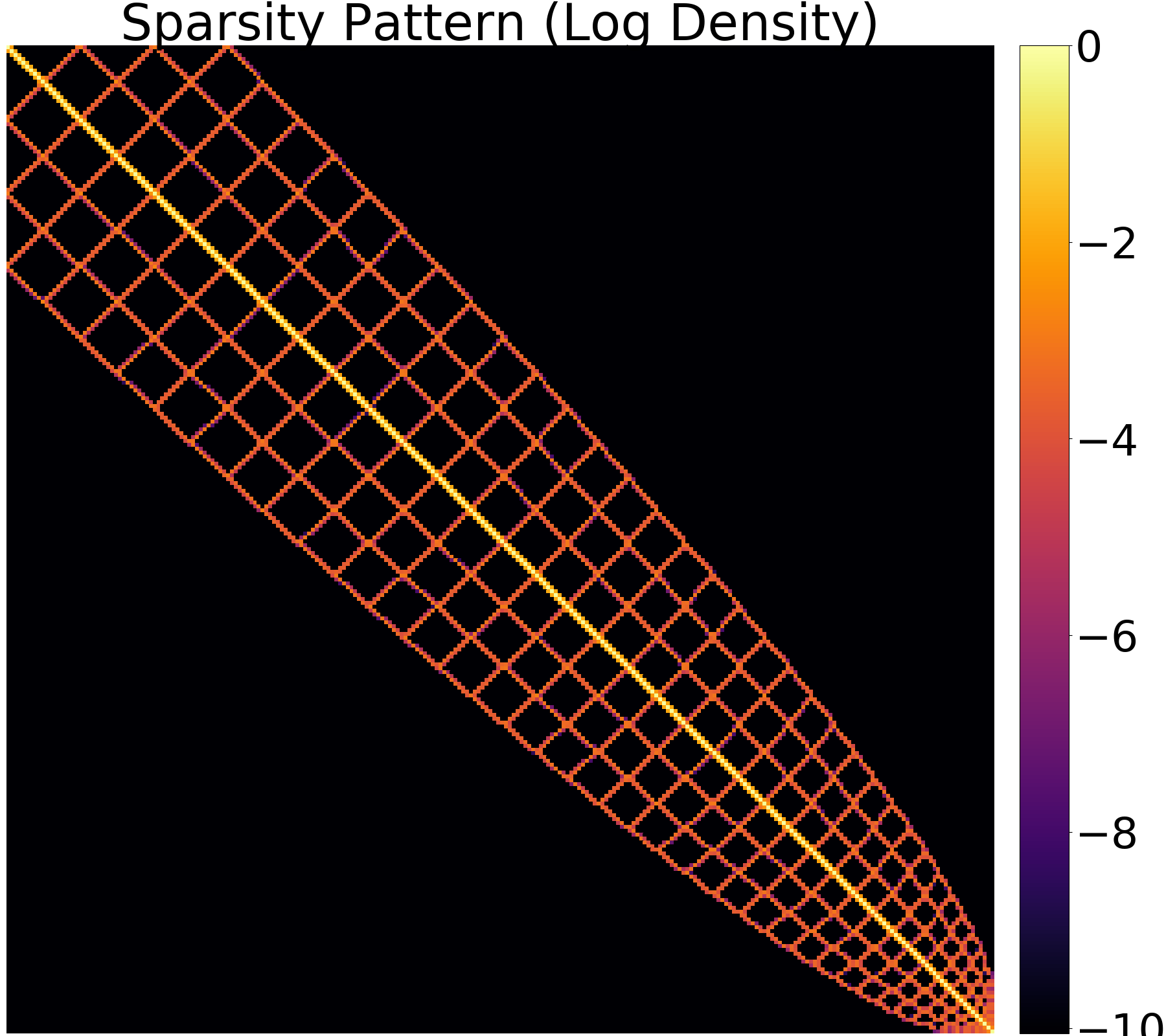}\quad\includegraphics[height=10cm]{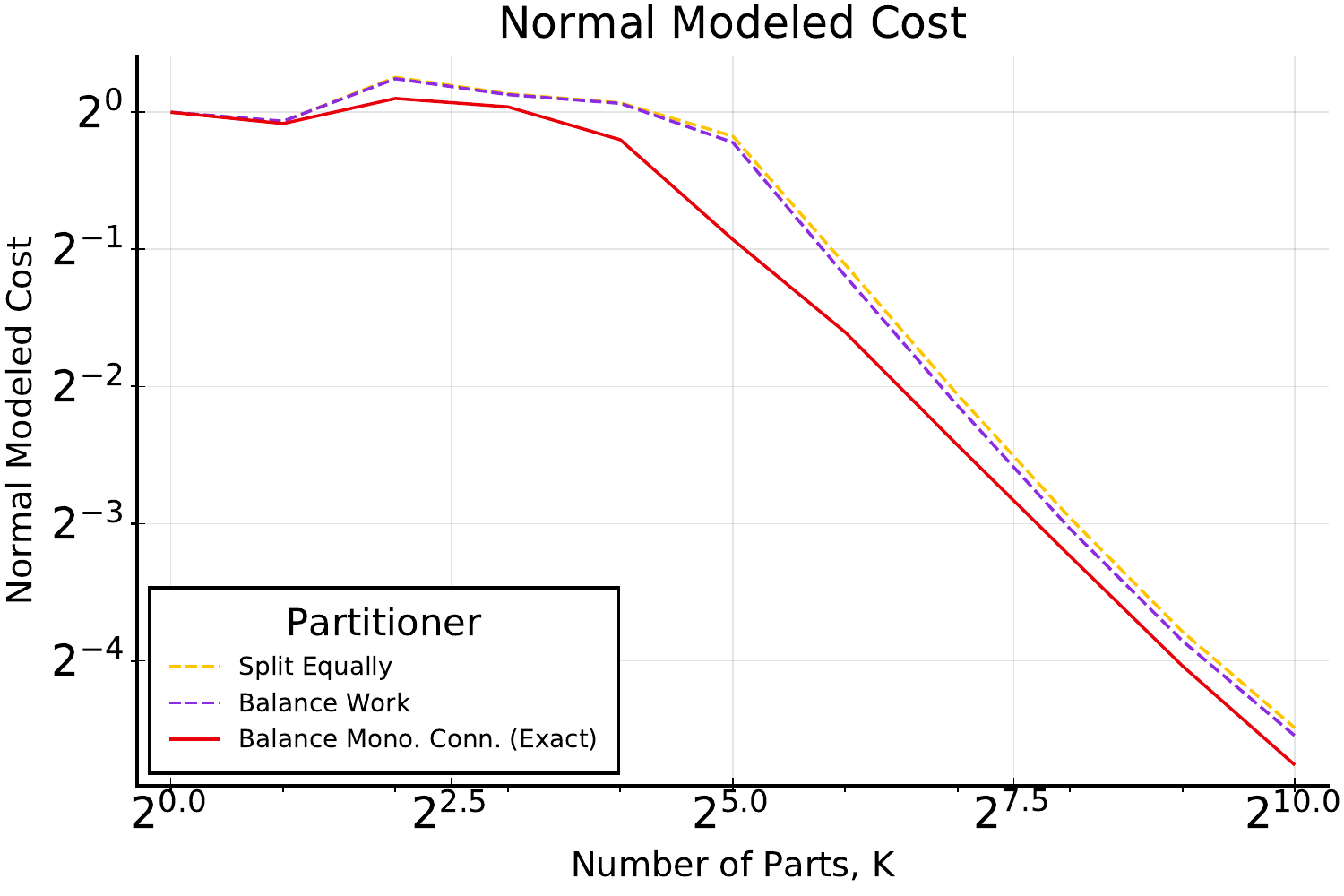}}

    Mallya/lhr34 (Column Partition)

    \resizebox{\linewidth}{!}{\includegraphics[height=10cm]{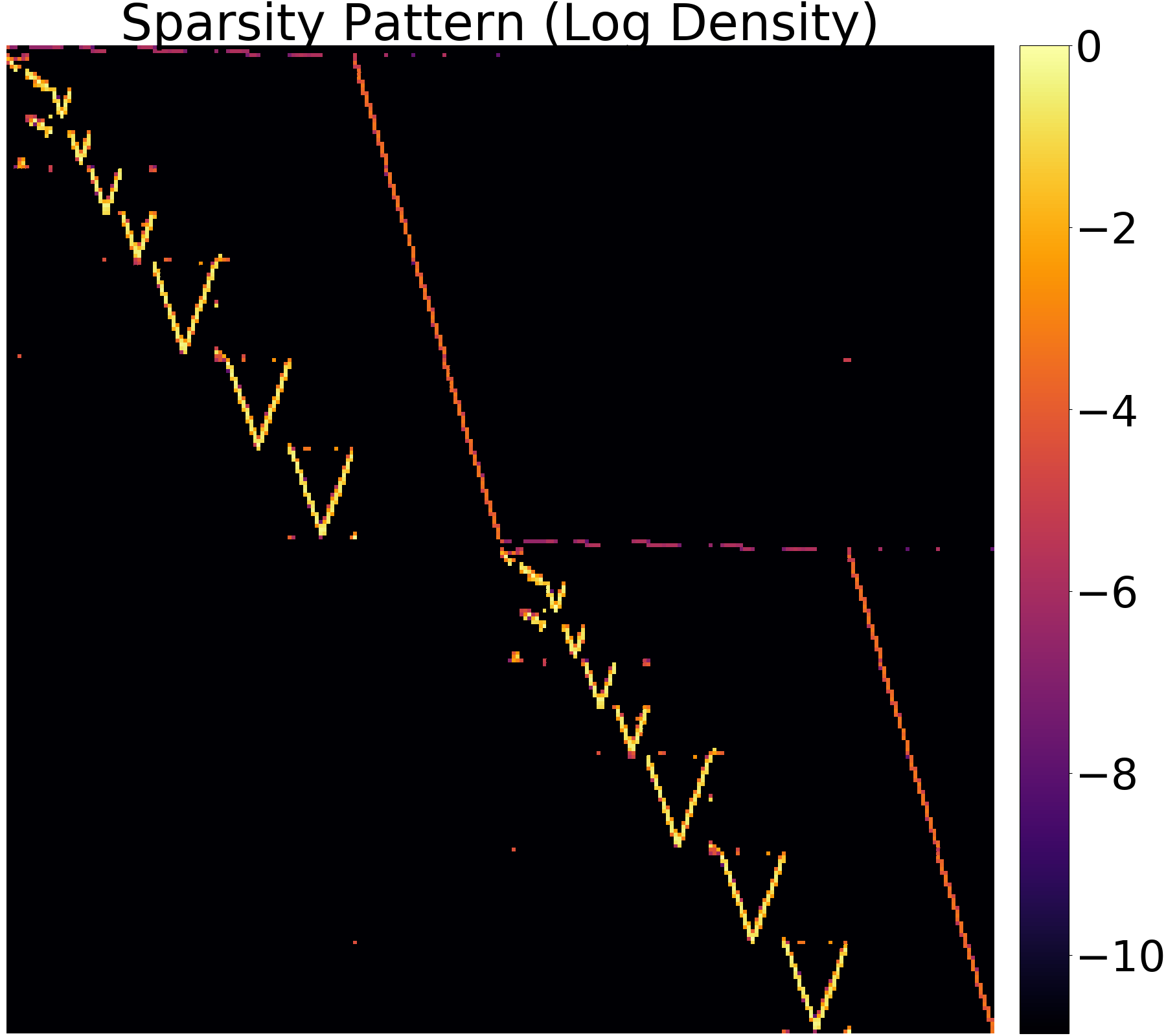}\quad\includegraphics[height=10cm]{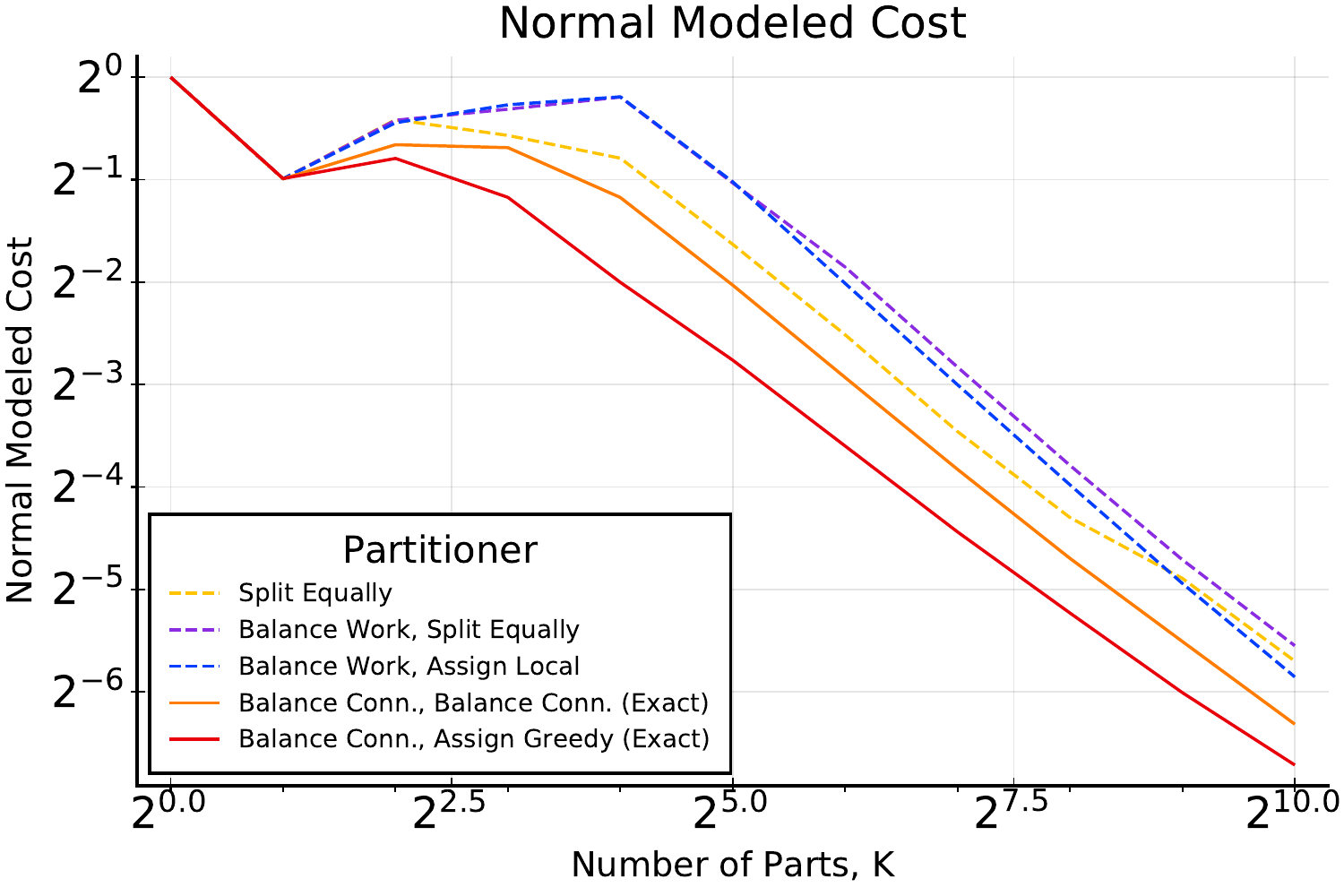}}

    \caption{Matrix sparsity patterns and the resulting partition quality of
    contiguous partitioners. Quality is measured with cost
    \eqref{eq:nonsymmetriccost}, using the coefficients $c_{\textbf{entry}} =
    1$, $c_{\textbf{row}} = 10$, and $c_{\textbf{message}} = 100$. Sparsity
    patterns show logarithmic density of nonzeros within each pixel in a 256
    by 256 grid. Partitioners and matrices are described in Tables
    \ref{tbl:matrices} and \ref{tbl:partitioners}.}\label{fig:spys}
\end{figure}

Figure \ref{fig:spys} is meant to help give intuition for how the sparsity
pattern of a matrix affects contiguous partitioning performance. Contiguous
partitioning returns for PARSEC/H2O, a quantum chemistry problem, show the
scale-dependence of structure utilization. For our naturally-ordered
partitions, we see that increasing the number of processors does not reduce
the cost until the partitions are skinny enough to resolve the structure in
the matrix. Since the pattern is roughly 32 large diamonds long along the
diagonal, we can only split all the diamonds at around 64 processors, which
is roughly when the cost of the partitions begins to decrease in earnest.
Mallya/lhr34 is a matrix arising in a nonlinear solver in a chemical process
simulation. Because this matrix has a $2\times2$ macro block structure, it is
easily split into two parts in its natural ordering. However, continued
bisection increases communication costs until there are enough parts to
effectively split the finer structures in the matrix. Since Mallya/lhr34 is
asymmetric, we can compare the effectiveness of greedy and contiguous
secondary partitioning. Nonzeros occur in roughly two bands; the greedy
strategy can split local rows among column parts in both bands, while the
contiguous strategy suffers because the nonzeros aren't clustered on the
diagonal.

\section{Conclusion} 

Traditional graph partitioning is NP-Hard, and only a limited set of objectives
have been considered. While the ordering of the rows and columns of a matrix
does not affect the meaning of the described linear operation, there are many
situations where it carries useful information about the problem structure.
Contiguous partitioning shifts the burden of reordering onto the user, asking
them to use domain-specific knowledge or known heuristics to produce good
orderings. In exchange, we show that the contiguous partitioning problem can be
solved optimally in linear or near linear time and space, provably optimizing
cost models which are close to the realities of distributed parallel computing.

Researchers point out that traditional graph partitioning approaches are
inaccurate, since they minimize the total communication, rather than the
maximum runtime of any processor under combined work and communication factors.
\cite{hendrickson_load_2000, ucar_encapsulating_2004,
bisseling_communication_2005, bader_parallel_2013}. We show that, in the
contiguous partitioning case, we can efficiently minimize the maximum runtime
under the more accurate combined cost models.

We present a rich framework for constructing and optimizing expressive cost
models for contiguous decompositions of iterative solvers. We describe a
taxonomy of cost model properties, including convexity, monotonicity, and
perhaps subadditivity. Using a set of efficiently computable ``atoms'', we can
construct complex ``molecules'' of cost functions which express complicated
nonlinear dynamics such as cache effects, memory constraints, and communication
costs. We adapt state-of-the-art load balancing and least-weight subsequence
algorithms to optimize our costs. In order to efficiently compute our
communication costs, we reduce our cost queries to dominance count queries and
generalize a classical dominance counting algorithm to reduce construction time
by increasing query time. Our new data structure can also be used to compute
sparse prefix sums.  We demonstrate that there are significant quality
variations among partitioners in the contiguous setting, and that all of our
algorithms efficiently produce high-quality partitions in practice.

There are several opportunities for future work. First, we discuss parallel
implementations of our partitioners. Because our lazy partitioner relies on
linear search to construct partitions, it does not appear to be amenable to
parallelization. On the other hand, the bulk of the runtime for our
binary-search-based partitioners (Algorithms \ref{alg:bisectpartition} and
\ref{alg:nicolpartition}) consists of dominance counting subroutines. As our
dominance counters (Algorithms \ref{alg:construct} and \ref{alg:query}) are
composed of decorated histogram sorts and one-dimensional prefix sums,
parallelizing our dominance counters with similar strategies looks promising.

Our approaches might also be accelerated with algorithmic improvements. Since
the dominance counters recursively decompose the index space, there may be
opportunities to integrate dominance counting with binary search subroutines.
Additionally, one might be able to approximate the computation by sampling edges
or nonzeros.

Finally, we hope to see investigations applying these techniques to
embedding-based or geometric multi-jagged partitioning approaches
\cite{chan_optimality_1997, aydin_distributed_2019,
acer_sphynx_2021,deveci_multi-jagged_2016}. Our dominance-counters may also have
applications for other geometric load-balancing problems
\cite{nicol_rectilinear_1994, saule_load-balancing_2012, yasar_symmetric_2020}.

\bibliographystyle{IEEEtran}
\bibliography{IEEEabrv,article}

%

\vfill
\begin{IEEEbiography}[{\includegraphics[width=1in,height=1.25in,clip,keepaspectratio]{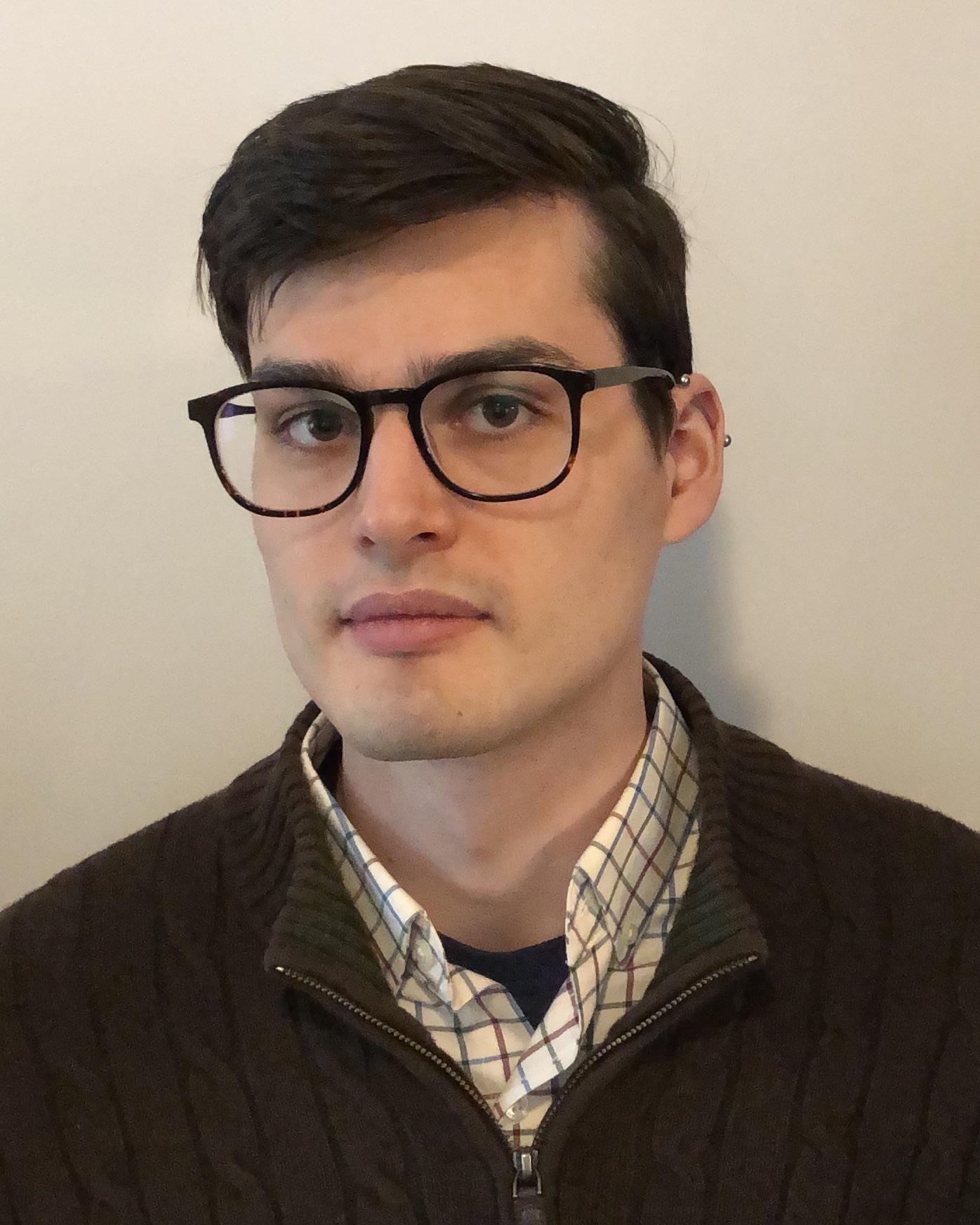}}]{Peter Ahrens}
    is Department of Energy Computational Science Graduate Fellow working
    towards a Ph.D. in computer science at the Massachusetts Institute of
    Technology under the supervision of Professor Saman Amarasinghe. Peter
    completed his B.S. in computer science and minor in mathematics at the
    University of California, Berkeley. His research interests include graph
    algorithms, compilers, numerical analysis, and high performance computing
    with applications in scientific computing.
\end{IEEEbiography}





\cleardoublepage
\appendices

\section{Lazy Probe Pseudocode}\label{app:lazy}
Algorithm \ref{alg:lazybisectpartition} details pseudocode for the lazy probe
algorithm described in Section \ref{sec:partitionmonotonic}. The algorithm computes the
atoms needed for all of our cost functions in primary or symmetric
partitioning settings. 

\setcounter{algorithm}{5}
\begin{algorithm}[Lazy BISECT Partitioner]
    Given a cost function(s) $f$ defined on the atoms
        $x_{\textbf{row}} = |\pi_k|$,
        $x_{\textbf{entry}} = \sum_{i \in \pi_k}|v_i|$,
        $x_{\Delta_{\textbf{entry}}} = \sum_{i \in \pi_k}\max(|v_i| - w_{\min}, 0)$,
        $x_{\textbf{incident}} = |\cup_{i \in \pi_k} v_i|$,
        $x_{\textbf{local}} = |\cup_{i \in \pi_k} v_i \cap \phi_{k}|$, and
        $x_{\textbf{diagonal}} = |\cup_{i \in \pi_k} v_i \cup \pi_{k}|$,
    which is monotonic increasing in $\pi_k$, find a contiguous $K$-partition $\Pi$ which minimizes
    \[
        c = \max\limits_{k} f_k(\pi_k)
    \]
    to a relative accuracy of $\epsilon$ within the range $c_{\low} \leq c \leq c_{\high}$,
    if such a partition exists.

    This algorithm differs from Algorithm \ref{alg:bisectpartition} only in
    the probe function, so we only describe the new \textproc{LazyProbe}
    pseudocode. We assume that $c_{\low} \geq \max\limits_{k}
    f_k(\emptyset)$.

    \label{alg:lazybisectpartition}
    \begin{algorithmic}
        \Function{LazyProbe}{$c$}
            \State $hst \gets \text{pre-allocated length $n$ vector filled with zero}$
            \State $drt \gets \text{pre-allocated length $K$ vector filled with zero}$
            \State $lcl \gets \text{pre-allocated length $K$ vector}$
            \State $i \gets 1$, $k \gets 1$, $x... \gets 0$
            \For{$i' \gets 1, 2, ..., m$}
                \State $x_{\textbf{row}} \gets x_{\textbf{row}} + 1$
                \State $x_{\textbf{entry}} \gets x_{\textbf{entry}} + pos_{i' + 1} - pos_{i'}$
                \State $x_{\Delta_{\textbf{entry}}} \gets x_{\Delta_{\textbf{entry}}} + \max(pos_{i' + 1} - pos_{i'} - w_{\min}, 0)$
                \For{$q \gets pos_{i'}, pos_{i'} + 1, ..., pos_{i' + 1} - 1$}
                    \State $j \gets idx_{q}$
                    \State Set $k'$ such that $j \in \phi_{k'}$
                    \If{$drt_{k'} < i'$}
                        \State $lcl_{k'} \gets 0$
                    \EndIf
                    \State $lcl_{k'} \gets lcl_{k'} + 1$
                    \State $drt_{k'} \gets i'$
                    \If{$hst_{j} < i$}
                        \State $x_{\textbf{incident}} \gets x_{\textbf{incident}} + 1$
                        \If{$k' = k$}
                            \State $x_{\textbf{local}} \gets x_{\textbf{local}} + 1$
                        \EndIf
                    \EndIf
                    \If{$(j < i$ \MyOr $i \leq j)$ \MyAnd $hst_{j} < i$}
                        \State $x_{\textbf{diagonal}} \gets x_{\textbf{diagonal}} + 1$
                    \EndIf
                    \State $hst_j \gets i'$
                \EndFor
                \If{$i' \leq n$ \MyAnd $hst_{i'} < i$}
                    \State $x_{\textbf{diagonal}} \gets x_{\textbf{diagonal}} + 1$
                \EndIf
                \While{$f(x..., k) > c$}
                    \If{$k = K$}
                        \State \Return $\false$
                    \EndIf
                    \State $s_{k + 1} \gets i'$
                    \State $i \gets i'$
                    \State $k \gets k + 1$
                    \State $x_{\textbf{row}} \gets 1$
                    \State $x_{\textbf{entry}} \gets pos_{i' + 1} - pos_i'$
                    \State $x_{\Delta_{\textbf{entry}}} \gets \max(pos_{i' + 1} - pos_i' - w_{\min}, 0)$
                    \State $x_{\textbf{incident}} \gets pos_{i' + 1} - pos_i'$
                    \State $x_{\textbf{diagonal}} \gets pos_{i' + 1} - pos_i'$
                    \If{$i' \leq n$ \MyAnd $hst_{i'} < i'$}
                        \State $x_{\textbf{diagonal}} \gets x_{\textbf{diagonal}} + 1$
                    \EndIf
                    \State $x_{\textbf{local}} \gets 0$
                    \If{$drt_{k} = i'$}
                        \State $x_{\textbf{local}} \gets x_{\textbf{local}} + lcl_k$
                    \EndIf
                \EndWhile
            \EndFor
            \While{$k \leq K$}
                \State $s_{k + 1} \gets m + 1$
            \EndWhile
            \State \Return $\true$
        \EndFunction
    \end{algorithmic}
\end{algorithm}

\end{document}


\subsection{Sparse Prefix Sums}


Like Chazelle's algorithm, our dominance counting algorithm can be extended
to compute prefix sums of non-Boolean values over a bounded integer grid. At
each level, we simply need to store the values associated with the current
ordering of points. When we query the structure, in the same way we count
$c$, the number of points in our bucket where $d' < d$, we would also need to
sum the values associated with these points. In order to maintain the same
asymptotic runtime, this necessitates the use of an array similar to $cnt$
which records a prefix sum of the values of previous points (rather than
their count) in the ordering at each level. These modifications would not
increase the runtime of construction or queries, but would increase the
storage by a factor of $H$, so that the storage requirement for a sparse
prefix sum would be (up to addition of a constant),
\begin{equation}\label{eq:chazelle_generic_storage}
    m + n + N(1 + H + H 2^{b - b'}).
\end{equation}
We also implemented this algorithm in our codebase.